\UseRawInputEncoding
\documentclass[bibyear]{aa}
\usepackage{graphicx}                
\usepackage{txfonts}                 
\newcommand{\aAur}{$\alpha$\,Aur}
\newcommand{\bAur}{$\beta$\,Aur}
\newcommand{\epsAur}{$\varepsilon$\,Aur}
\newcommand{\zAur}{$\zeta$\,Aur}
\newcommand{\etaAur}{$\eta$\,Aur}
\newcommand{\tAur}{$\theta$\,Aur}
\newcommand{\nAur}{$\nu$\,Aur}
\newcommand{\iAur}{$\iota$\,Aur}

\newcommand{\bTau}{$\beta$\,Tau}
\newcommand{\kCet}{$\kappa^1$\,Cet}
\newcommand{\HR}{V711\,Tau}
\newcommand{\iqAur}{IQ\,Aur}

\newcommand{\Halpha}{H$\alpha$}

\newcommand{\kms}{km\,s$^{-1}$}
\newcommand{\ms}{m\,s$^{-1}$}

\begin{document}

\title{BRITE photometry and STELLA spectroscopy of bright stars in Auriga: Rotation, pulsation, orbits, and eclipses\thanks{Based on data obtained with the BRITE-Constellation satellite and the STELLA robotic telescope in Tenerife. BRITE (BRIght Target Explorer) Constellation was designed, built, launched, operated, and supported by the Austrian Research Promotion Agency (FFG), the University of Vienna, the Technical University of Graz, the University of Innsbruck, the Canadian Space Agency (CSA), the University of Toronto Institute for Aerospace Studies (UTIAS), the Foundation for Polish Science \& Technology (FNiTP MNiSW), and National Science Centre (NCN).} \thanks{The data used in this paper is available in electronic form at the CDS  via  anonymous  ftp  to cdsarc.u-strasbg.fr (130.79.128.5)  or  via http://cdsweb.u-strasbg.fr/cgi-bin/.}}

\author{K. G. Strassmeier\inst{1,2}, T. Granzer\inst{1}, M. Weber\inst{1}, R. Kuschnig\inst{3,4}, A.~Pigulski\inst{5}, A.~Popowicz\inst{6}, A. F. J. Moffat\inst{7}, \\ G. A. Wade\inst{8}, K. Zwintz\inst{9}, \and G. Handler\inst{10}}

\institute{
    Leibniz-Institute for Astrophysics Potsdam (AIP), An der Sternwarte 16, D-14482 Potsdam, Germany; \\ \email{kstrassmeier@aip.de},
    \and
    Institute for Physics and Astronomy, University of Potsdam, D-14476 Potsdam, Germany
    \and
    Institute for Astrophysics, University of Vienna, T\"urkenschanzstrasse 17, A-1180 Vienna, Austria
    \and
    Institut f\"ur Kommunikationsnetze und Satellitenkommunikation, Technical University Graz, Inffeldgasse 12, A-8010 Graz, Austria
    \and
    Instytut Astronomiczny, Uniwersytet Wroc{\l}awski, Kopernika 11, 51-622 Wroc{\l}aw, Poland
    \and
    Faculty of Automatic Control, Electronics and Computer Science, Akademicka 16, 44-100 Gliwice, Poland
    \and
    D\'ept. de Physique and Centre de Recherche en Astrophysique du Qu\'ebec (CRAQ), Universit\'e de Montr\'eal, C.P.\,6128, Succ. Centre-Ville, Montr\'eal, H3C\,3J7, Canada
    \and
    Department of Physics and Space Science, Royal Military College of Canada, P.O. Box 17000, Station Forces, ON, K7K4B4, Canada
    \and
    Institut f\"ur Astro- und Teilchenphysik, Universit\"at Innsbruck, Technikerstra{\ss}e 25, A-6020 Innsbruck, Austria
    \and
    Nicolaus Copernicus Astronomical Center, Polish Academy of Sciences, ul. Bartycka 18, 00-716, Warsaw, Poland
}

\date{Received ... ; accepted ...}

\abstract{Knowing rotational and pulsational periods across the Hertzsprung-Russell diagram is of top priority for understanding stellar activity as a function of time.}{We aim to determine periods for bright stars in the Auriga field that are otherwise not easily accessible for ground-based photometry.}{Continuous photometry with up to three BRITE  satellites was obtained for 12 targets and subjected to a period search. Contemporaneous high-resolution optical spectroscopy with STELLA was used to obtain radial velocities through cross correlation with template spectra as well as to determine astrophysical parameters through a comparison with model spectra.}{The Capella red light curve was found to be constant over 176 days with a root mean square of 1\,mmag, but the blue light curve showed a period of 10.1$\pm$0.6\,d, which we interpret to be the rotation period of the G0 component. From STELLA we obtained an improved orbital solution based on 9600 spectra from the previous 12.9 yr. We derive masses precise to $\approx$0.3\%\ but 1\%\ smaller than previously published. The BRITE light curve of the F0 supergiant \epsAur\ suggests 152\,d as its main pulsation period, while the STELLA radial velocities reveal a clear 68 d period. An ingress of an eclipse of the \zAur\ binary system was covered with BRITE and a precise timing for its eclipse onset derived. A possible 70 d period fits the proposed tidal-induced, nonradial pulsations of this ellipsoidal K4 supergiant. \etaAur\ is identified as a slowly pulsating B (SPB) star with a main period of 1.29\,d and is among the brightest SPB stars discovered so far. The rotation period of the magnetic Ap star \tAur\ is detected from photometry and spectroscopy with a period of 3.6189\,d and 3.6177\,d, respectively, likely the same within the errors. The radial velocities of this star show a striking non-sinusoidal shape with a large amplitude of 7\,\kms . Photometric rotation periods are also confirmed for the magnetic Ap star \iqAur\ of 2.463\,d and for the solar-type star \kCet\ of 9.065\,d, and also for the B7 HgMn giant \bTau\ of 2.74\,d. Revised orbital solutions are derived for the eclipsing SB2 binary \bAur , which replaces the initial orbit dating from 1948 for the 27-year eclipsing SB1 \epsAur , and for the RS~CVn binary \HR,\ for which a spot-corrected orbital solution was achieved. The two stars \nAur\ and \iAur\ are found to be long-term, low-amplitude RV and brightness variables, but provisional orbital elements based on a period of 20 yr and an eccentricity of 0.7 could only be extracted for  \nAur. The variations of \iAur\ are due to oscillations with a period of $\approx$4~yr.}{}

\keywords{Stars: binaries -- stars: rotation -- stars: activity -- stars: fundamental parameters -- stars: pulsations -- stars: late-type -- stars: eclipsing}

\authorrunning{Strassmeier, Granzer, Weber et al.}

\titlerunning{BRITE and STELLA}

\maketitle
\section{Introduction}

Photometric time series of stars have been an essential tool for stellar astrophysics for over a century (e.g., Guthnick \cite{gut}, Stebbins \cite{stebb}), but really became top gear with the advent of ultra-precise photometry from space. Time series can resolve the complex light curves due to stellar pulsations or follow the ever-changing tracers that govern stellar magnetic activity. Among the physical parameters that can be derived from photometry is stellar rotation and its subtle latitudinal dependence called differential rotation (e.g., Hall \cite{hall}, Lanza et al. \cite{lanza}). Rotation drives the interaction of turbulent plasma motions and large-scale shearing forces in the convective envelope of cool stars and thus can be an important physical quantity for stellar evolution. By determining rotation periods throughout the Hertzsprung-Russell diagram (H-R diagram), we can determine stellar ages through gyrochronology (e.g., Barnes \& Kim \cite{barnes}) and indirectly constrain the stellar dynamo itself (e.g., R\"udiger \& Hollerbach \cite{rue:hol}, Brun \& Browning \cite{bb}).

Ground-based observations of very bright stars with very small photometric amplitudes however are a painstaking challenge, if possible at all. Among other aficionados, we had obtained differential photometry of the $V$=0.1-mag \aAur\ = Capella system, as well as \epsAur, both from 1996-2000. These data were taken with one of the two 0.75 m Vienna Automatic Photoelectric Telescopes (APTs) at Fairborn Observatory in southern Arizona equipped with a photomultiplier tube, a 5 mag neutral density filter, and a narrowband \Halpha\ filter. We believe we had detected photometric periods for both stellar components of Capella (Strassmeier et al. \cite{cap1}, and references therein). If interpreted as rotation periods, it appeared that the brighter G8 giant has a rotation period of $\approx$106~days and the fainter, but more active, early-G giant of $\approx$8.6~days. However, the quality and sampling of the ground-based data was such that the periods had uncertain error bars (if known at all) or could even be questioned as artifacts; the detection of differential rotation was completely out of reach.

While we generally measure stellar parameters to a better precision when the target is brighter, this is not the case for stellar rotation periods. Photometric rotation periods are typically 10 to 100 times more precise than spectroscopically determined periods, but usually fail for very bright stars owing to detector saturation and nonlinearities at high photon flux and other technical shortcomings or the simple lack of appropriate comparison stars. Employing small aperture telescopes does not do the trick because ground-based photometry is scintillation-noise limited. This is not so in space. Therefore, several space missions were flown in the past that had centimeter-sized telescopes on board to do bright-star photometry. Most of these were simply star trackers for bigger satellites such as HST/FGS (Zwintz et al. \cite{fgs}), INTEGRAL/OMC (Alfonso-Garcon et al. \cite{omc}), or for the WIRE infrared spacecraft (Bruntt \& Buzasi \cite{wire}). Some of these space missions were dedicated to stellar photometry such as Coriolis/SMEI (Eyles et al. \cite{smei}) or MOST (Walker et al. \cite{most}). The most recent addition to this flock is the fleet of five nano satellites called BRITE-Constellation (Weiss et al. \cite{brite}) launched by Canada, Austria, and Poland between 2013 and 2014.

In this paper, we present BRITE photometry of nine very bright stars within the Auriga-Perseus field (BRITE ID: 20-AurPer-I-2016) from 2016-2017, for two stars in the Cetus-Eridanus field (21-CetEri-I-2016) as well as one star at the border of Auriga and Taurus in the 31-Tau-I-2017 and 48-OriTau-II-2019 fields. For nine of these stars, we also present accompanying contemporaneous high-resolution, time-series spectroscopy. The basic science aim is the characterization of stellar rotation and its surface latitude dependence, but there are also three eclipsing binaries in the sample. We regret to report only a marginal period detection for Capella from BRITE photometry as a result of the combination of a very small intrinsic amplitude together with comparably large detector-related photometric noise. However, the time coverage and sampling for most of these targets is the best ever obtained. One of the targets (\nAur ) is likely a new spectroscopic binary from its radial velocity (RVs) variations and one (\etaAur ) is identified as a slowly pulsating B (SPB) variable. This source is among the brightest SPB stars discovered so far. As a highlight, we present a new improved spectroscopic orbit of the Capella binary system. The targets are briefly summarized in Sect.~\ref{S2} and the observations described in Sect.~\ref{S3}. Section~\ref{S4} is our data analysis of both the photometry and the spectroscopy, and Sect.~\ref{S5} is a summary of the results.

\begin{figure}[tbh]
\includegraphics[clip,angle=0,width=89mm]{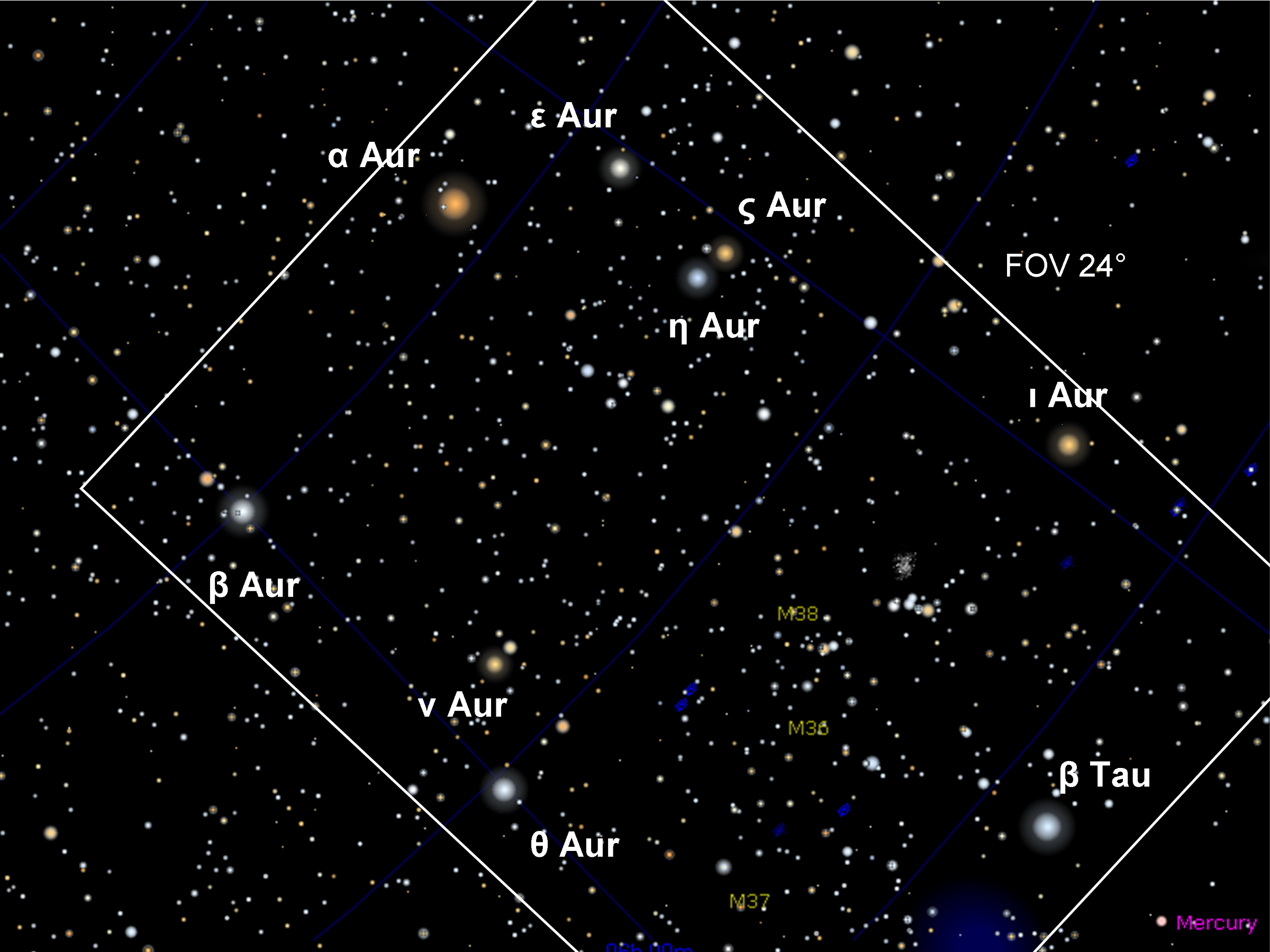}
\caption{Auriga field as seen from BRITE. This field contains nine variable stars brighter than fourth magnitude. The brightest is the active binary Capella, but each bright star represents a distinctive variable-star class of its own with a large variety of astrophysical challenges.}\label{F1}
\end{figure}

\begin{table*}[!tbh]
\caption{BRITE target list.}\label{T1}
\begin{flushleft}
\begin{tabular}{llllllllllll}
\hline\hline
\noalign{\smallskip}
Name           &  HR  & HD    & $B$ & $V$ & $R$ & $I$ &  Spec. type &\#\,bib&  Distance &  Ref & Target characteristic \\
               &      &       & \multicolumn{4}{c}{(mag)} &         &       & (pc)      &      & \\
\noalign{\smallskip}\hline\noalign{\smallskip}
\aAur          & 1708 & 34029 & 0.88 & 0.08 &-0.52 &-0.96 &G0III+G8III   &1123&  13.1$\pm$0.06   &(1) & Binary with two giant stars\\
\bAur          & 2088 & 40183 & 1.93 & 1.90 & 1.82 & 1.83 &A1IV-Vp       & 396&  24.87$\pm$0.14  &(1) & Detached eclipsing system\\
\epsAur        & 1605 & 31964 & 3.53 & 2.99 & 2.47 & 2.02 &F0Ia          & 550&  415$\pm$73      &(2) & 27-yr eclipsing system\\
\zAur          & 1612 & 32069 & 4.97 & 3.75 & 2.62 & 1.75 &K4Ib+B5V      & 431&  241$\pm$16      &(1) & Atmospheric eclipses\\
\etaAur        & 1641 & 32630 & 3.00 & 3.18 & 3.23 & 3.40 &B3V           & 352&  63.1$\pm$1.9    &(2) & CP B star\\
\tAur          & 2095 & 40312 & 2.54 & 2.62 & 2.62 & 2.68 &A0VpSi        & 259&  50.8$\pm$1.5    &(1) & Magnetic CP star \\
\nAur          & 2012 & 39003 & 5.09 & 3.95 & 3.15 & 2.59 &K0.5IIICN0.5  & 183&  66.0$\pm$1.4    &(2) & Radial velocity standard\\
\iAur          & 1577 & 31398 & 4.22 & 2.69 & 1.63 & 0.81 &K3II-III      & 260&  151$\pm$8       &(1) & Hybrid giant and susp. variable\\
\bTau          & 1791 & 35497 & 1.52 & 1.65 & 1.66 & 1.76 &B7III         & 344&  41.0$\pm$0.6    &(1) & CP HgMn star\\
\kCet          & 996  & 20630 & 5.52 & 4.85 & 4.27 & 3.91 &G5V           & 797&  9.14$\pm$0.02   &(1) & Spotted BY Dra variable \\
\HR            & 1099 & 22468 & 6.63 & 5.71 & 5.40 & 5.00 &K2IV+G5IV     &1110&  29.63$\pm$0.07  &(2) & Spotted RS\,CVn binary \\
\iqAur         & 1732 & 34452 & 5.20 & 5.37 & \dots & \dots & A0p        &269 & 145$\pm$3.5      &(2) & Magnetic CP star \\
\noalign{\smallskip}
\hline
\end{tabular}
\tablefoot{\#\,bib is the number of papers on this star and refers to the SAO/NASA ADS system dated 5/2019. Ref.: (1) Hipparcos (van\,Leeuwen \cite{hip}); (2) Gaia-DR2 (Gaia~\cite{gaia-DR2}).}
\end{flushleft}
\end{table*}

\section{Target sample}\label{S2}

Table~\ref{T1} gives an overview of the target sample. It also lists a CDS/Simbad entry of the number of papers that contain information on these stars. This and the following description are meant as a quick orientation.

\object{$\alpha$ Aur}. Capella is the brightest chromospherically active binary star in the sky. Its absolute properties are well known because of its precise astrometric and spectroscopic orbits (e.g., Torres et al. \cite{torres1}, Weber \& Strassmeier \cite{capella}, Torres et al. \cite{torres}, Takeda et al. \cite{takeda}). It consists of two giants in a 104-day non-eclipsing orbit: one star still in the Hertzsprung gap and of spectral type G0III, and the other in the clump region and of spectral type G8III. The orbital solution of this system provides the most precise masses for giant stars at two very specific positions in the H-R diagram. Just recently, high-resolution spectroscopy revised the $^{12}$C/$^{13}$C ratio of the primary and finally put it into agreement with current evolutionary tracks (Sablowski et al. \cite{sab}). Capella is also an interactive binary system. Shcherbakov et al. (\cite{shch}) discovered a modulation of the He\,{\sc i} 1.08\,$\mu$m equivalent width with the 104-day orbital period. This variation and its periodicity was later confirmed by Katsova \& Shcherbakov (\cite{kats}). The latter authors also concluded that the main He\,{\sc i} absorption must originate in the chromosphere of the quiescent cool component, while the residual helium emission and the highly ionized iron lines are thought to be connected with a magnetized wind that originates from plages on the active hot component and forms a shock wave in the corona of the cool component. Therefore, chromospheric emission is expected from both components and  both components appeared photometrically variable on a milli-mag level at \Halpha\ (Strassmeier et al.~\cite{cap1})
.

\object{$\beta$ Aur}.  This source is a detached double-lined (SB2) eclipsing binary used to determine nonlinear limb-darkening coefficients for both of its components based on high-precision space photometry from the WIRE satellite (Southworth et al. \cite{south}). The system consists of two narrow-lined A1 subgiants in a 3.96 d orbit. The previously best spectroscopic orbital solution is still that from Smith (\cite{smith}).

\object{$\varepsilon$ Aur}. This source is a single-lined (SB1) spectroscopic and eclipsing binary system with a two-year-long eclipse every 27.1 yr (e.g., Stefanik et al. \cite{stef}). The system is unique in many respects. It is among the brightest eclipsing binaries in the sky and has one of the longest orbital periods known among any of the eclipsing binaries. It is composed of a semi-regular pulsating F0\,Iab supergiant in orbit with a massive upper main-sequence star that is enshrouded in a dusty disk and not directly seen. The recent eclipse took place in 2009-2011 and a suite of modern instrumentation was pointed to this star (see, e.g., Griffin \& Stencel \cite{gri:ste}, Strassmeier et al. \cite{epsaur}). Interferometric imaging with the CHARA array resolved the opaque dusty disk as it passed in front of the F0 star during the eclipse (Kloppenborg et al. \cite{klopp}). The light curve is dominated by the pulsations of the supergiant.

\object{$\zeta$ Aur}. This source is also a long-period SB1 eclipsing binary with a K supergiant and a mid-B main-sequence star. An orbit was determined by Griffin (\cite{griff}) with a period of 972 days and confirmed by Eaton et al. (\cite{eat}). Of note are the eclipses of the B star when it shines from behind the chromosphere of the K star, events which last for some days either side of the eclipse. The B star has been used to probe the atmosphere of the K supergiant as a function of height above the limb. Griffin (\cite{griff}) also noted that the rotation period of the K supergiant appears to be close to the orbital period. We note that this binary has two HD numbers (HD\,32068, HD\,32069).

\object{$\eta$ Aur}. This source is a chemically peculiar (CP) B3 main-sequence star. It appears to be a single star from its constant RVs. Crawford et al. (\cite{craw}) lists this star as one of their photometric standard stars, while Simbad lists it as a suspected variable based on Guthnick \& Pavel (\cite{gut:pav}) and its entry in the NSV catalog (Kazarovets et al. \cite{nsv}).

\object{$\theta$ Aur}. This source is a well-known, chemically peculiar A0p star and the brighter member of the close visual binary ADS\,4566 (the B component is 4.5\,mag fainter; Edwards \cite{edwards}). The surface elemental distribution of this A0p star had been spatially resolved by means of Doppler imaging by several authors (e.g., Rice et al. \cite{rice},  Kochukhov et al. \cite{koch}). \tAur\ exhibits 0\fm035 photometric amplitude variations with a period of 3.6188\,d in phase with effective magnetic field measurements (Adelman \cite{adel}). The light minima are broad and require the contributions of at least two surface features. The most recent and precise period determination is from a combination of photometric and spectroscopic tracers (3.61866$\pm$0.00001\,d; Krticka et al. \cite{krt}).

\object{$\nu$ Aur}. Numerous photometric observations of this star exist but were mostly taken in the context of standard-star all-sky photometry (e.g., Oja \cite{oja}, Eggen \cite{eggen}). Adelman (\cite{adel01}) listed it among the least variable stars in the Hipparcos catalog with a standard deviation of only 0.5\,mmag from 69 visits. The Aitken Double Star catalog lists this source as the brighter component of a wide binary. Beavers \& Eitter (\cite{fick}) obtained a series of RVs and found $\nu$ Aur to be constant with a standard deviation of 1.1~\kms\ from 24 measures and suggested that this star is effectively a single star.

\object{$\iota$ Aur}. This star was suspected for variability by Guthnick \& Prager (\cite{gut:pra}) and Simbad/CDS titles it a variable star, yet Adelman (\cite{adel01}) listed it among the least variable stars in the Hipparcos catalog with a 1.4~mmag root mean square (rms) from 39 visits. Its entry came from the NSV catalog (Kazarovets et al. \cite{nsv}). The star is a single K3 bright giant, which is also called a hybrid giant because of its location in the H-R diagram. From the RV monitoring program at Lick\ Observatory, Hekker et al. (\cite{hek}) reported periods of 767\,d and 1586\,d. L\'ebre et al. (\cite{lebre}) determined its lithium abundance and gave a summary of its rotational properties.

\object{$\beta$ Tau}. This source is also named $\gamma$~Aur. No continuous photometry of this very bright star seems to exist. However, it has received substantial attention after it was listed among the few mercury-manganese stars by Adelman et al. (\cite{adel06}).

\object{$\kappa^1$\,Cet}. This source is a single solar-like star with a spotted photosphere and an active chromosphere. Its light variability is rotationally modulated with an average $\approx$9 d period. Rucinski et al. (\cite{ruc}) and  Walker et al. (\cite{walker}) were able to determine its differential surface rotation from the period range observed by MOST. These authors had isolated the photometric signatures from individual spots and obtained rotation periods from 8.7 d to more than 9.3\,d. Spectroscopic observations of the chromospheric Ca\,{\sc ii} K-line emission by Shkolnik et al. (\cite{shk}) were modulated with a period of about 9.3\,d.

\object{V711\,Tau = HR\,1099}. This source is one of the most active and most well-studied RS\,CVn binaries. It consists of two subgiants in a close two-day orbit with a double-lined spectrum. The K1 component is the active star and more rapidly rotating, the G5-component comparably inactive with narrow spectral lines. The light variations are attributed to the rapidly rotating and heavily spotted K star (see, e.g., Vogt et al. \cite{vogt}, Strassmeier \& Bartus \cite{str:bar}). It is one of the faintest stars observed by BRITE.

\object{IQ\,Aur = HR\,1732}. This star is among the hottest known Ap Si star (Bohlender et al. \cite{bohl}). \iqAur\ was discovered to be an X-ray emitter and explained by a magnetically confined wind-shock model (Babel \& Montmerle \cite{bab:mon}). Its period of 2.466\,d is the rotation period of the star and was initially obtained from spectroscopic observations by Deutsch (\cite{deu}) and confirmed by many others in the meantime. Rakos (\cite{rak}) obtained the same period from intermediate-band photometric observations, which was again confirmed later by Pyper \& Adelman (\cite{pyp:ade}) among others. However, the photometric amplitude never exceeded 0\fm02 and thus the periods from photometry remain utterly uncertain.

\begin{table*}[!tbh]
\caption{STELLA observing log and results.}\label{T2}
\begin{flushleft}
\begin{tabular}{llllllllll}
\hline\hline
\noalign{\smallskip}
Name         & Time range & $N_{\rm Spec}$ & $t_{\rm exp}$ & S/N    & $\langle RV\rangle$ & $T_{\rm eff}$ & $\log g$ & $v\sin i$ & [M/H] \\
             & (ddmmyy) &                & (s)           & range  & (\kms )       & (K)           & (cgs)    & (\kms )   & (solar) \\
\noalign{\smallskip}\hline\noalign{\smallskip}
\aAur$^1$    & ongoing & \dots & 8 & 220-700 & SB2 & 5150 & 1.7 & 4.5 & --0.55 \\
\bAur$^2$    & 260516-020417 & 127 & 60 & 120-510 &SB2 & \dots & \dots & 29.3$\pm$1.3& \dots \\
\epsAur      & ongoing & \dots & 120 & 100-360 & --3 var & 7600$\pm$110 & 0.0$\pm$0.3 & 32$\pm$4 & 0.0$\pm$0.4 \\
\etaAur      & 110416-010417 & 109 & 60 & 80-270 & 10.2 var & \dots & \dots & 66$\pm$14 & \dots \\
\tAur        & 040916-300317 & 163 & 250 & 110-680 & 30.5 var & \dots & \dots & 54.6$\pm$1.0 & \dots \\
\nAur        & 130916-020417 & 103 & 60 & 80-220 & 13.0 var & 4610$\pm$50 & 2.2$\pm$0.1 & 2.7$\pm$0.2 & --0.12$\pm$0.03 \\
\iAur        & 080416-010417 & 104 & 60 & 100-400 & 18.1 var & 4130$\pm$60 & 1.25$\pm$0.3 & 5.5$\pm$1.0 & --0.20$\pm$0.03 \\
\kCet        & 091216-180217 & 28 & 1800 & 200-670 & 18.95$\pm$0.03 & 5664$\pm$13 & 4.46$\pm$0.02 & 4.6$\pm$0.2 & --0.12$\pm$0.03 \\
\HR$^1$      & 091216-030417 & 116 & 1800 & 100-480 &SB2 & 4420 var &2.5$\pm$0.7 & 45.0$\pm$1.0 & --0.66$\pm$0.33\\
\noalign{\smallskip}
\hline
\end{tabular}
\tablefoot{$\langle RV\rangle$ are mean values. Time range is in standard UT time. ``Ongoing'' means that the target is continued to be observed whenever possible. The notation {\sl var} indicates that the value is systematically changing. Error bars are rms. S/N is given per pixel; $^1$for the primary; $^2$same value for both components. }
\end{flushleft}
\end{table*}

\section{Observations}\label{S3}

\subsection{BRITE photometry}

Observations of the Aur-Per field (Fig.~\ref{F1}) took place from September~9, 2016 to March~9, 2017. Two of the five satellites were used: BRITE-Toronto (BTr) for red light and the full time of 176~days, and BRITE-Lem (BLb) for blue light but only for the first 35~days and the last 27~days. The two targets in the Cet-Eri field (\kCet\ and \HR ) were observed with just a single satellite (BRITE-Heweliusz; BHr) in the red and for 90~days between JD 2,457,668 and 2,457,757. One satellite orbit is approximately 99 minutes and observations were made for periods of 5-10 minutes per orbit when the field was visible. The BRITE telescopes have 3~cm apertures recording ``red'' images at 550-700\,nm and ``blue'' images at 390-460\,nm. The filters are similar to the Geneva~$B$ and Sloan~$r$ filters. The field of view of the satellites is 24\degr$\times$20\degr. Pixel sampling on their Kodak KAI-11002M CCDs is 27\arcsec/pixel.

The BRITE data are obtained in different satellite configurations, which are numbered consecutively (e.g., BTr1, BTr2 a.s.o.). The BTr observations were conducted with two types of setups: one for the brightest stars to avoid saturation and the other setup for the fainter objects to enhance the signal to noise. The measurements were made with 0.2\,s and 3\,s exposures for the two types of setups, respectively. Some stars appear in both short and long exposure time setups. Severe issues with the stability of BTr pointing caused many outliers for the whole time range except for the start (first ten orbits) and the end (after TJD\footnote{Truncated Julian date: TJD = HJD -- 2,456,000.0.}~1665.6). Orbits between TJD\,1665.9-1667.6 and 1665.3-1665.6 were rejected. Also, in the 20-Aur-Per field, the Moon got close enough for at least one part of the field and caused $\approx$28\,d regular increases of the background. These appear as periodic broad bumps in the raw light curves. The BLb experienced a large data gap (TJD\,1692-1789) due to its inability to achieve fine pointing. An eclipse-like event appeared for all stars at TJD\,1662.166 with strong variations starting around TJD\,1660 and continued until TJD\,1680. The overall photometric precision was below expectation owing to a continuous satellite drift and we decided not to use most of the BLb data. Three setups with BHr were used (BHr photometry is red sensitive as BTr) and all three appeared appropriate.

\begin{figure}[tbh]
\includegraphics[angle=0,width=95mm,clip]{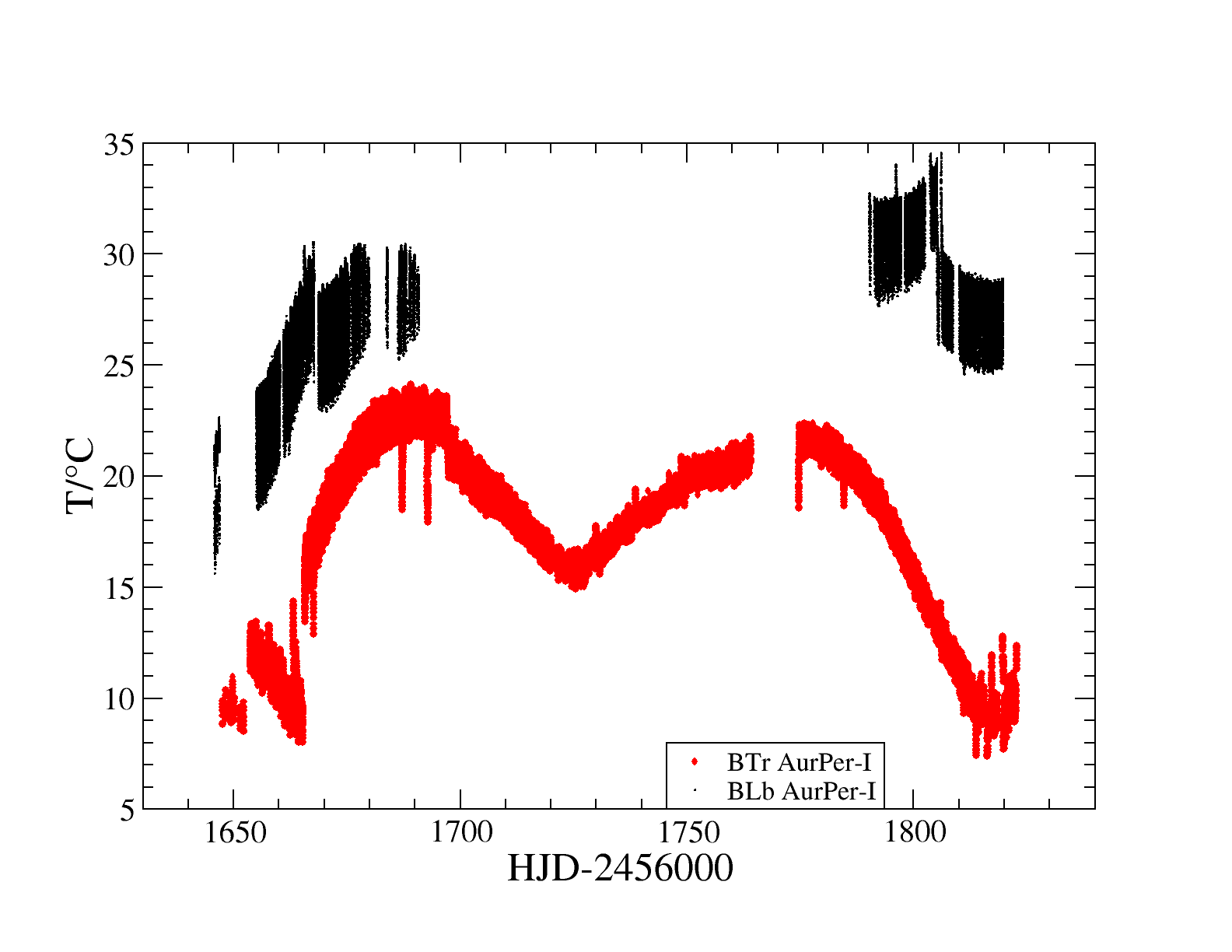}
\caption{CCD temperature in degree Celsius on board BRITE-Toronto (red points; BTr) and BRITE-Lem (black points; BLb) during the observing period of Capella in 2016/17. }\label{Fccd}
\end{figure}

Because of the problems with high dark current introduced by the proton radiation (Popowicz \cite{popo1}), a chopping mode of observing was introduced. No bias, dark, and background corrections are necessary to
do differential photometry. The main remaining data-reduction step is the definition of the optimal apertures and extraction of the flux within these apertures (Popowicz et al. \cite{popo}). The data resulting from this pipeline reduction were provided by the BRITE consortium, which eliminated on average 30\%\ of the raw data. Further details about the detector and data acquisition of BRITE Constellation are described by Pablo et al. (\cite{pablo16}). The CCD temperature variations shown in Fig.~\ref{Fccd} and other instrumental effects were described by Pigulski et al. (\cite{pig18a}) and were removed with the de-correlation technique presented in detail in the BRITE Cookbook 2.0 (Pigulski et al. \cite{pig18b}). We note that we co-added all CCD frames per orbit (orbital binning) to increase the photometric precision. An average error bar for the binned Capella data from BTr was 0.4~mmag.

\begin{table*}[t]
\caption{Photometric periods from BRITE.}\label{T3}
\begin{flushleft}
\begin{tabular}{lllllllll}
\hline\hline
\noalign{\smallskip}
Name & Satellite& rms$_{\rm all}$ & $P$ & $\Delta P$& FAP           & Amp         & $T_0$     & rms$_{\rm fit}$ \\
     &          & (mmag)          & (d) & (d)       & (-)       & (mmag)    & (TJD)     & (mmag) \\
\noalign{\smallskip}\hline\noalign{\smallskip}
\aAur   &BTr    &   1.02        &  4.38         & 0.11      & 0.014     &   0.36  &  1813.1686    &   1.01\\
        &BLb    &   0.96        & 10.1018       & 0.57      & 0.0001    &   0.89  &  1803.9445    &   0.90\\
\bAur   &BTr    &  18.06        &  3.9601       & 0.0001        & $<$10$^{-20}$&  82.06  &  1816.9535    &   2.80\\
        &BLb    &  16.17        &  3.9601       & 0.0001        & $<$10$^{-20}$&  81.24  &  1816.9535    &   1.87\\
\epsAur &BTr    &  31.05        & 152.661       & 0.064     & $<$10$^{-20}$     & 121.68&  1515.3200 &   4.78\\
        &BLb    &  28.58        &   \dots   &   \dots   &   \dots   &   \dots   &   \dots   &   \dots   \\
\zAur   &BTr    &  26.83        & 69.949        & 0.037     & $<$10$^{-20}$&  95.72  &  1659.2235    &   9.52\\
        &BLb    &  41.70        & \dots         & \dots         & \dots          & \dots  & \dots & \dots   \\
\etaAur &BTr    &   5.05        &  1.2891       & 0.0001        & $<$10$^{-20}$&   7.34  &  1819.8943    &   4.34\\
        &BLb    &   8.54        &  1.2777       & 0.0014        & 2.09 $10^{-18}$       &  10.99  &  1819.8943    &   7.56\\
\tAur   &BTr    &  12.30        &  3.6189       & 0.0001        & $<$10$^{-20}$&  34.71  &  1817.6543    &   1.24\\
        &BLb    &  12.19        &  3.6192       & 0.0002        & $<$10$^{-20}$&  34.02  &  1817.6543    &   2.66\\
\nAur   &BTr    &   2.71        & 19.159        & 0.085     & 2.86 $10^{-12}$&   1.81  &  1795.7694    &   2.66\\
\iAur   &BTr    &   3.14        &  9.0712       & 0.0019        & $<$10$^{-20}$&   4.27   &  1800.6762   &   2.80\\
        &BLb    &   5.11        &   \dots   &   \dots   &   \dots   &   \dots   &   \dots   &   \dots   \\
\bTau   &BTr    &   5.37        &   \dots   &   \dots   &   \dots   &   \dots   &   \dots   &   \dots   \\
        &BLb1   &   1.78        &  2.7371       & 0.0044        & 0.19          &   \dots   &   \dots   &   \dots   \\
        &BLb2   &   1.78        &  2.7414       & 0.0028        & 0.0069        &   0.54  &  2868.7218&   1.16\\
\kCet   &BHr    &   4.31        &  9.0648       & 0.018         & $<$10$^{-20}$&  12.76  &  1744.8921    &   2.72\\
\HR     &BHr    &  26.68        &  2.8359       & 0.0003        & $<$10$^{-20}$&  79.95  &  1751.8432    &   4.77\\
\iqAur  &BTr    &  11.08        &  2.4630   & 0.0092    & 2.56 $10^{-5}$&  20.13&  1694.1688      &   9.31\\
\noalign{\smallskip}
\hline
\end{tabular}
\tablefoot{Column rms$_{\rm all}$ is the standard deviation from the entire data set and just indicates the level of photometric variability in the respective satellite bandpass (BTr and BHr are red sensitive, BLb is blue sensitive). Column $\Delta P$ is the error from a Monte Carlo simulation of 10,000 synthetic data sets. Column Amp is the peak-to-valley amplitude and $T_0$ is the phase zero point in heliocentric truncated Julian days (HJD = TJD + 2,456,000). Column rms$_{\rm fit}$ is the rms of the data after removal of the (multi)harmonic fit. }
\end{flushleft}
\end{table*}

\subsection{STELLA spectroscopy}

All but three (\bTau , \zAur,\ and \iqAur) of our BRITE targets were monitored contemporaneously from the ground using the robotic 1.2\,m STELLA-I telescope at the Izan\~{a} Observatory on Tenerife (Strassmeier et al.~\cite{malaga}). The fiber-fed Echelle Spectrograph (SES) of STELLA with an e2v 2k$\times$2k CCD detector was used to cover the wavelength range from 390--880\,nm at a resolving power of $R$=55,000 (3-pixel sampling) corresponding to a spectral resolution of 0.12~\AA \ at 650\,nm. A recent observatory and instrument status report is given by Weber et al. (\cite{spie12}).\ The SES spectra were automatically reduced using the IRAF-based STELLA data-reduction pipeline (Weber et al. \cite{spie08}). The images were corrected for bad pixels and cosmic-ray impacts. Bias levels were removed by subtracting the average overscan from each image followed by the subtraction of the mean of the (already overscan-subtracted) master bias frame. After removal of scattered light, the one-dimensional spectra were extracted using an optimal-extraction algorithm. The blaze function was then removed from the target spectra, followed by a wavelength calibration using consecutively recorded Th-Ar spectra. Finally, the extracted spectral orders were continuum normalized by dividing by a matching flux-normalized synthetic spectrum. For the present program, a total of 750 spectra were acquired for 9 of the 12 stars. Two targets were re-observed in 2019 (\nAur\ and \iAur ). Those spectra acquired contemporaneously with the BRITE observations are reported in this work. Two other targets are engaged in an ongoing long-term monitoring program (\aAur\ and \epsAur ). For \aAur , we combined all existing STELLA RVs for a new orbit employing $\approx$9600 spectra over 12.9  consecutive years. For \epsAur , we combined all existing STELLA and literature RVs ($\approx$3000) for a new spectroscopic orbit.

\begin{figure*}[tbh]
{\bf a.} \hspace{91mm} {\bf b.}\\
\includegraphics[clip,angle=0,width=87mm]{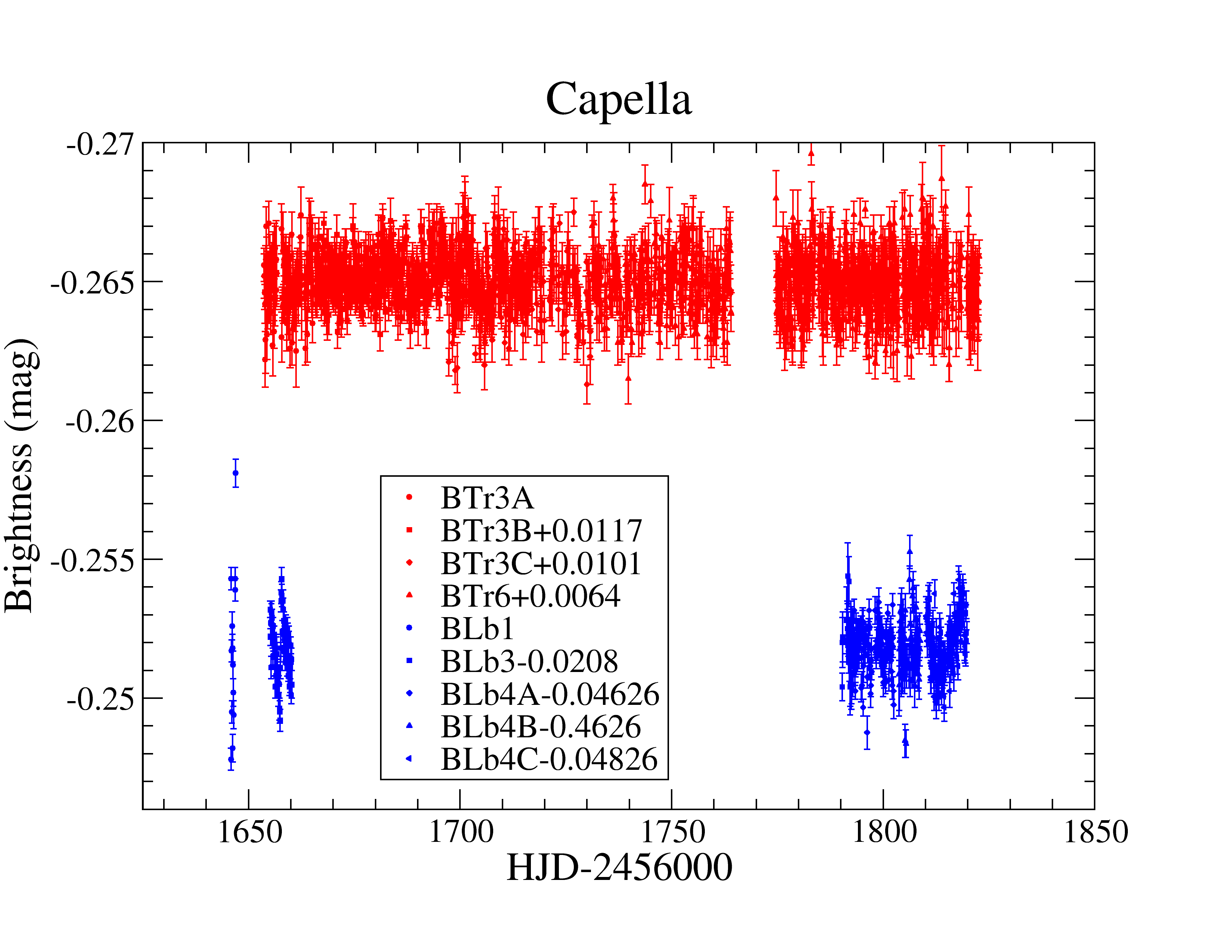} \hspace{5mm}
\includegraphics[clip,angle=0,width=87mm]{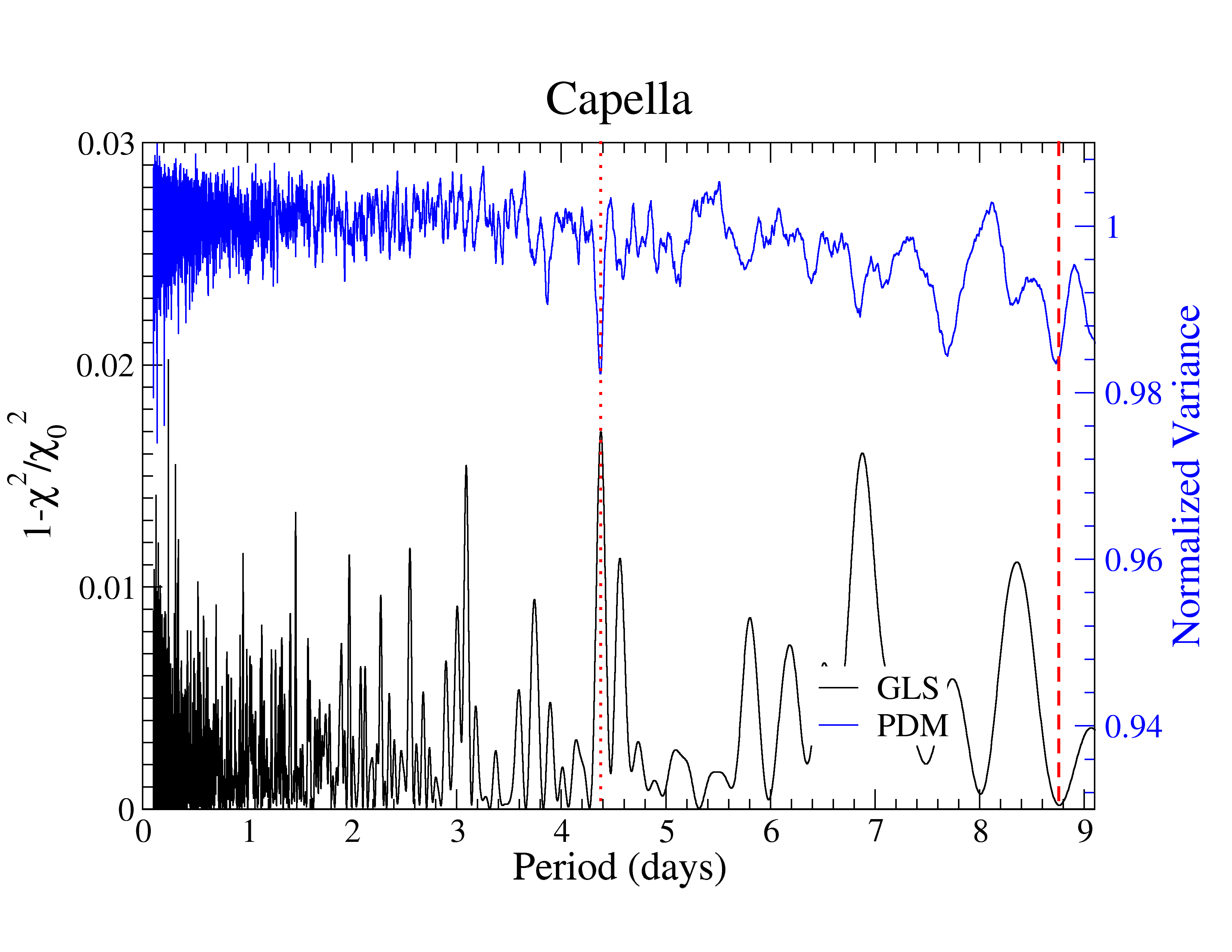}\\
{\bf c.} \hspace{91mm} {\bf d.}\\
\includegraphics[clip,angle=0,width=87mm]{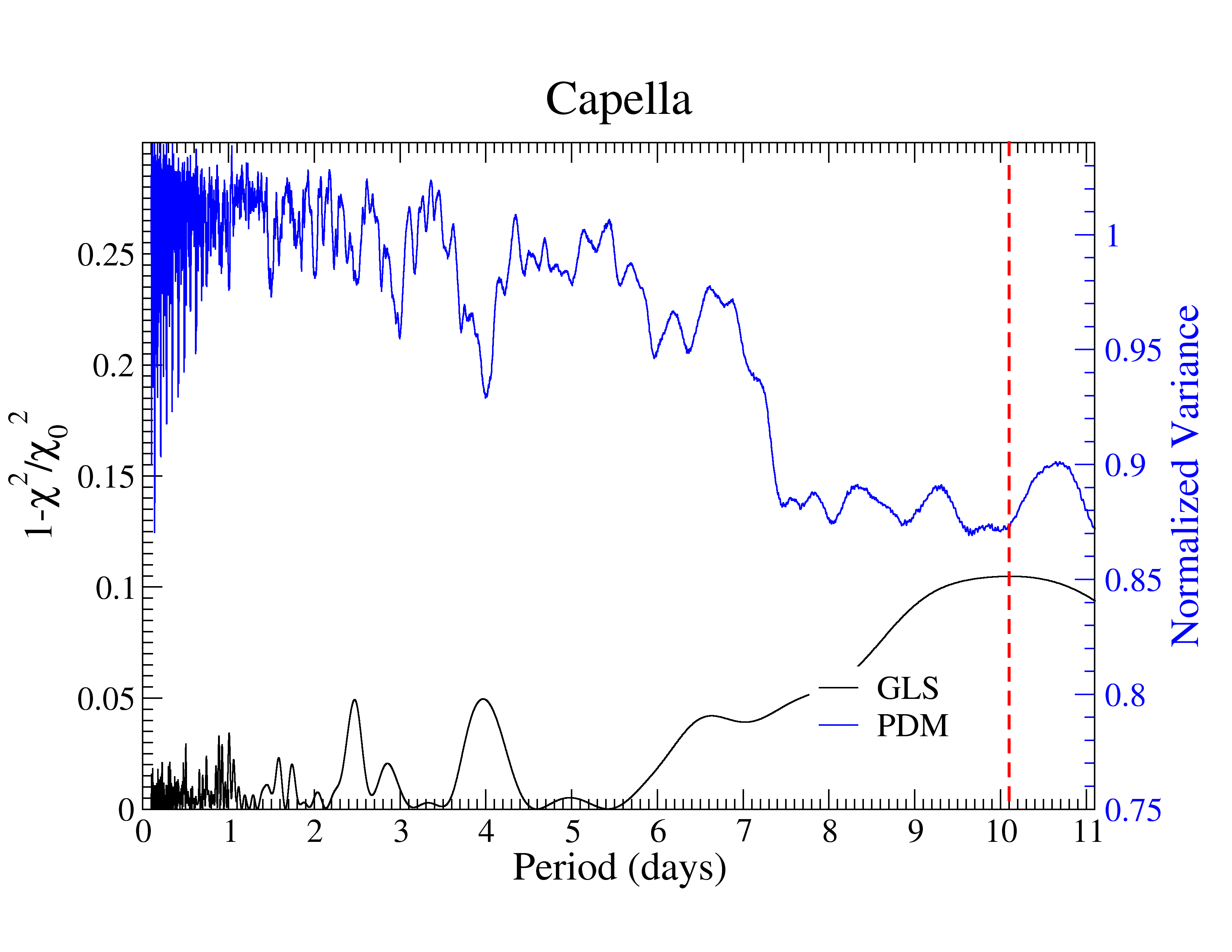}\hspace{5mm}
\includegraphics[clip,angle=0,width=87mm]{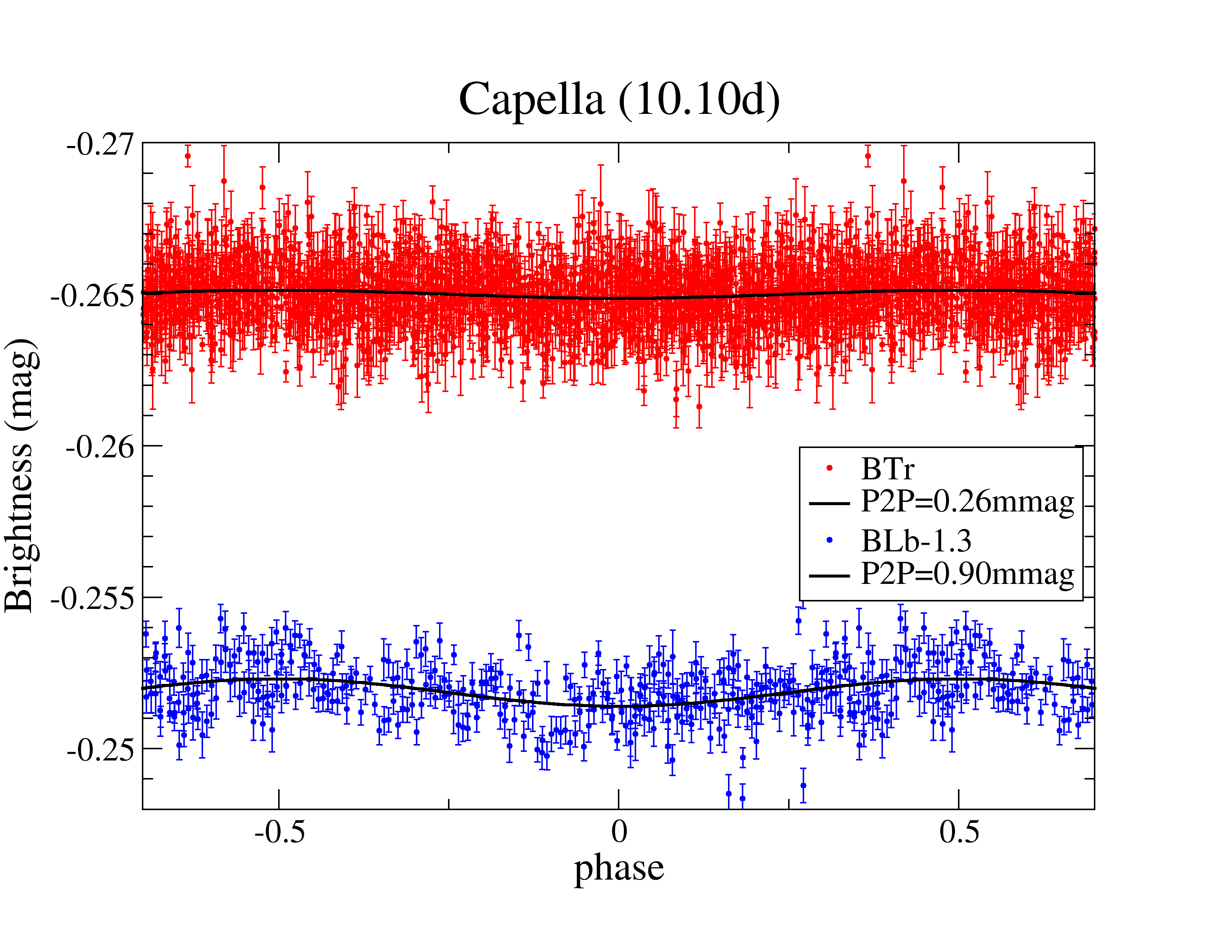}
\caption{BRITE photometry of Capella. Panel a) Full data set. The top light curve is the red BTr data (red dots); the bottom light curve is the blue BLb data (blue dots). The insert identifies the magnitude shifts applied to each individual CCD setup per satellite. Panel b) Periodograms of the red data. The top (blue) line is from the PDM given in units of its normalized variance; the bottom (black) line is the generalized LS (GLS) given in units of $1-\chi^2 / \chi_0^2$. The vertical dashed lines are the periods identified (see text). Panel c) Periodograms of the blue data, otherwise as in panel b. Panel d) Phase plots for the red data (red dots, top curve) and blue data (blue dots, bottom curve) with the best-fit 10.10 d period. Phases outside $\pm$0.5 are just repeated.}\label{Fcap}
\end{figure*}

\section{Data analysis}\label{S4}

\subsection{BRITE data preparation}\label{S41}

The pipeline data from BRITE must be corrected for the presence of moonlight as well as for the many systematic outliers due to telescope pointing errors and long-term CCD sensitivity changes due to temperature variations. Data phases with moonlight contamination were simply excluded from the analysis but only amount to 1-2 days every 28 days. Pointing jitter is seen in the data as random outliers that were removed with the GESD algorithm (see Pigulski et al. \cite{pig18b}). The correction of a long-term trend due to CCD temperature changes (and/or charge transfer efficiency changes) is the most severe impact and was removed in first order using a one-dimensional linear as well as a two-dimensional de-correlation of flux values. The procedure was basically the same as in Pigulski et al. (\cite{pig18a}, \cite{pig18b}) and similar to that described by Kallinger et al. (\cite{Kall}). The final data are further referred to as the de-correlated data. Figure~\ref{Fccd} shows the run of the average CCD temperature of BTr and BLb during the epoch of our Capella observations. Its double-peaked curve with a peak separation of $\approx$100\,d is unfortunately of nearly identical duration as the orbital period of Capella and the (synchronized) rotation period of its cooler component (both $\approx$104\,d). This prevents a conclusive analysis of low-amplitude variability of similar timescale.

Photometric zero points of each BRITE satellite instrumental setup were determined and adjusted before they were subjected to the period search. We took the average brightness of each setup and shifted all setups to match the first one. For cases with a clear brightness gradient within the setup only the start and end points were used. The individual shifts in magnitudes are identified in the insert of the respective light-curve plots (e.g., in Fig.~\ref{Fcap}a).

\subsection{Period determinations from BRITE data}

Period searches include, in almost any case, the choice of a (periodic) data model and the least-squares fit of the light curve to the underlying model. This model is chosen to help in the extraction of features the light curve is expected to show. For example, for exoplanet transit searches the expected signal is box shaped, and consequentially the method of Kov{\'a}cs et al. (\cite{kov:zuc}) tries to fit a box-shaped transit signal to the light curve. The Lomb-Scargle (LS) periodogram (Lomb \cite{lomb}, Scargle \cite{sca}) as a variant of a Fourier transformation is the least-squares fitting of a harmonic (sinusoidal) function to the light curve. A seemingly model-free method like phase dispersion minimization (PDM; Stellingwerf~\cite{Stel}) can be seen as the least-squares fit of a piece-wise constant function to the (phase-folded) light curve, rendering the PDM method similar to the analysis of $\chi^2$ reduction in methods fitting data models.

For the period determinations in this paper, we used PDM in the original formulation of Stellingwerf~(\cite{Stel}) and complement it with the GLS periodogram as formulated by Zechmeister \& K{\"u}rster (\cite{zech}). Other variants of PDM, such as the string-length maximization method of Dworetsky~(\cite{dwor}) or the binless method of Plavchan et al. (\cite{pla}), did not improve the period detection from our data and were thus not followed further. A multicomponent variant of the LS periodogram, as described by Defa{\"y} et al. (\cite{def}), was also deemed inappropriate for our type of data.

For all stars, frequencies were searched at an equidistant sampling starting at $f_{\rm min}=0.01$~c\,d$^{-1}$ and extending to $f_{\rm max}=10$~c\,d$^{-1}$, such that $10^5$ frequencies were scanned. A false-alarm probability (FAP) according to Zechmeister \&  K{\"u}rster (\cite{zech}) was calculated for all peaks in the LS version (but see Baluev \cite{bal} for an enlightening discussion on FAPs). Significance levels for the PDM periods as in Stellingwerf~(\cite{Stel}) were assigned. The PDM results, and to a lesser extend also the GLS results, were contaminated with aliases of the orbital period of the BRITE satellites (at $f_0\approx 14.5$~c\,d$^{-1}$). These peaks were ignored. For stars with significant periods, GLS and PDM were combined to reach a single, most probable period. This period was then used in the phase plots and is listed in Table~\ref{T3}.
Errors of the periods were estimated by generating synthetic data sets on all stars. Each synthetic data set had magnitude values with a random, Gaussian-distributed offset according to the error of the magnitude as initially estimated. A period search in a small region around the detected period was then performed using GLS. The standard deviation of the periods found was used as the error of the actual best period and is given in Table~\ref{T3}.

\subsection{Detailed BRITE results}

\begin{figure*}[tbh]
{\bf a.} \hspace{60mm} {\bf b.} \hspace{60mm} {\bf c.}\\
\includegraphics[clip,angle=0,width=60mm]{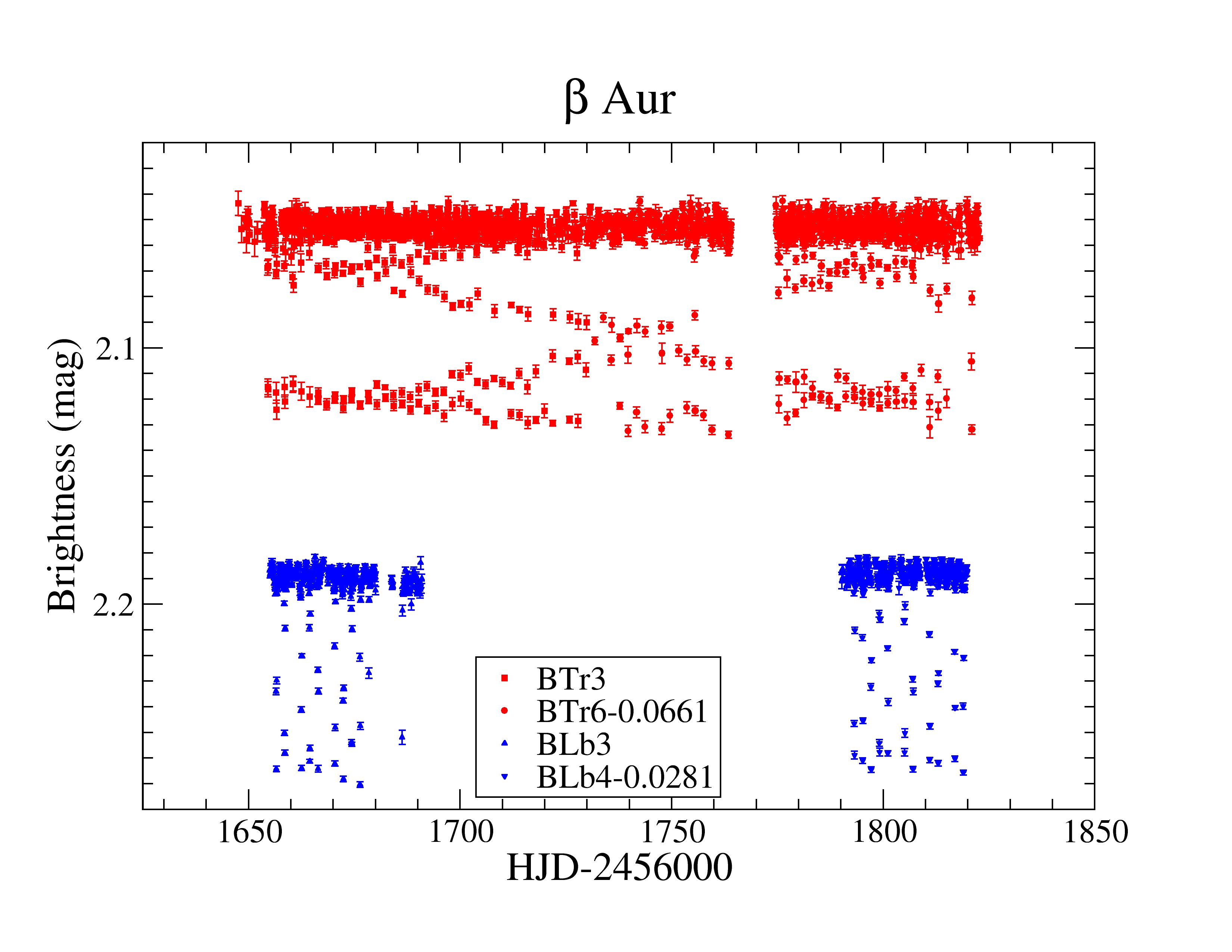} \hspace{5mm}
\includegraphics[clip,angle=0,width=60mm]{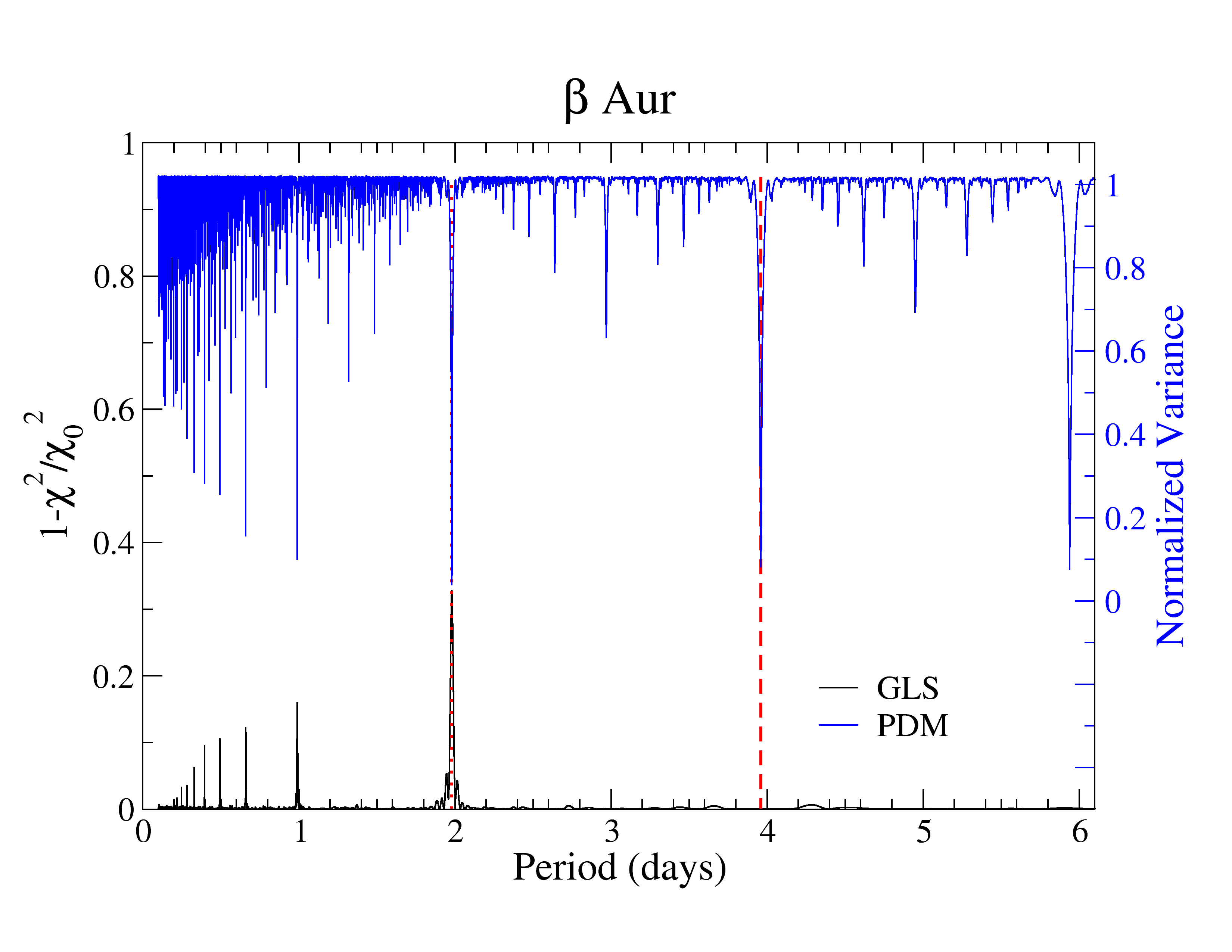}\includegraphics[clip,angle=0,width=60mm]{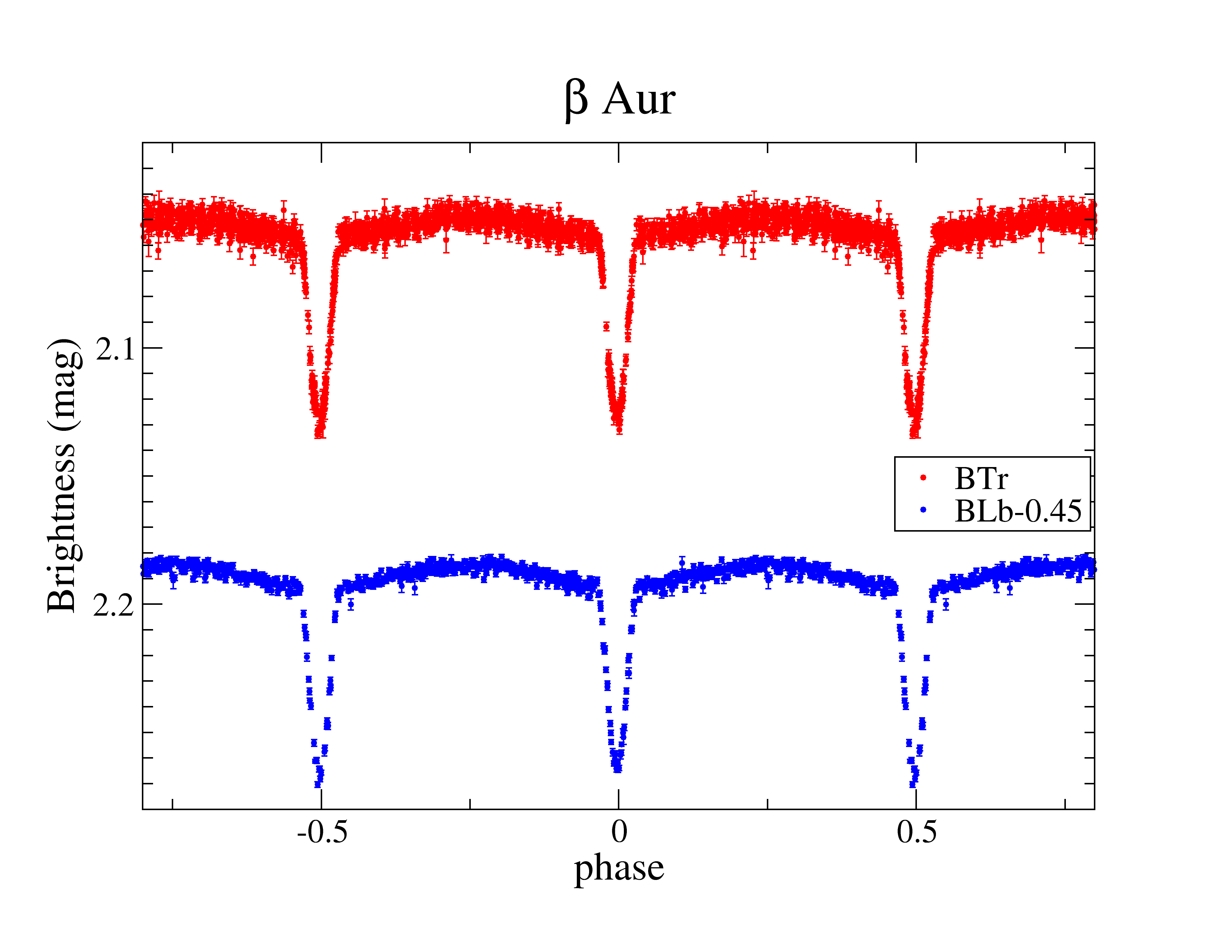}
\caption{Photometry of \bAur \ from BRITE. Panel a) Full data set. Panel b) Periodograms. Panel c) Phased light curve with the orbital period. Otherwise as in Fig.~\ref{Fcap}. }\label{FbAur}
\end{figure*}
\begin{figure*}[tbh]
{\bf a.} \hspace{60mm} {\bf b.} \hspace{60mm} {\bf c.}\\
\includegraphics[clip,angle=0,width=60mm]{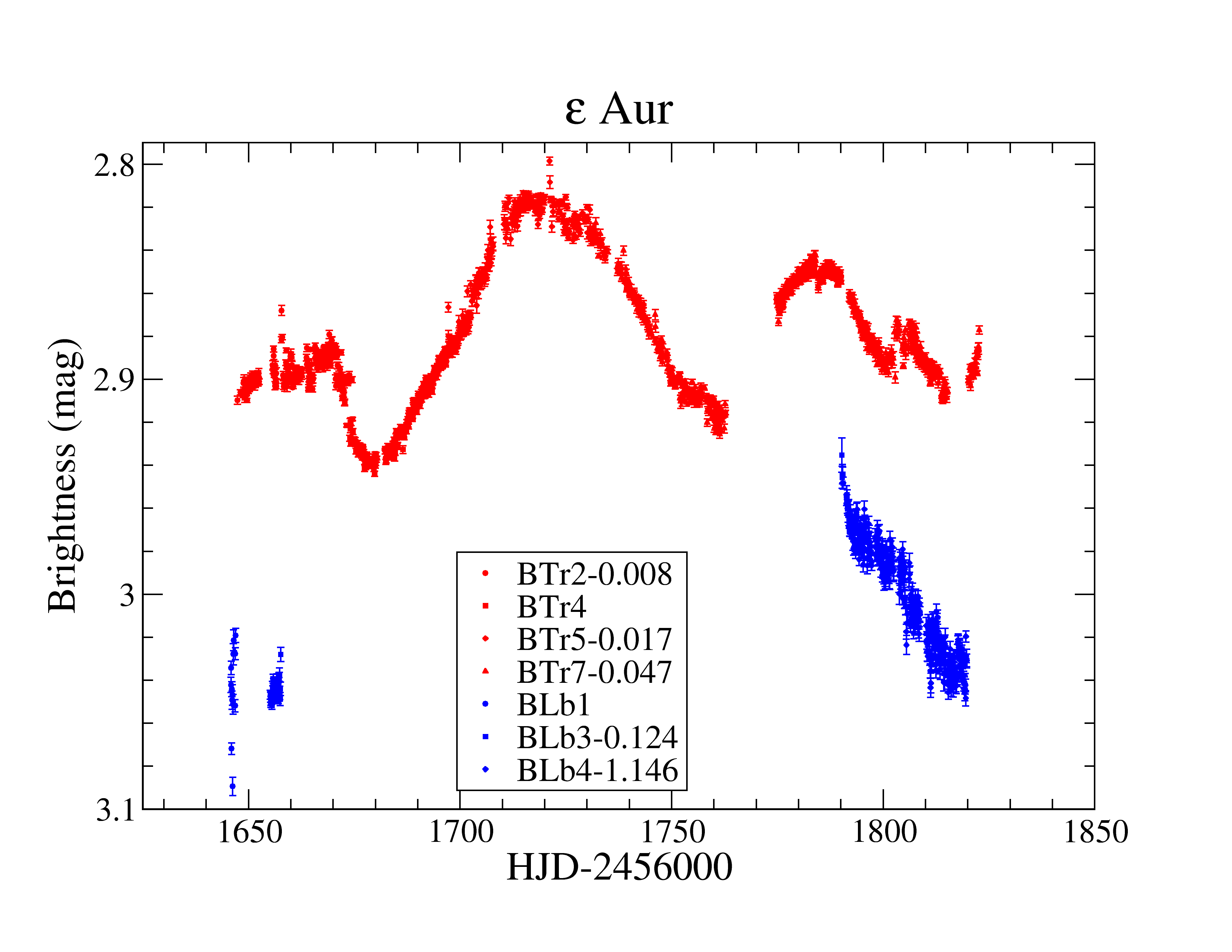}
\includegraphics[clip,angle=0,width=60mm]{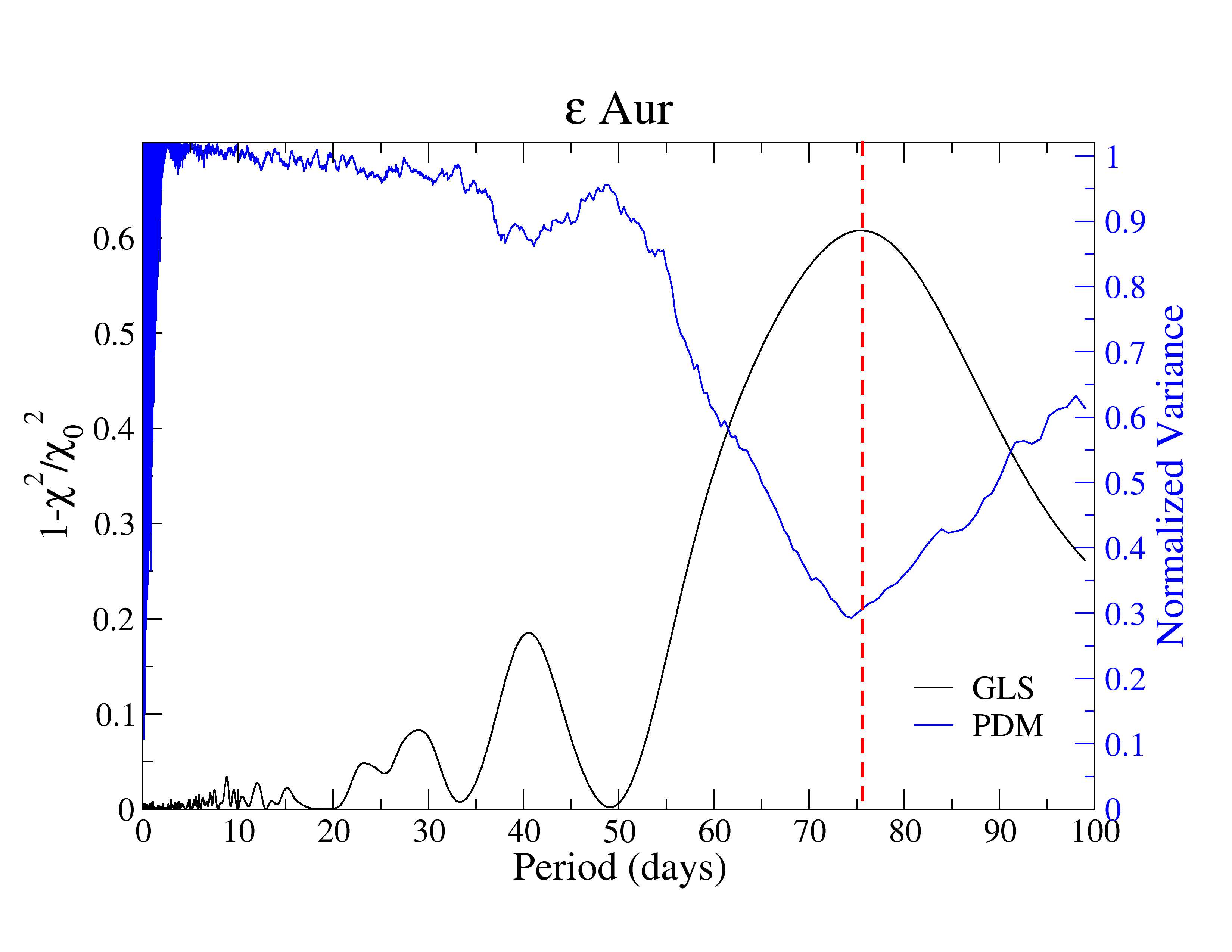}
\includegraphics[clip,angle=0,width=60mm]{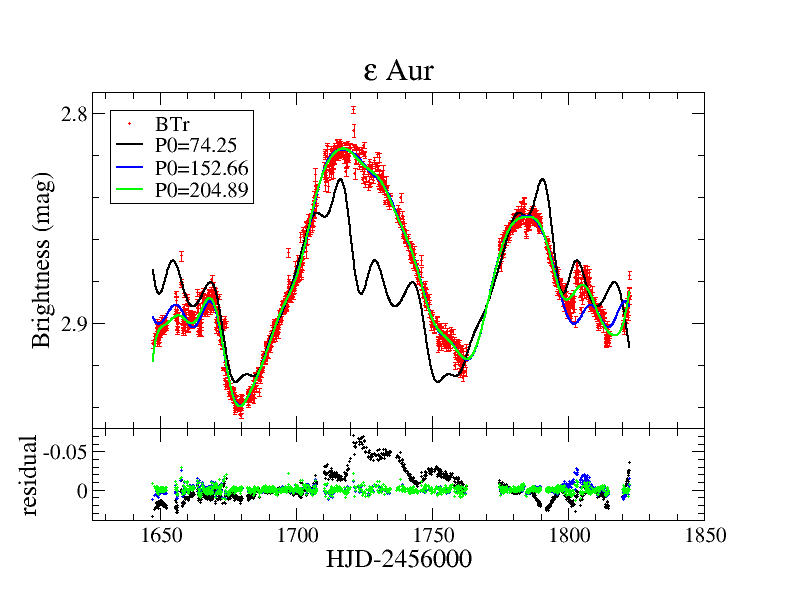}
\caption{Photometry of \epsAur \  from BRITE. Panel a) Full data sample. Panel b) Periodograms. Panel c) Multi-harmonic fits with fundamental periods of 74.25, 152.66, and 204.89\,d, respectively (colored full lines identified in the inset). The 152 d and 205 d fits are nearly equally good and mostly overlay in the plot; a 74 d fundamental period does not fit at all. The bottom panel shows the residuals from these fits. Otherwise as in Fig.~\ref{Fcap}.  }\label{FepsAur}
\end{figure*}
\begin{figure*}[tbh]
{\bf a.} \hspace{60mm} {\bf b.} \hspace{60mm} {\bf c.}\\
\includegraphics[clip,angle=0,width=60mm]{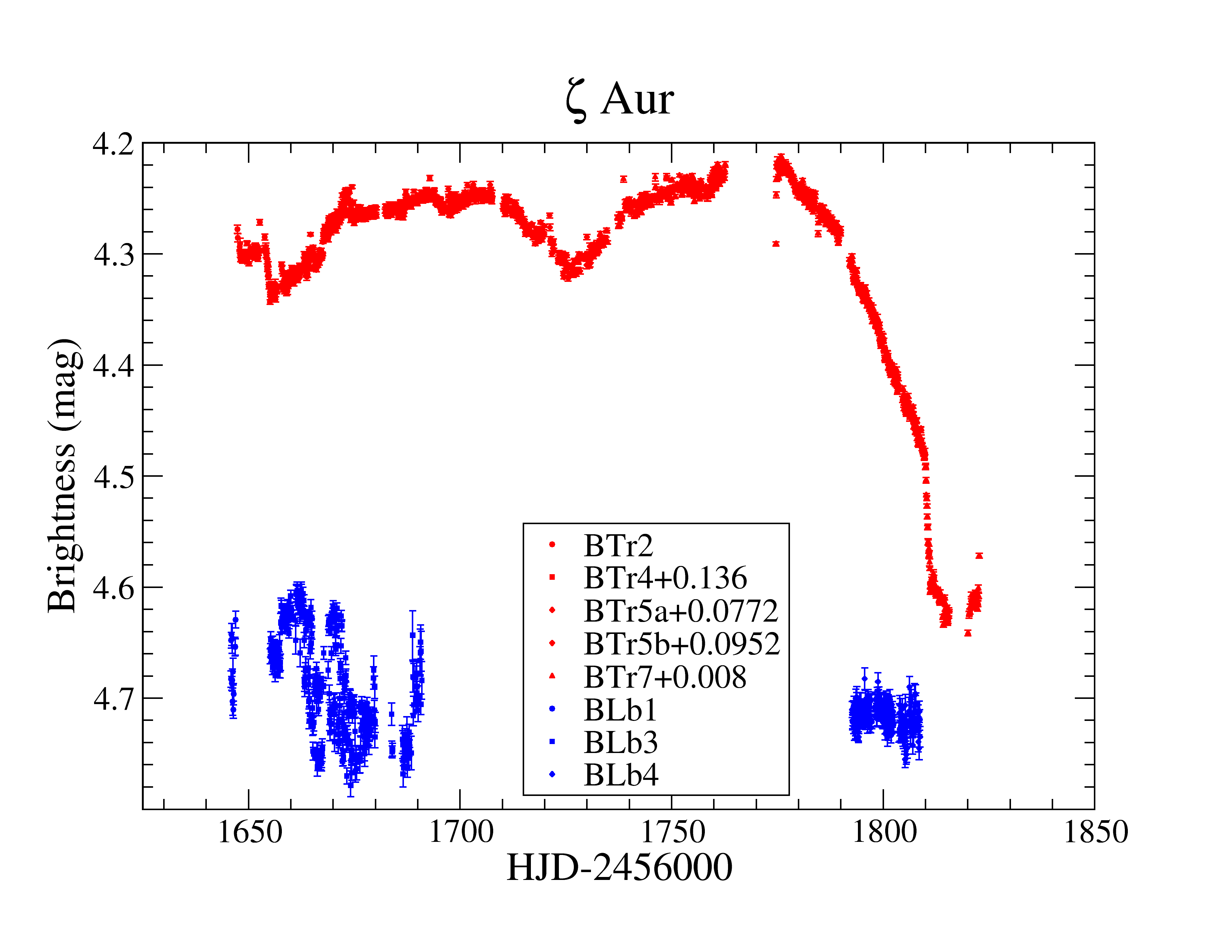}
\includegraphics[clip,angle=0,width=60mm]{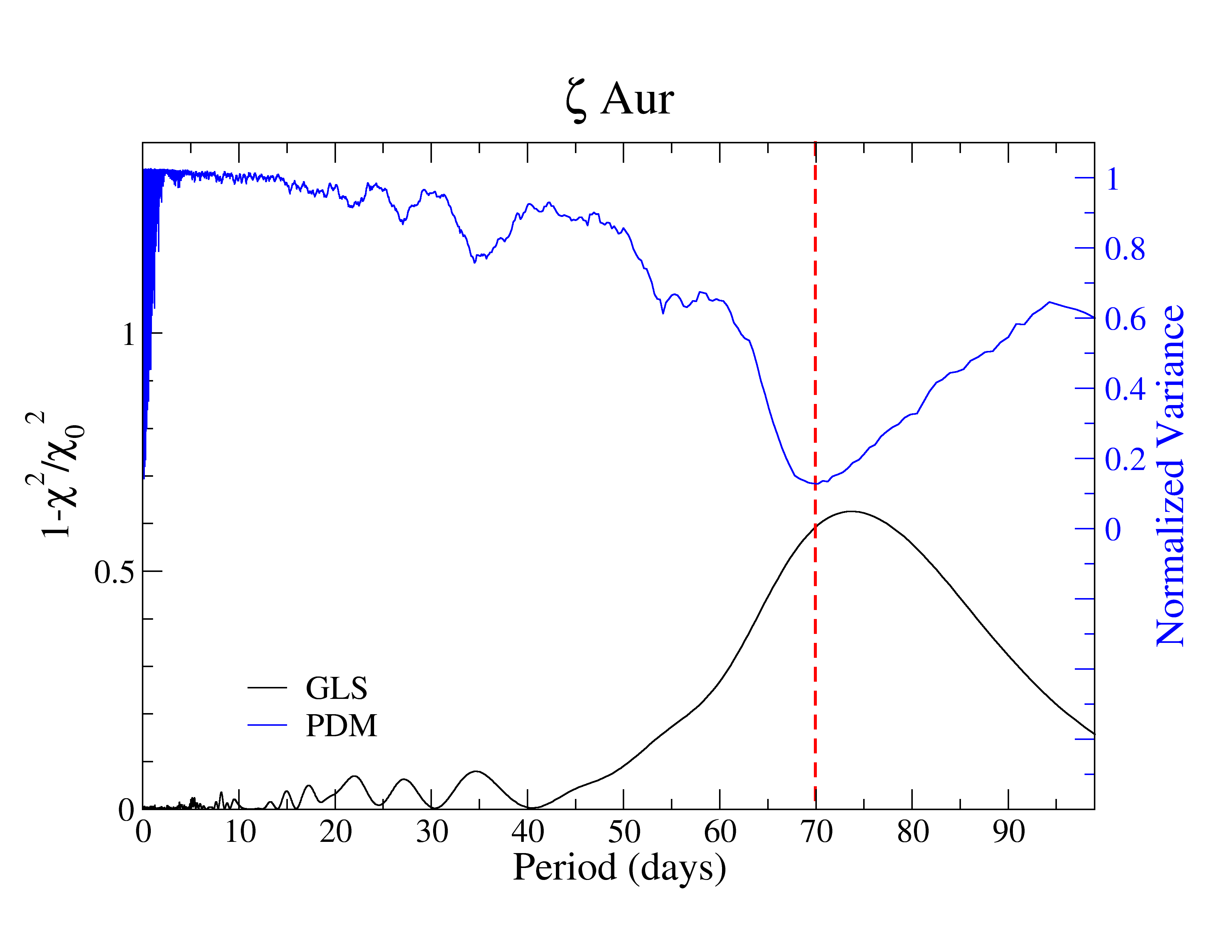}\includegraphics[clip,angle=0,width=60mm]{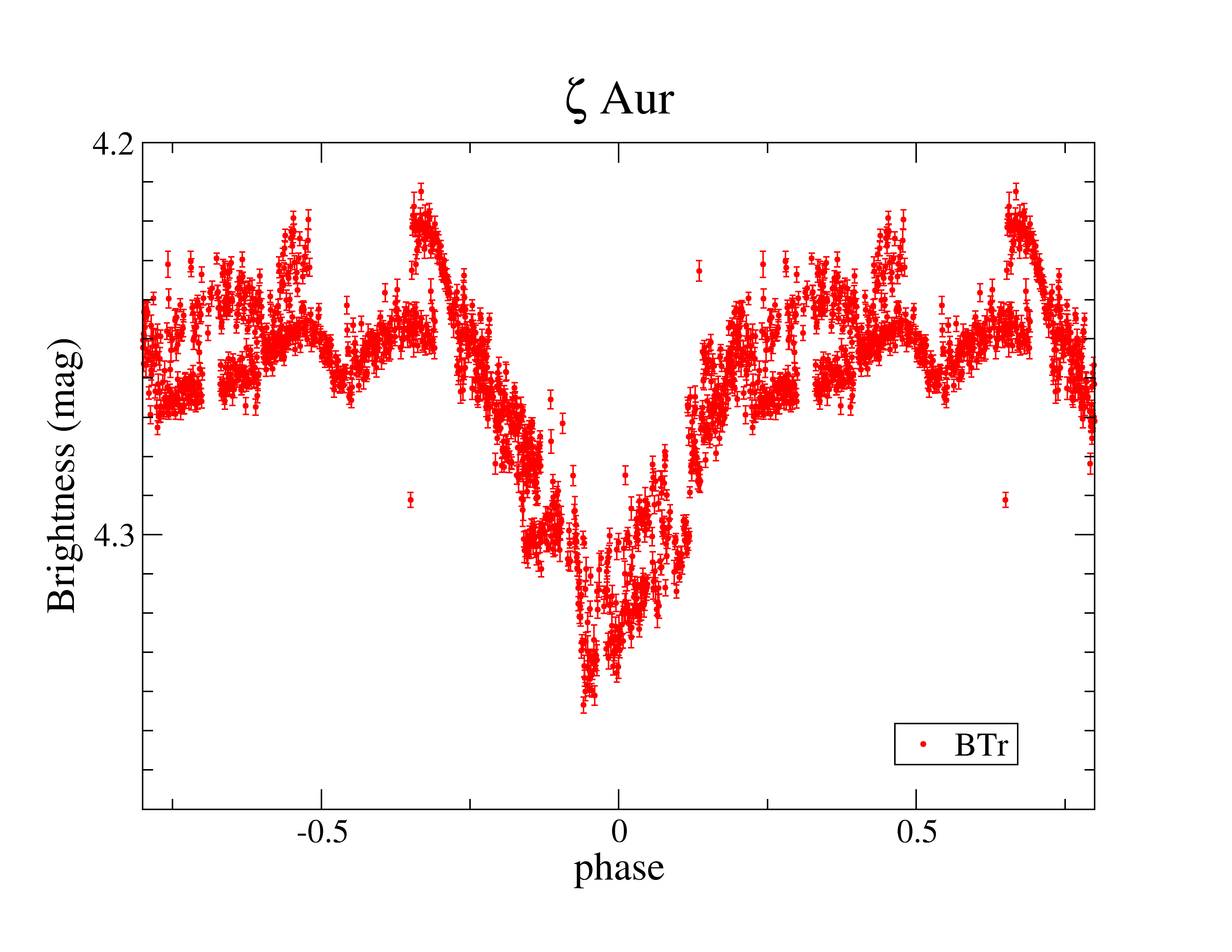}
\caption{Photometry of \zAur \  from BRITE. Panel a) Full data sample. Panel b) Periodograms. Panel c) Phased light curve with the 70 d period. Otherwise as in Fig.~\ref{Fcap}.  }\label{FzAur}
\end{figure*}

{$\alpha$ Aur = Capella}. Figure~\ref{Fcap}a shows the BRITE Capella light curves. Only the red-filter data (two setups of BTr observations) have the long and continuous 176 d coverage. The BTr3 setup had to be split into three pieces owing to some small instrumental effects that were not the same in all data parts. BLb blue data were mostly rejected but covered approximately 10 days at the beginning of the BTr observations and 30 days at the end. Much to our surprise, and disappointment, the 176 d long red light curve of Capella did not show any significant variability. The light curve is basically constant to within $\approx$1\,mmag. A PDM and a GLS periodogram of the red data both show a 4.38\,d period with a FAP of 10$^{-2}$ (Fig.~\ref{Fcap}b). A least-squares fit suggests a full amplitude of just 0.36$\pm$0.09\,mmag. For the G0 component we expect a rotation period of $\approx$8.6\,d from $v\sin i$ and the stellar radius, which is almost exactly twice as long as this weak signal. We applied a forced fit with 2$\times$4.38 = 8.76\,d, which then resulted in a comparable best-fit amplitude of 0.33\,mmag, still well below the long-term rms level. We do not claim this period to be real but take it as evidence that the $\approx$8.6 d period from ground-based \Halpha\ monitoring of Capella is at least not contradicted by the red-filter data. On the contrary, the blue BLb light curve showed a slightly longer period of 10.1$\pm$0.6\,d with a larger amplitude of 0.90\,mmag (Fig.~\ref{Fcap}c,d). If we phase the red data with this 10.1 d period the amplitude is a mere 0.26\,mmag (Fig.~\ref{Fcap}c). Because the blue BRITE data (from BLb) are dominated by the hotter of the two Capella components, that is the G0 star, the red data (from BTr) are more dominated by the cooler G8 component and we thus should see the G0-star rotation period better in the blue data than in the red. We conclude that the 10.1 d period is our only real period detection in the Capella data. We see no evidence for the 106$\pm$3\,d rotation period of the cooler G8 component as claimed by Strassmeier et al.~(\cite{cap1}) from \Halpha , but see Sect.~\ref{S41}.

\begin{figure*}[tbh]
{\bf a.} \hspace{60mm} {\bf b.} \hspace{60mm} {\bf c.}\\
\includegraphics[clip,angle=0,width=60mm]{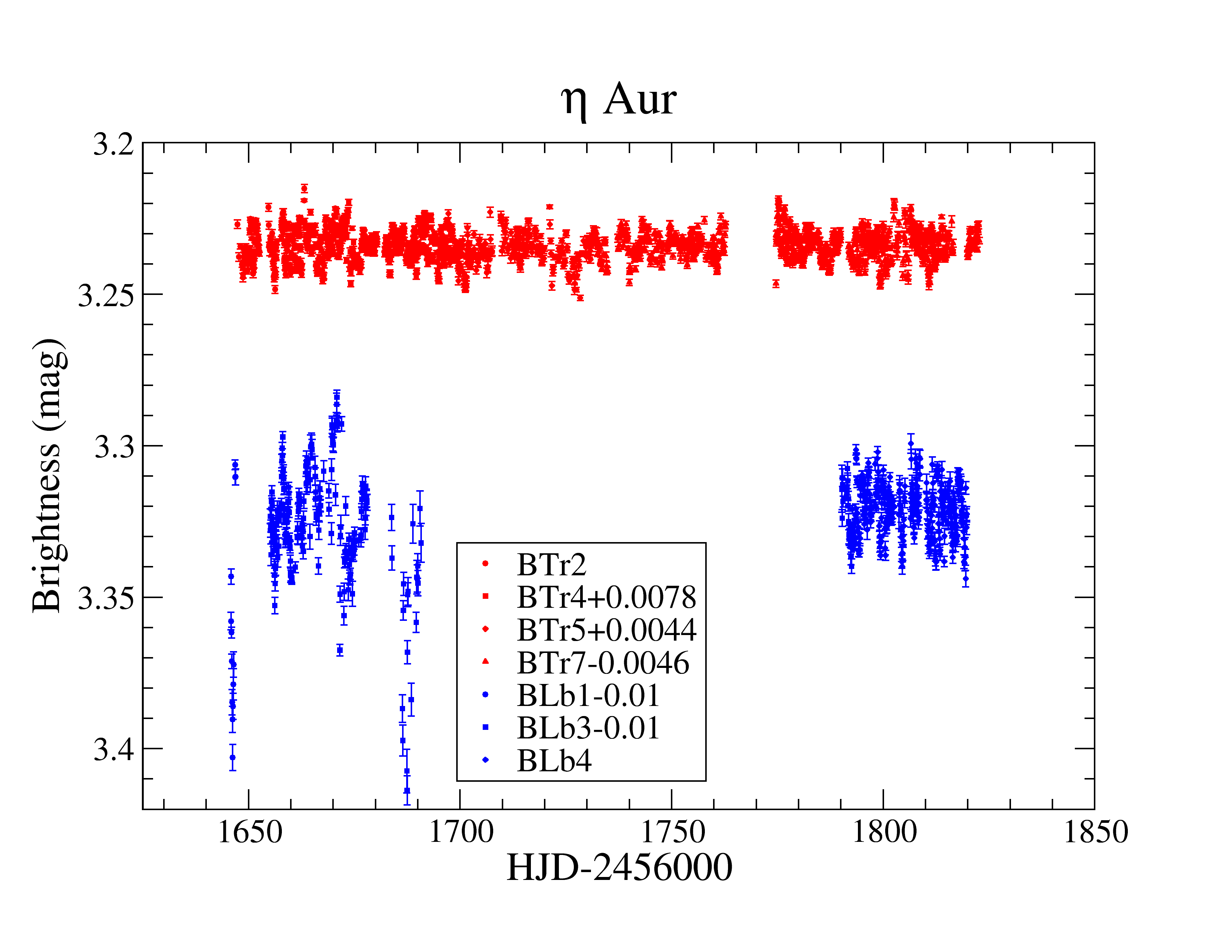}
\includegraphics[clip,angle=0,width=60mm]{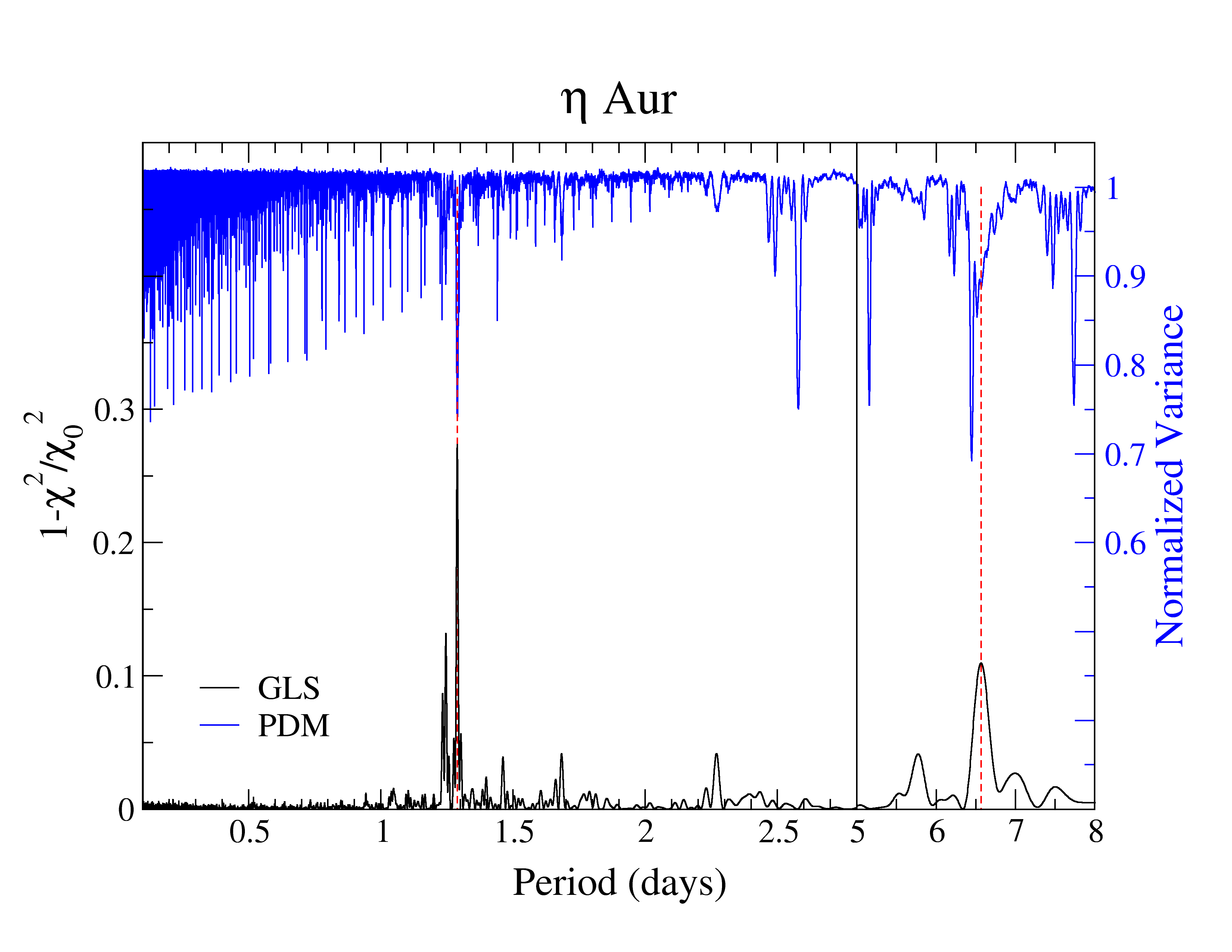}
\includegraphics[clip,angle=0,width=60mm]{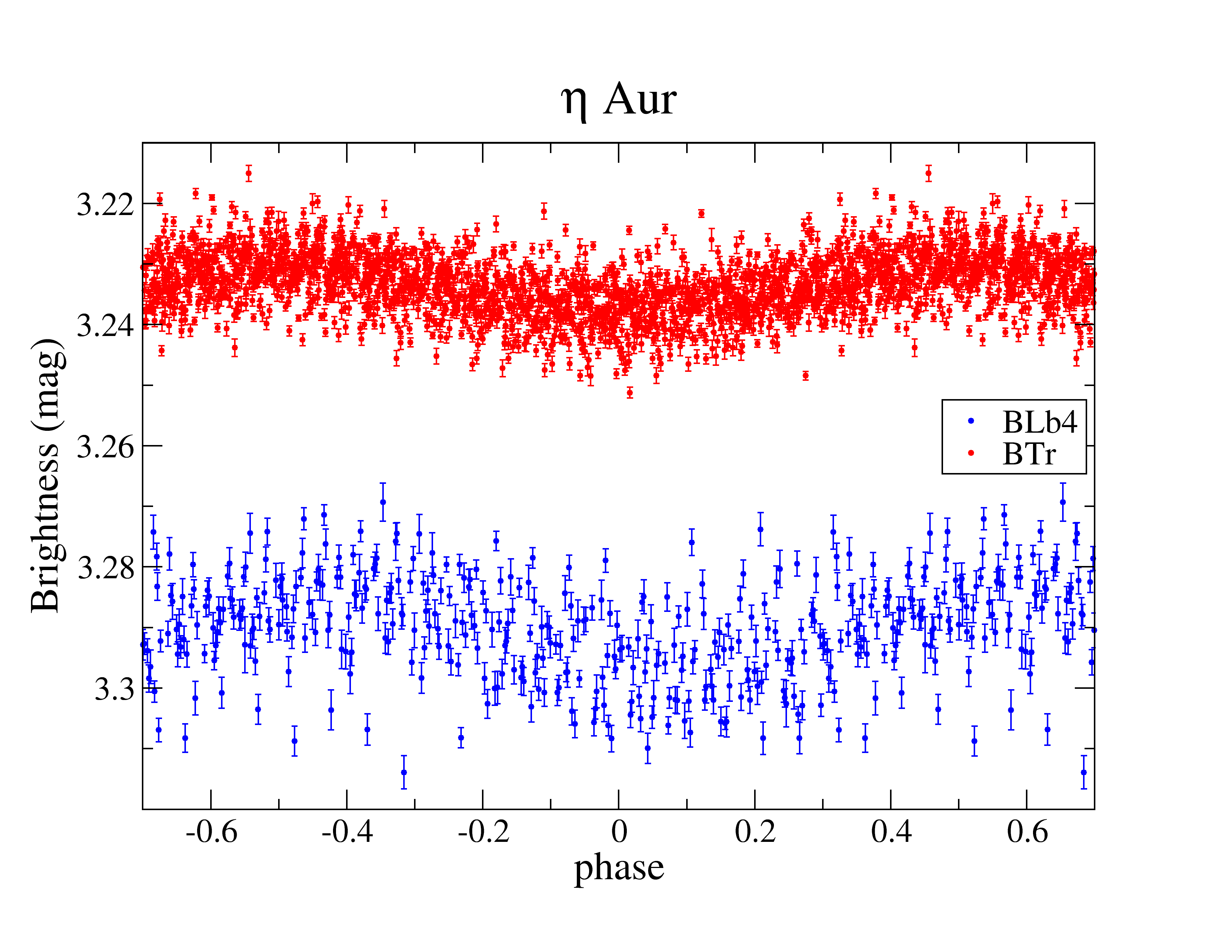}
\caption{Photometry of \etaAur \  from BRITE. Panel a) Full data sample. Panel b) Periodograms. The period axis is split into two parts for the ranges 0--2.8 d\,and- 5--8\,d. The main peak is at a period at 1.28\,d, while other significant peaks are seen at 1.23\,d and 6.6\,d. Panel c) Phased light curve with the 1.28 d period. Otherwise as in Fig.~\ref{Fcap}.  }\label{FetaAur}
\end{figure*}
\begin{figure*}[tbh]
{\bf a.} \hspace{60mm} {\bf b.} \hspace{60mm} {\bf c.}\\
\includegraphics[clip,angle=0,width=60mm]{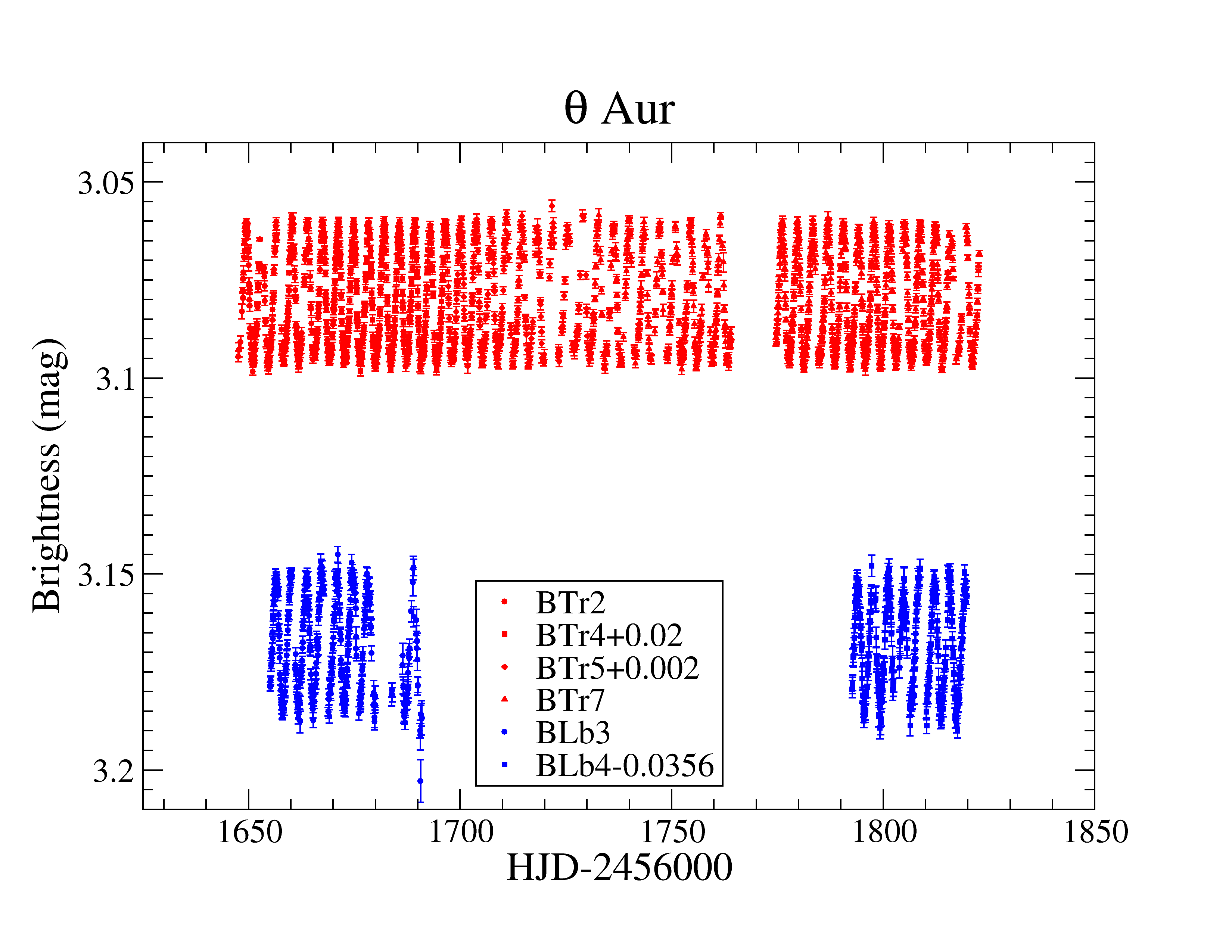}
\includegraphics[clip,angle=0,width=60mm]{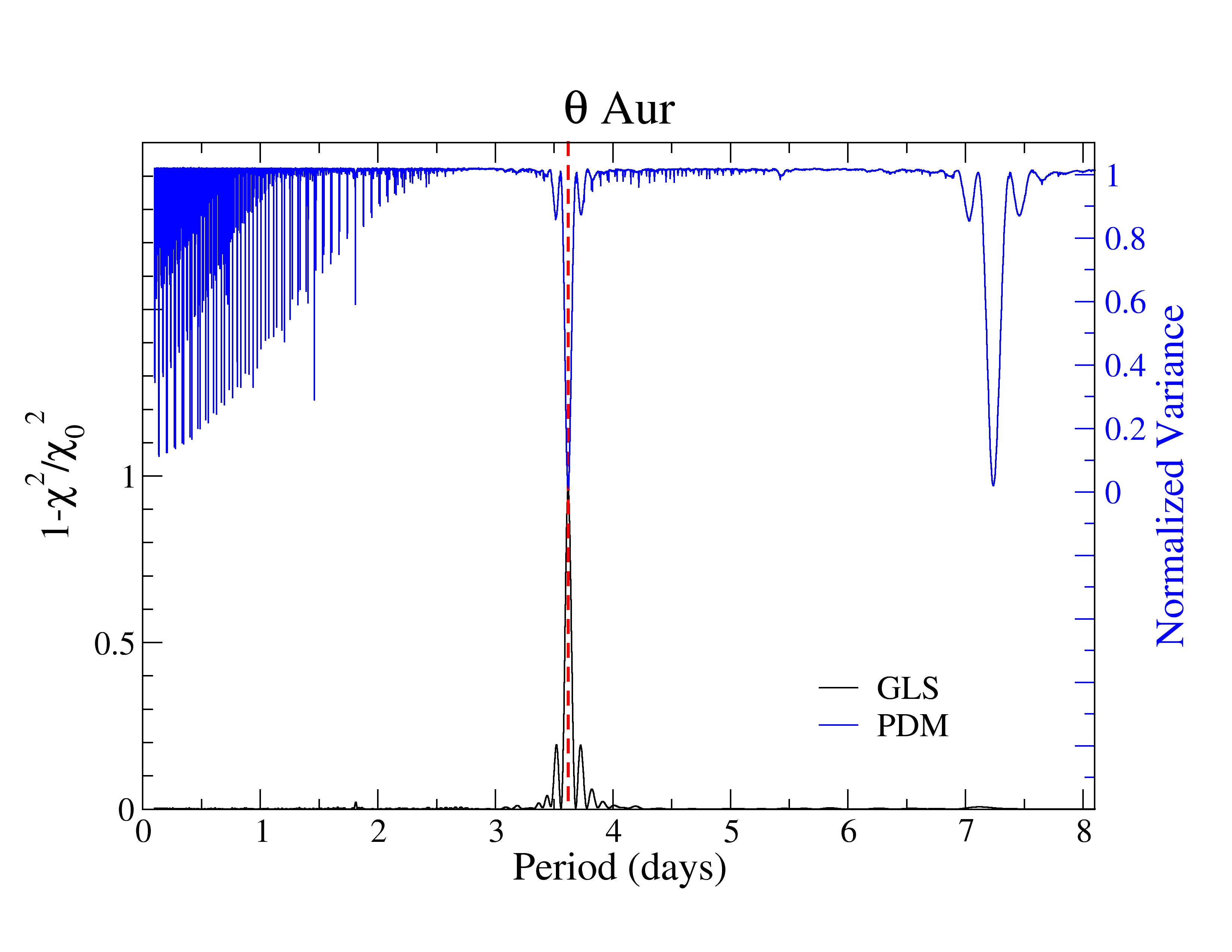}
\includegraphics[clip,angle=0,width=60mm]{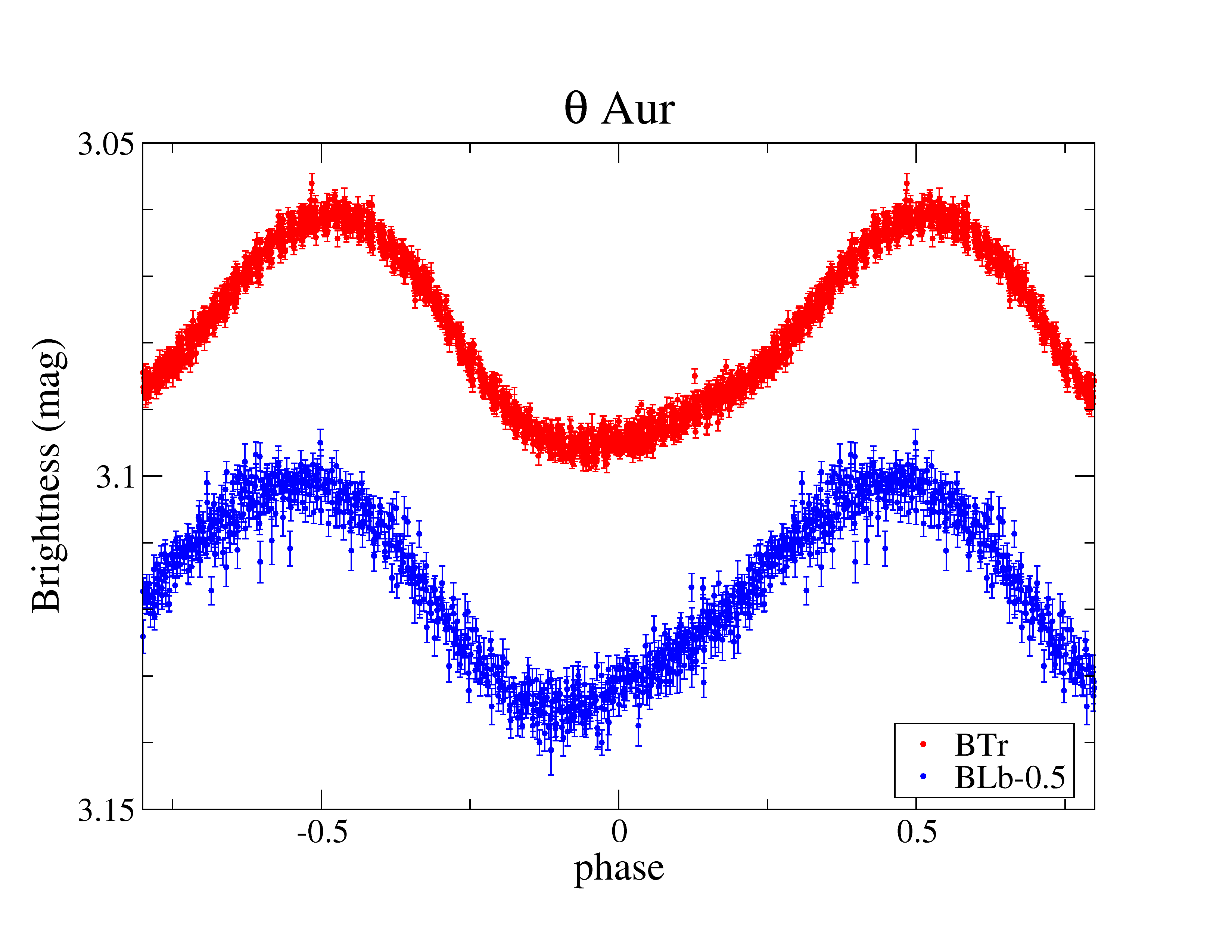}
\caption{Photometry of \tAur\ from BRITE. Panel a) Full data sample. Panel b) Periodograms. Panel c) Phased light curve with the dominant 3.6189 d period. Otherwise as in Fig.~\ref{Fcap}.  }\label{FtAur}
\end{figure*}
\begin{figure*}[tbh]
{\bf a.} \hspace{60mm} {\bf b.} \hspace{60mm} {\bf c.}\\
\includegraphics[clip,angle=0,width=60mm]{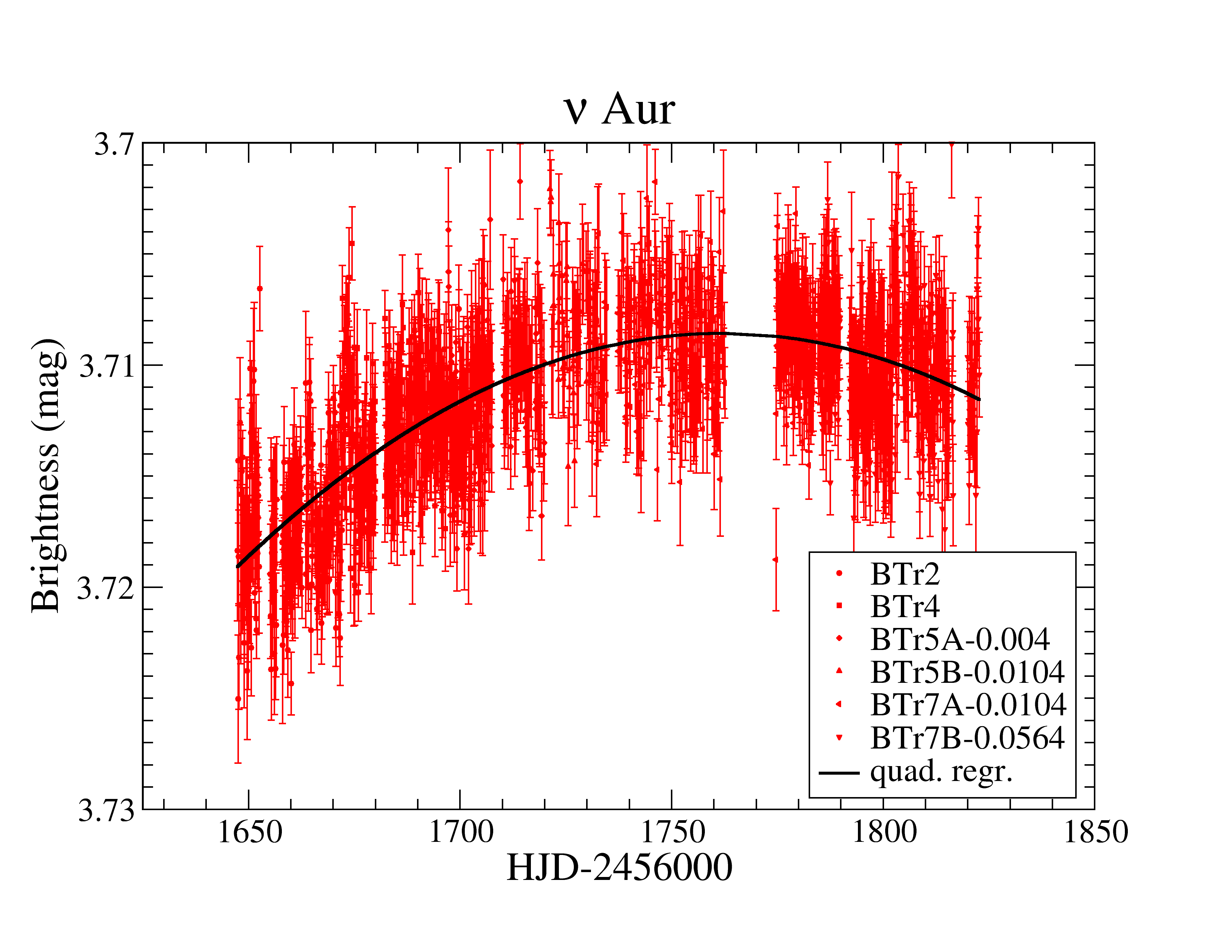}
\includegraphics[clip,angle=0,width=60mm]{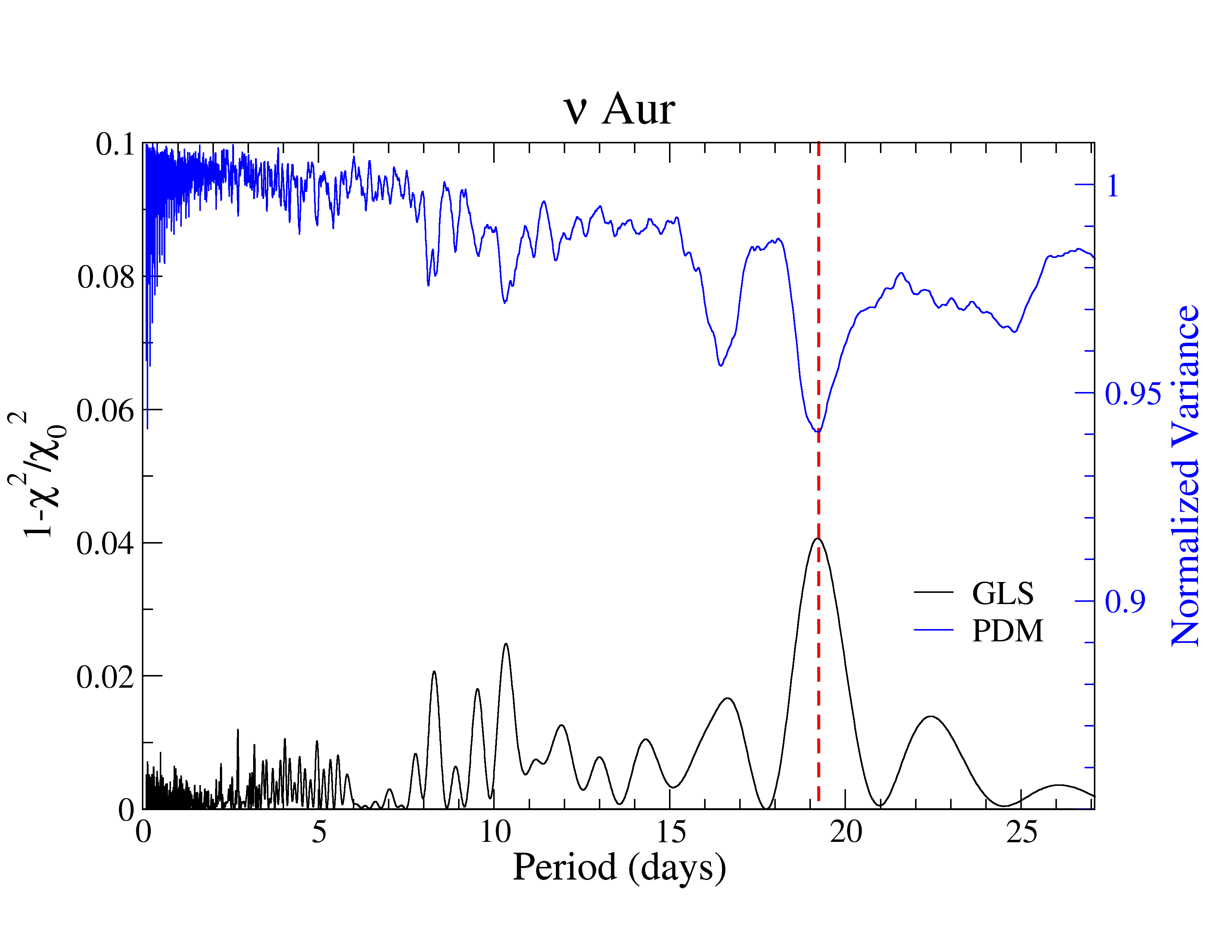}
\includegraphics[clip,angle=0,width=60mm]{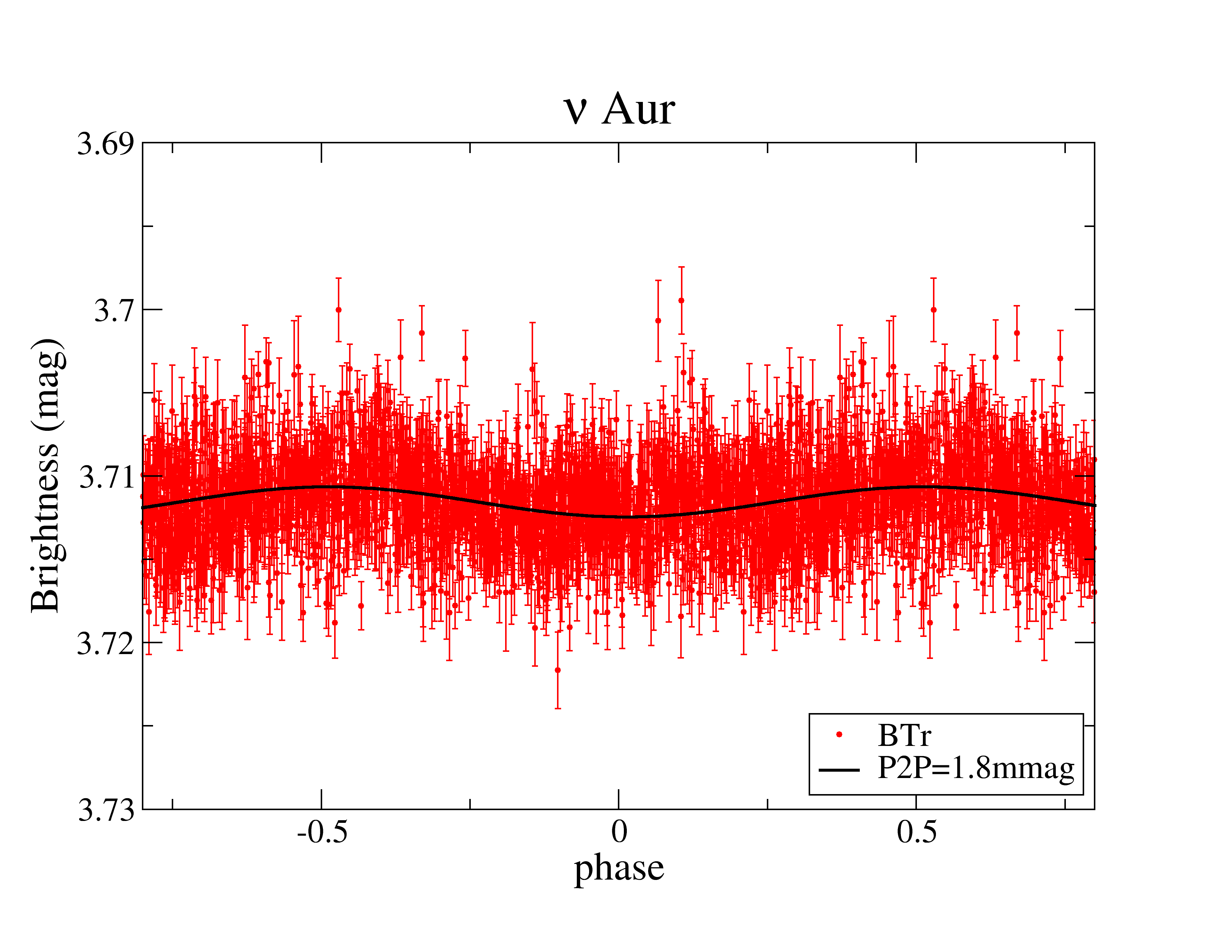}
\caption{Photometry of \nAur\ from BRITE. Panel a) Full data sample, only BTr data were acquired. The line is a quadratic polynomial fit that was then removed from the data for the periodograms. Panel b) Periodograms. Panel c) Phased light curve with the 19.16 d period. Otherwise as in Fig.~\ref{Fcap}.  }\label{FnuAur}
\end{figure*}

{$\beta$ Aur}. This target was also observed for 176 days by BTr with only a single 10 d gap around TJD\,1770 (Fig.~\ref{FbAur}a). The BLb observed the target for approximately 30 days at the beginning and again 30 days at the end of the BTr coverage. The eclipse data points in both filters reach a maximum depth of 80\,mmag with respect to the average out-of-eclipse light. As for the red Capella data, the strongest peak in both periodograms is at one-half the true period (Fig.~\ref{FbAur}b), in the case of \bAur\ this value is very close to 2.0 days. This is easily identified for the eclipsing binary \bAur\ because its orbital period is known from RVs as well and is very close to 4.0 days (3.96\,d). Both of our periodograms identify the true period as the second strongest peak. Figure~\ref{FbAur}c is a phase plot with the 3.96 d (orbital) period from the BRITE data.  We note that the BRITE data could be used to refine the nonlinear component of the limb-darkening laws of the two stars. Southworth et al. (\cite{south}) had employed the integral-light WIRE data and the eclipsing light-curve modeling code EBOP to infer nonlinear coefficients, which then led to an increase of the measured stellar radii by 0.4\%.

\begin{figure*}[tbh]
{\bf a.} \hspace{60mm} {\bf b.} \hspace{60mm} {\bf c.}\\
\includegraphics[clip,angle=0,width=60mm]{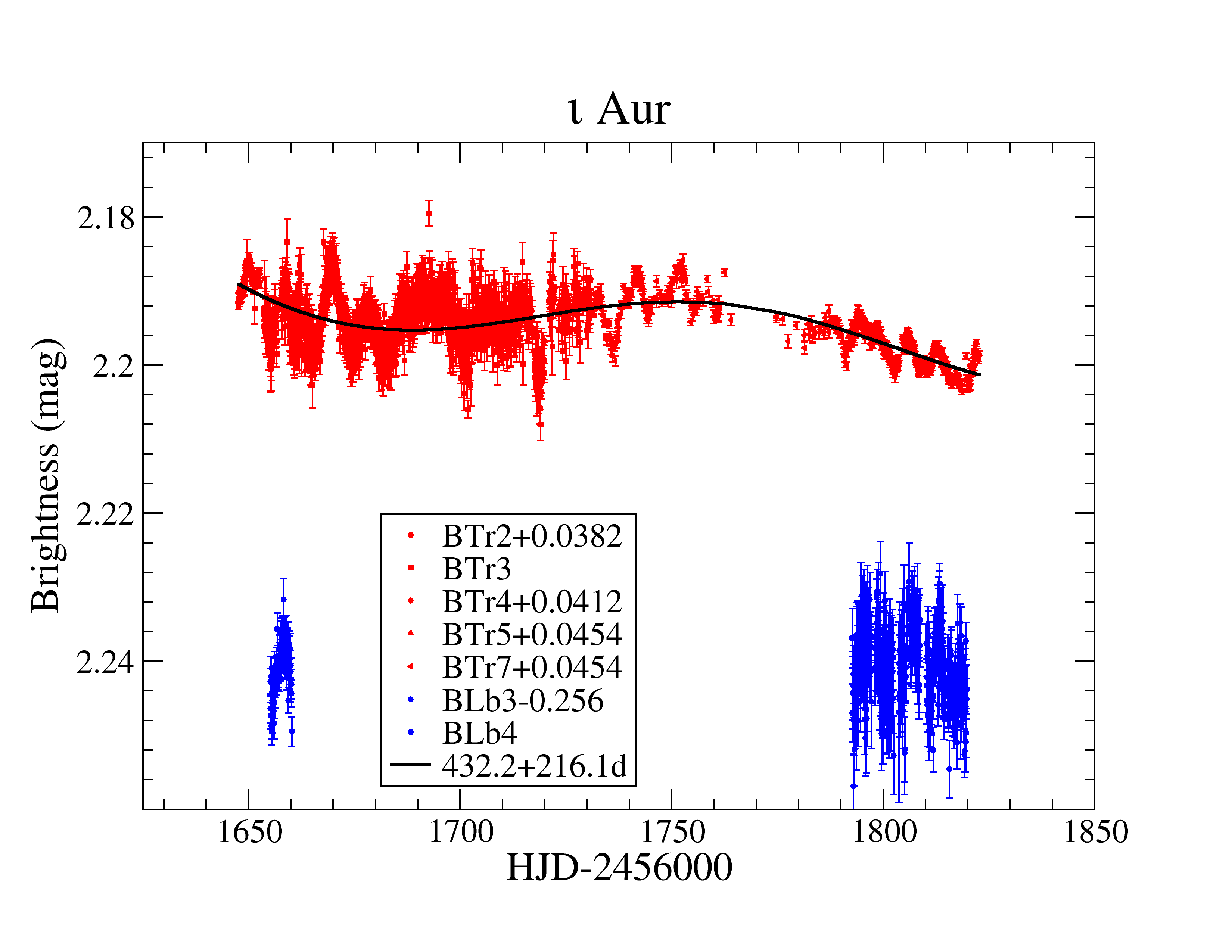}
\includegraphics[clip,angle=0,width=60mm]{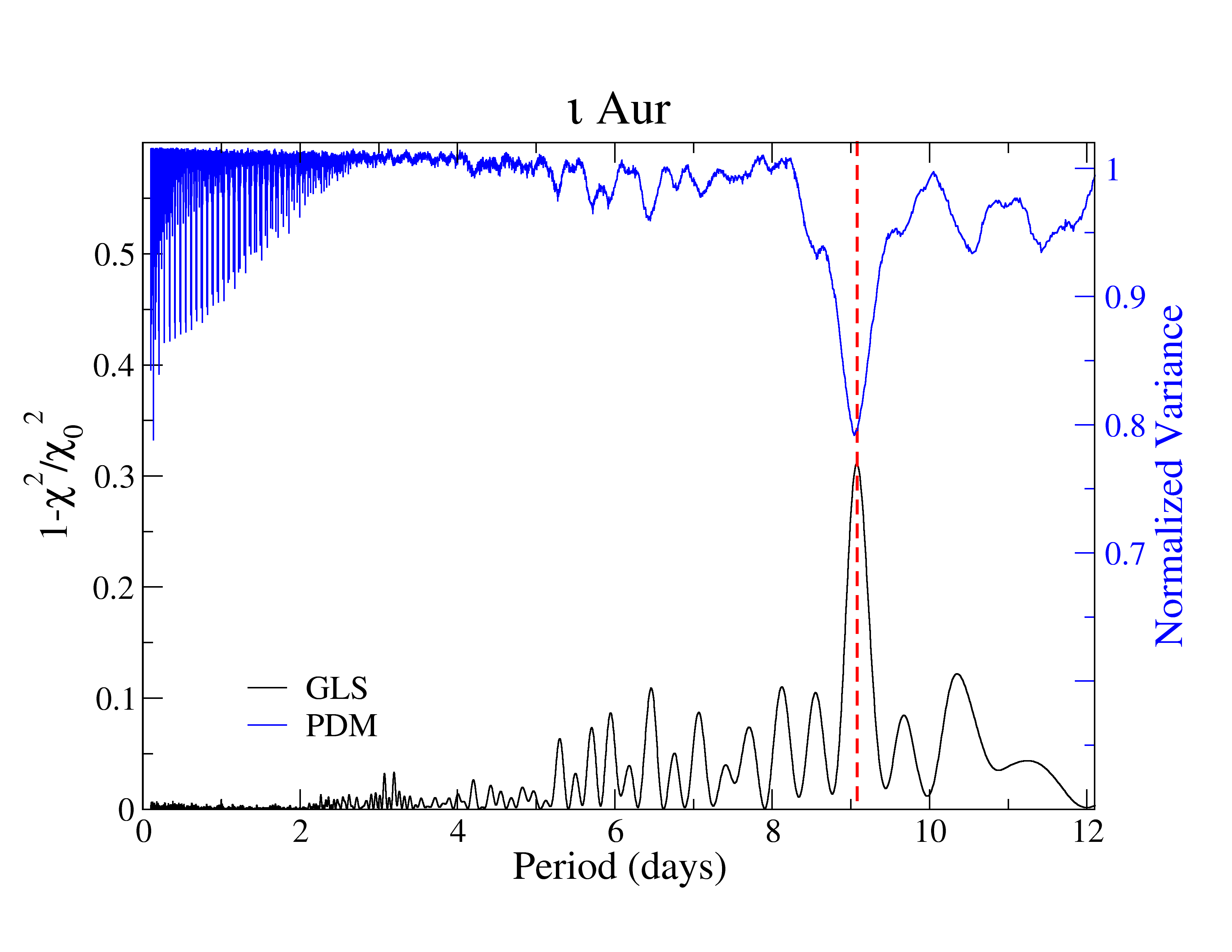}\includegraphics[clip,angle=0,width=60mm]{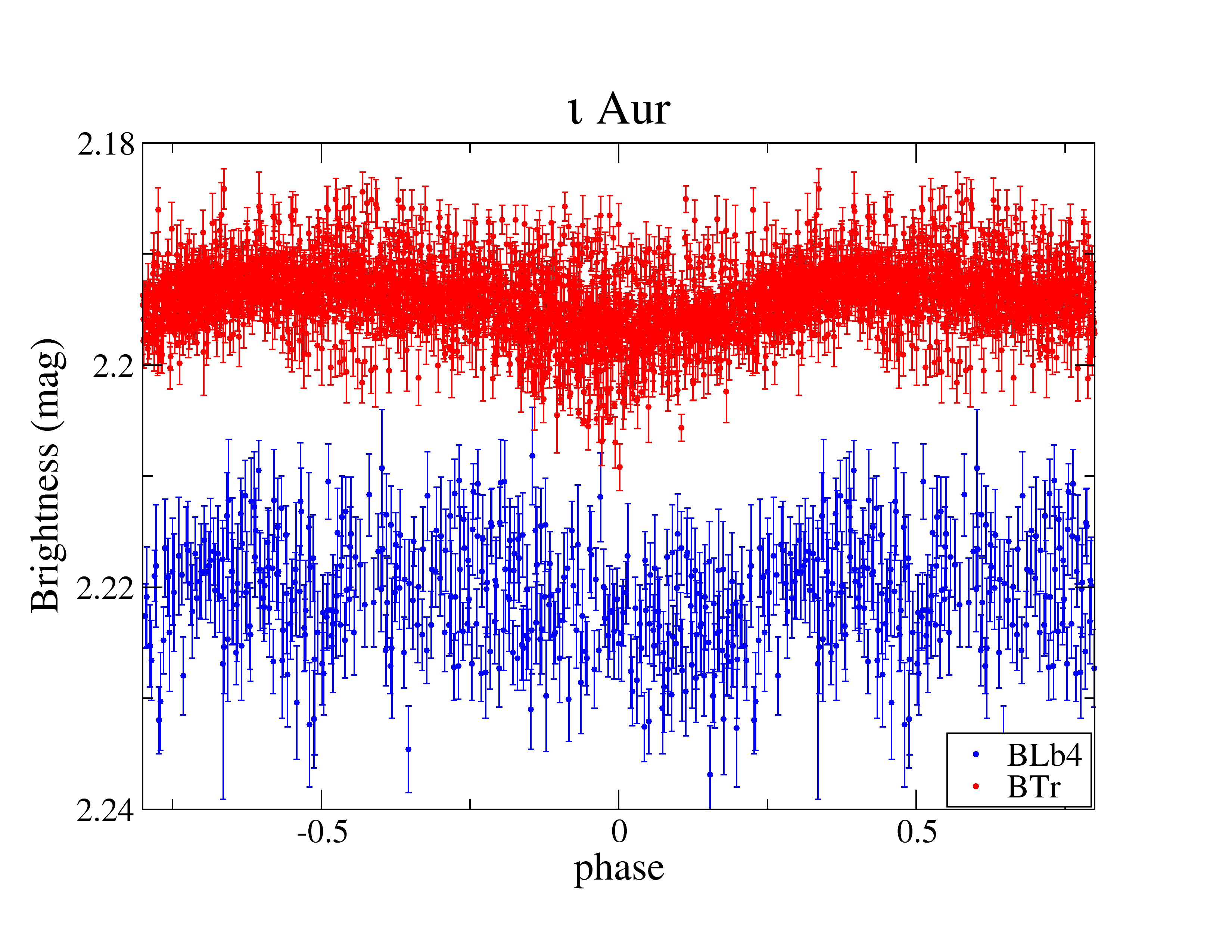}
\caption{Photometry of \iAur \ from BRITE. Panel a) Full data sample. The line in the BTr data (top) is a harmonic fit to remove the trend for the periodogram analysis. Panel b) Periodograms. Panel c) Phased light curve with the dominant 9.07 d period. Otherwise as in Fig.~\ref{Fcap}.  }\label{FiAur}
\end{figure*}
\begin{figure*}[tbh]
{\bf a.} \hspace{60mm} {\bf b.} \hspace{60mm} {\bf c.}\\
\includegraphics[clip,angle=0,width=60mm]{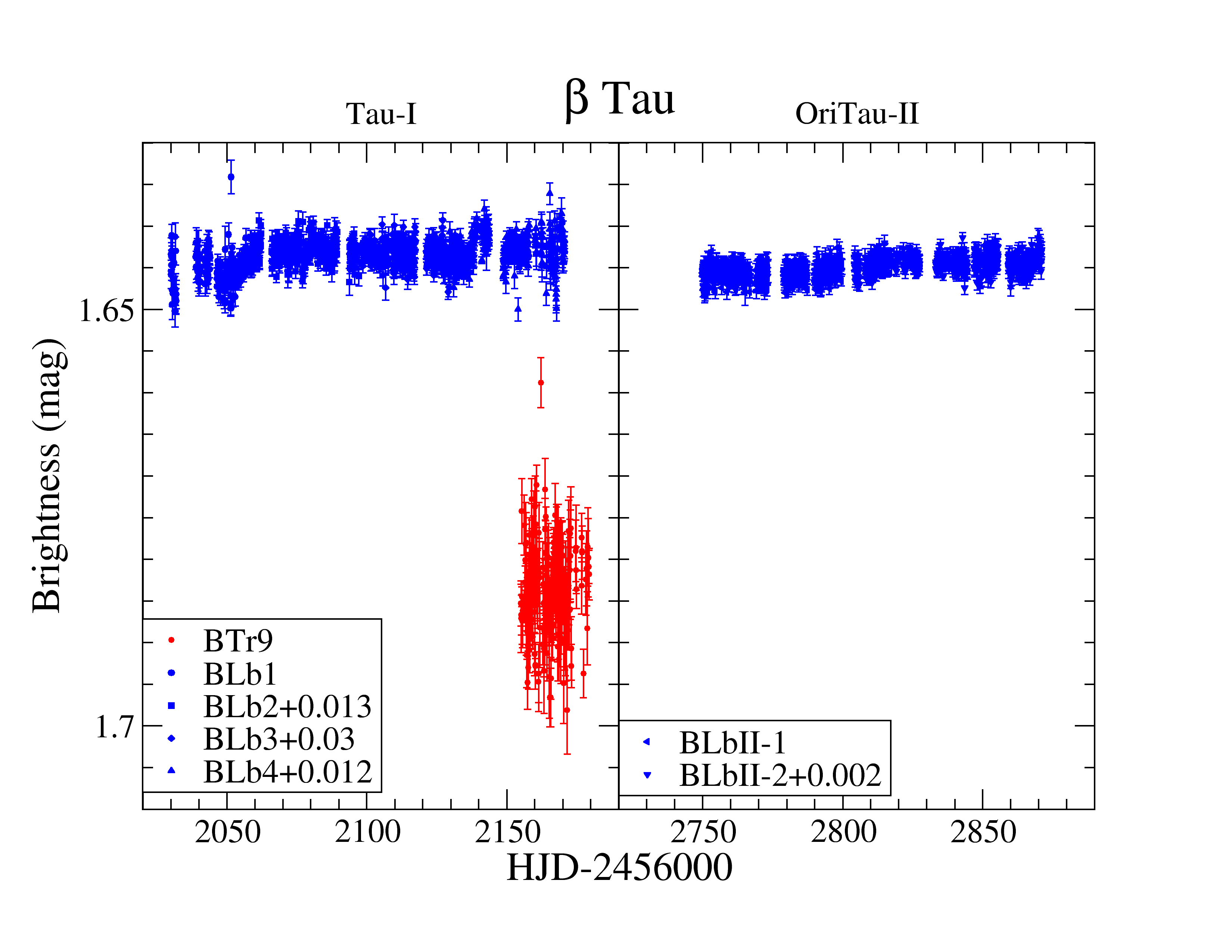}
\includegraphics[clip,angle=0,width=60mm]{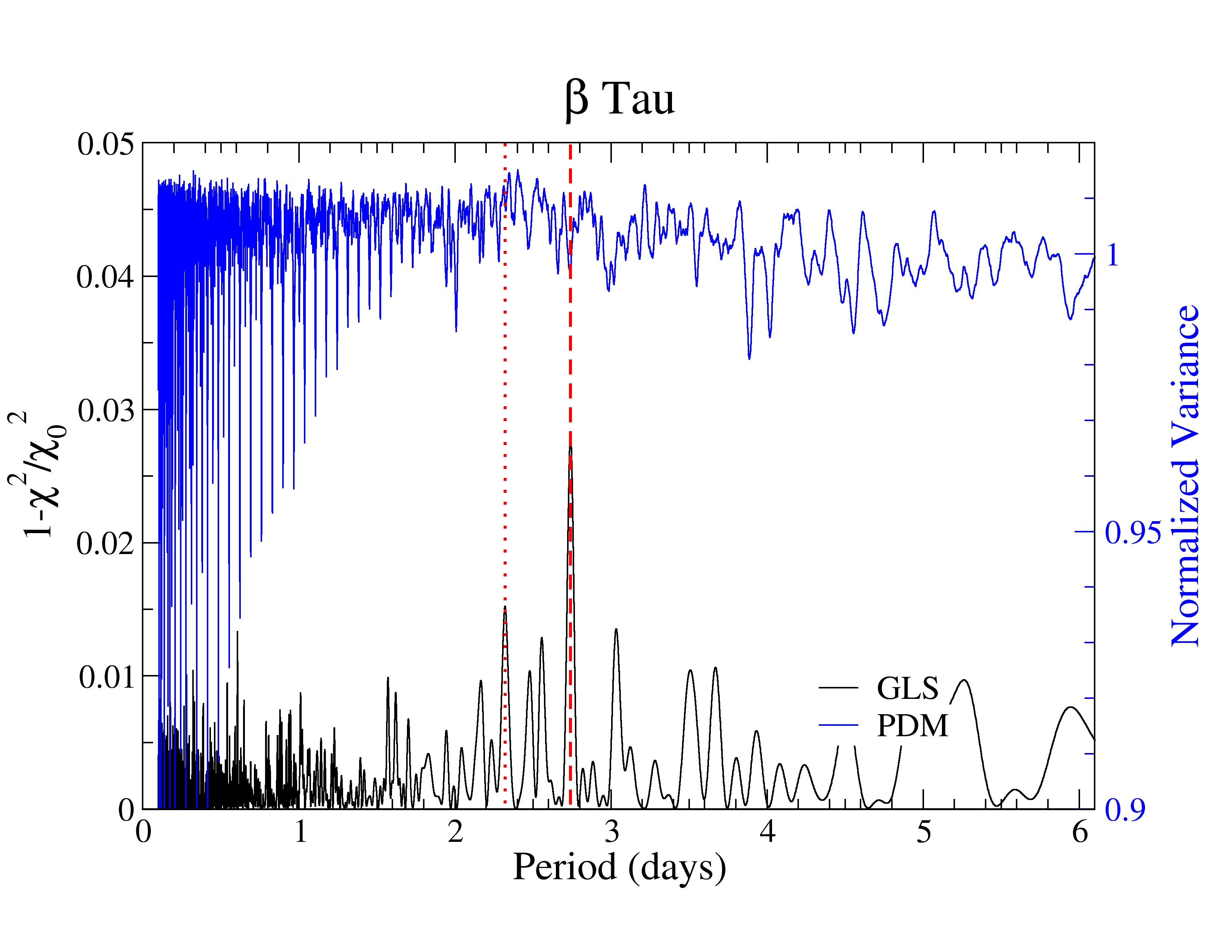}\includegraphics[clip,angle=0,width=60mm]{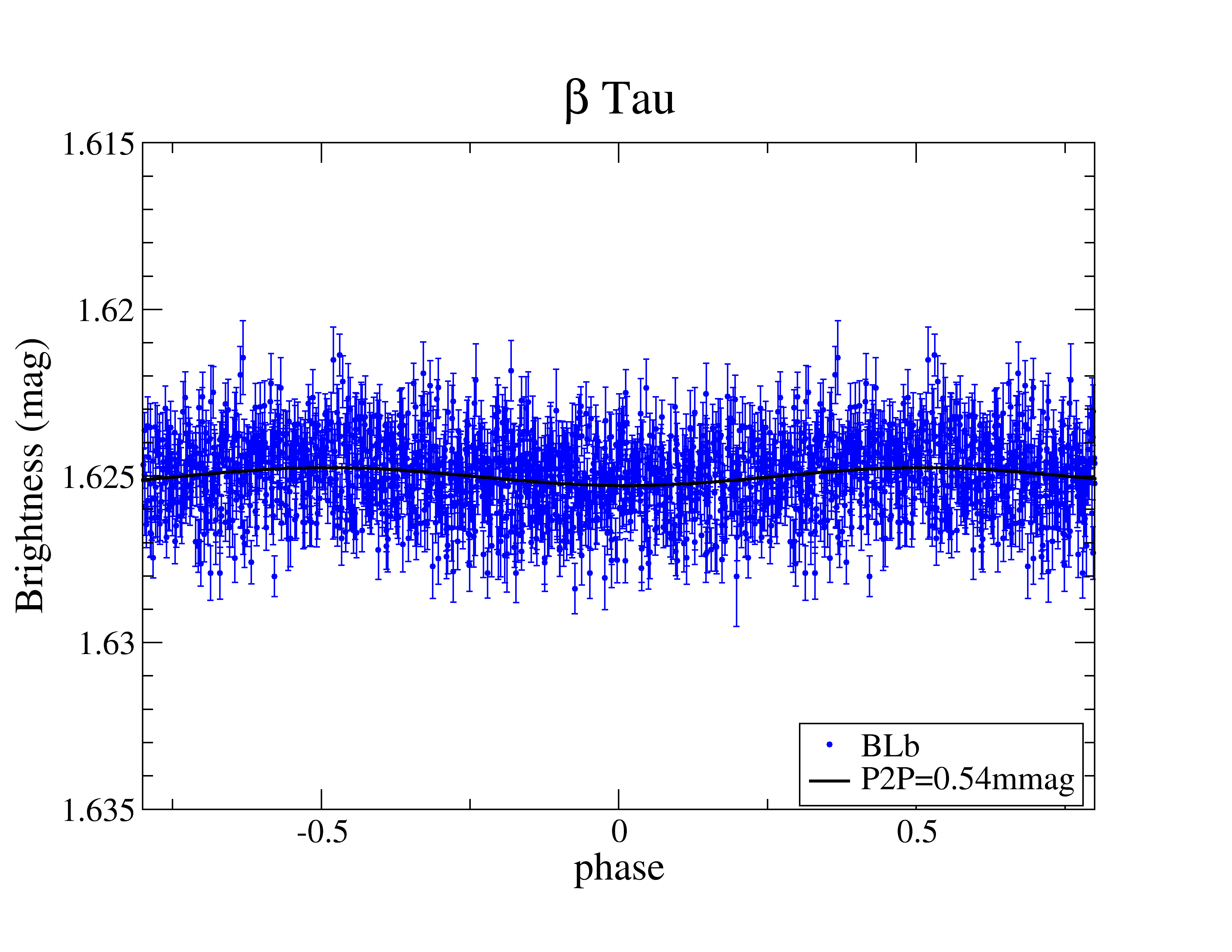}
\caption{Photometry of \bTau\ from BRITE. Panel a) Full data sample for the two visits in 2017 (left) and 2019 (right). Panel b) Periodograms. Panel c) Phased light curve with the 2.74 d period. Otherwise as in Fig.~\ref{Fcap}.  }\label{FbTau}
\end{figure*}
\begin{figure*}[tbh]
{\bf a.} \hspace{60mm} {\bf b.} \hspace{60mm} {\bf c.}\\
\includegraphics[clip,angle=0,width=60mm]{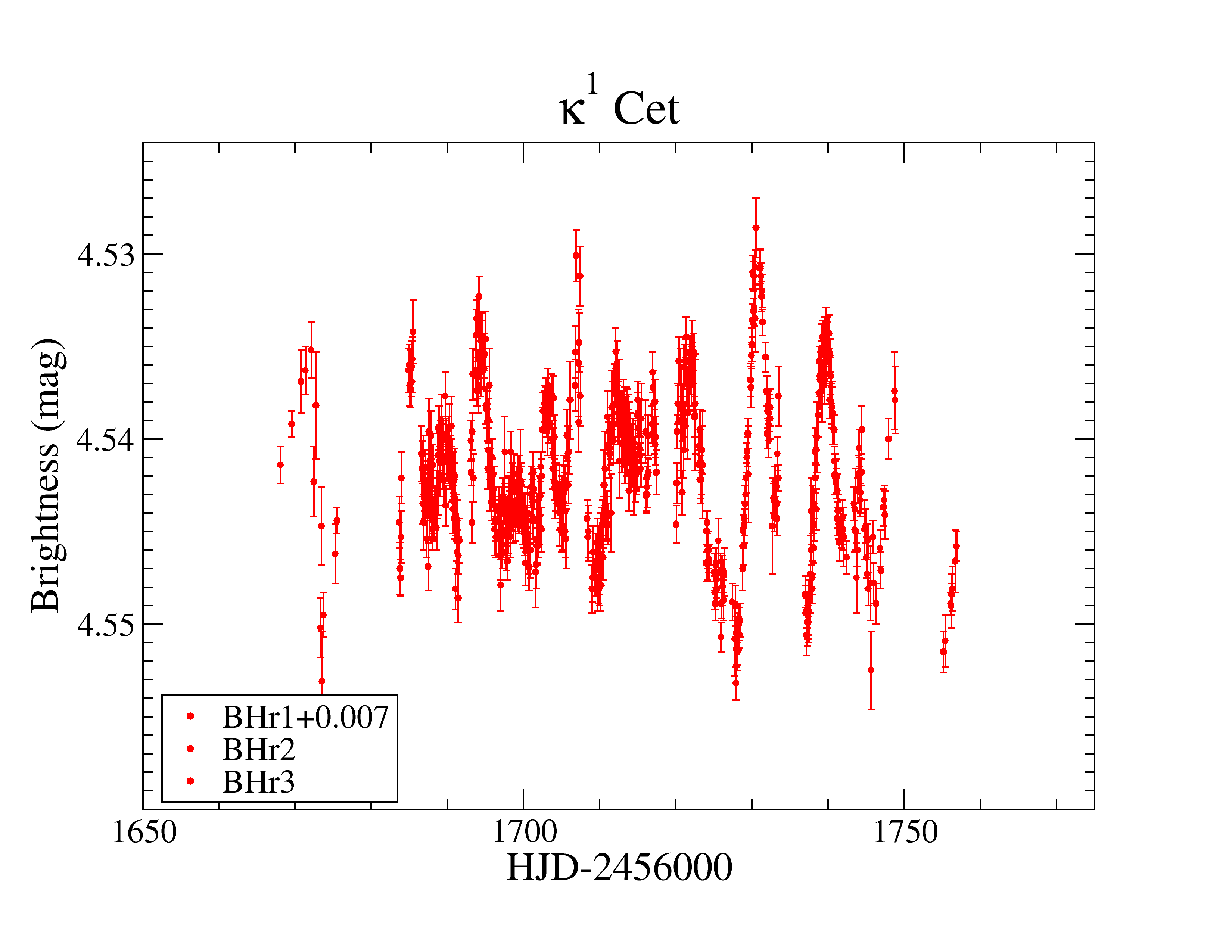}
\includegraphics[clip,angle=0,width=60mm]{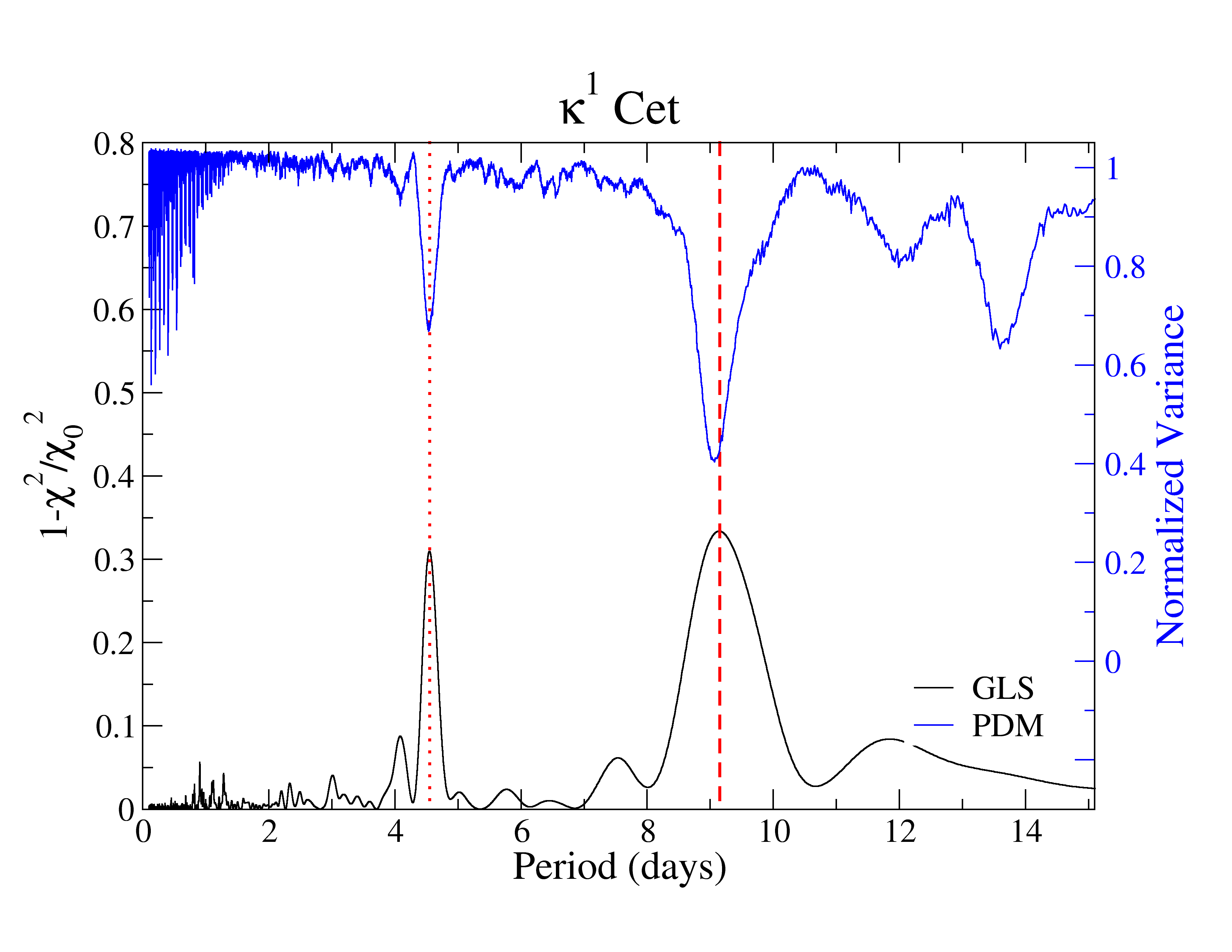}
\includegraphics[clip,angle=0,width=60mm]{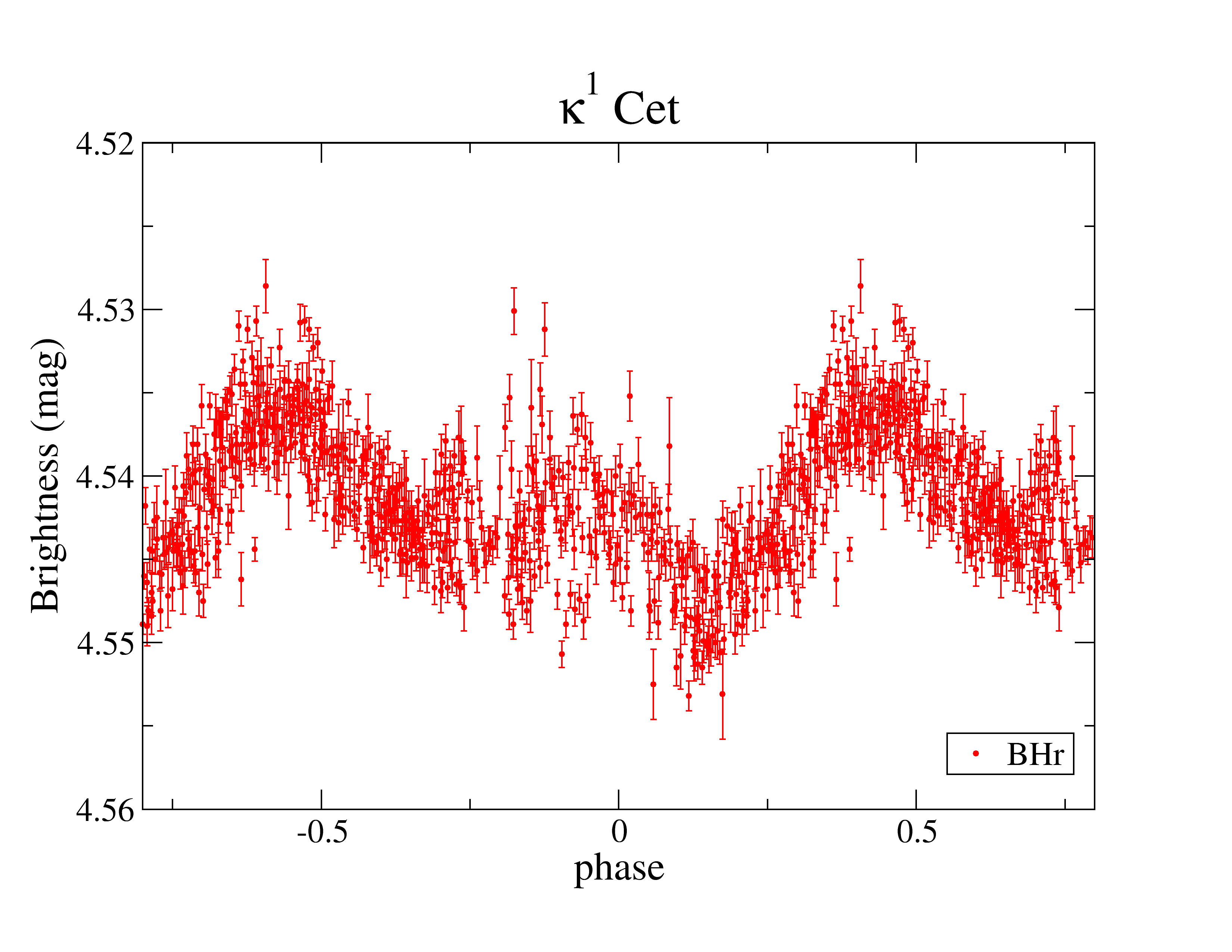}
\caption{Photometry of \kCet\ from BRITE. Panel a) Full data sample from BHr. Panel b) Periodograms. Panel c) Phased light curve with the dominant 9.065 d period. Otherwise as in Fig.~\ref{Fcap}.  }\label{FkCet}
\end{figure*}

{$\varepsilon$ Aur}. All BTr setups are relatively good quality. The blue-filter BLb observations are checkered; BLb1 duration was short and was rejected, BLb3 was affected by instrumental effects and was partly rejected. BLb4 was fine but data appear scattered with respect to BTr. Figure~\ref{FepsAur}a shows the full data set. The red-filter data show a pronounced brightness increase of 125\,mmag between TJD\,1680 and 1715, that is within 35 days, and a decrease back to the previous brightness at TJD\,1760, that is another 45 days later. Both periodograms in Fig.~\ref{FepsAur}b suggest a 74.25 d variability but can not exclude longer harmonics due to the limited length of the data set. However, if we allow for a multi-harmonic fit with a total of three periods based on $P_0=152.66$\,d, that is 152.66, 152.66/2, and 152.66/3\,d, we arrive at the best overall fit of the light curve as shown in Fig.~\ref{FepsAur}c. Similar fits with a period of 74.25\,d do not match the light curve at all, while a multi-harmonic fit with a fundamental period of 204.89\,d also matches the light curve reasonably well (see Fig.~\ref{FepsAur}c). We note that the fits worsen for multi-harmonics with $n>3$ simply because of the limited length of the data set. Any individual frequency up to 15$f_0$ could not alone fit the light curve. Therefore, we suggest that the period of 152.66\,d is likely the true fundamental pulsation period of this F0 supergiant.

\begin{figure*}[tb]
{\bf a.} \hspace{60mm} {\bf b.} \hspace{60mm} {\bf c.}\\
\includegraphics[clip,angle=0,width=60mm]{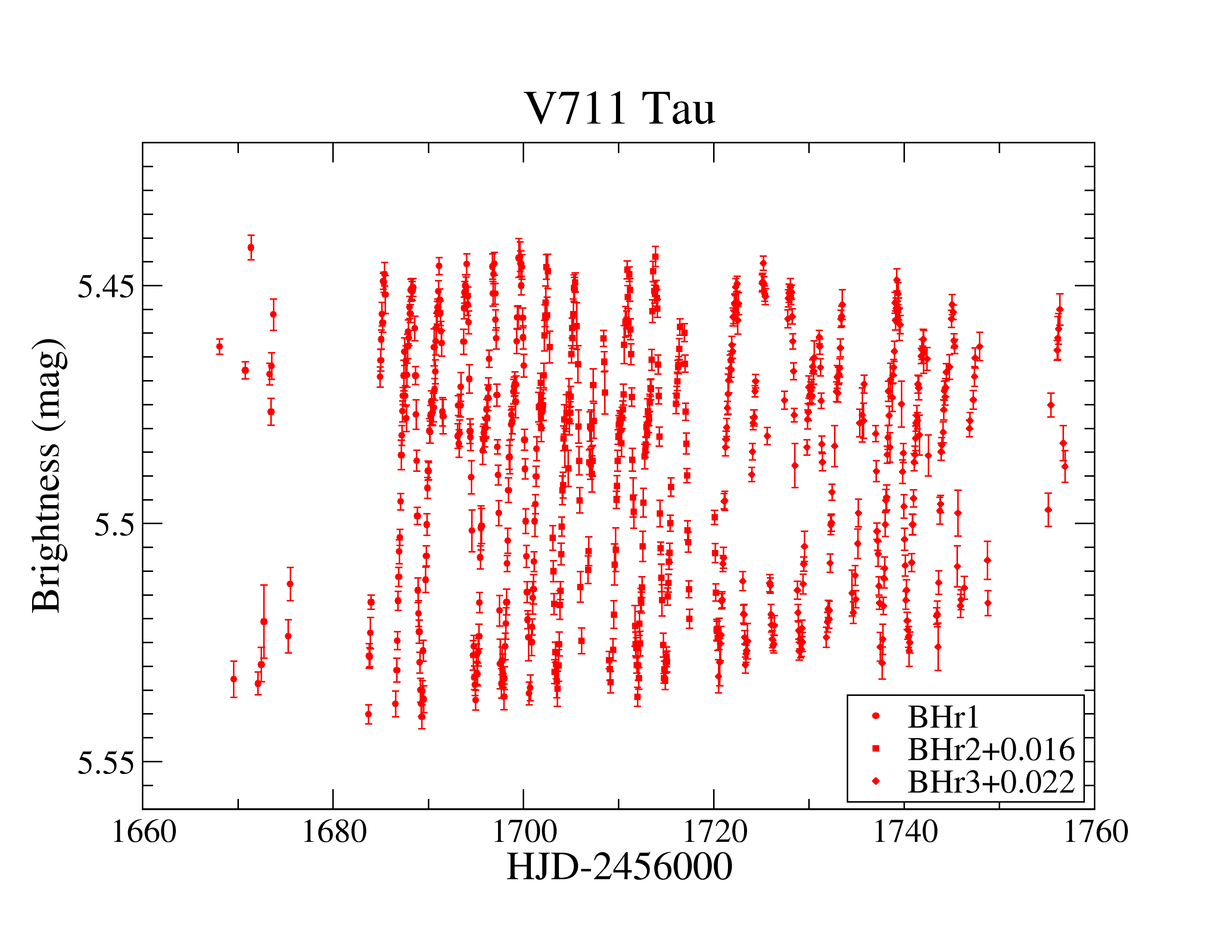}
\includegraphics[clip,angle=0,width=60mm]{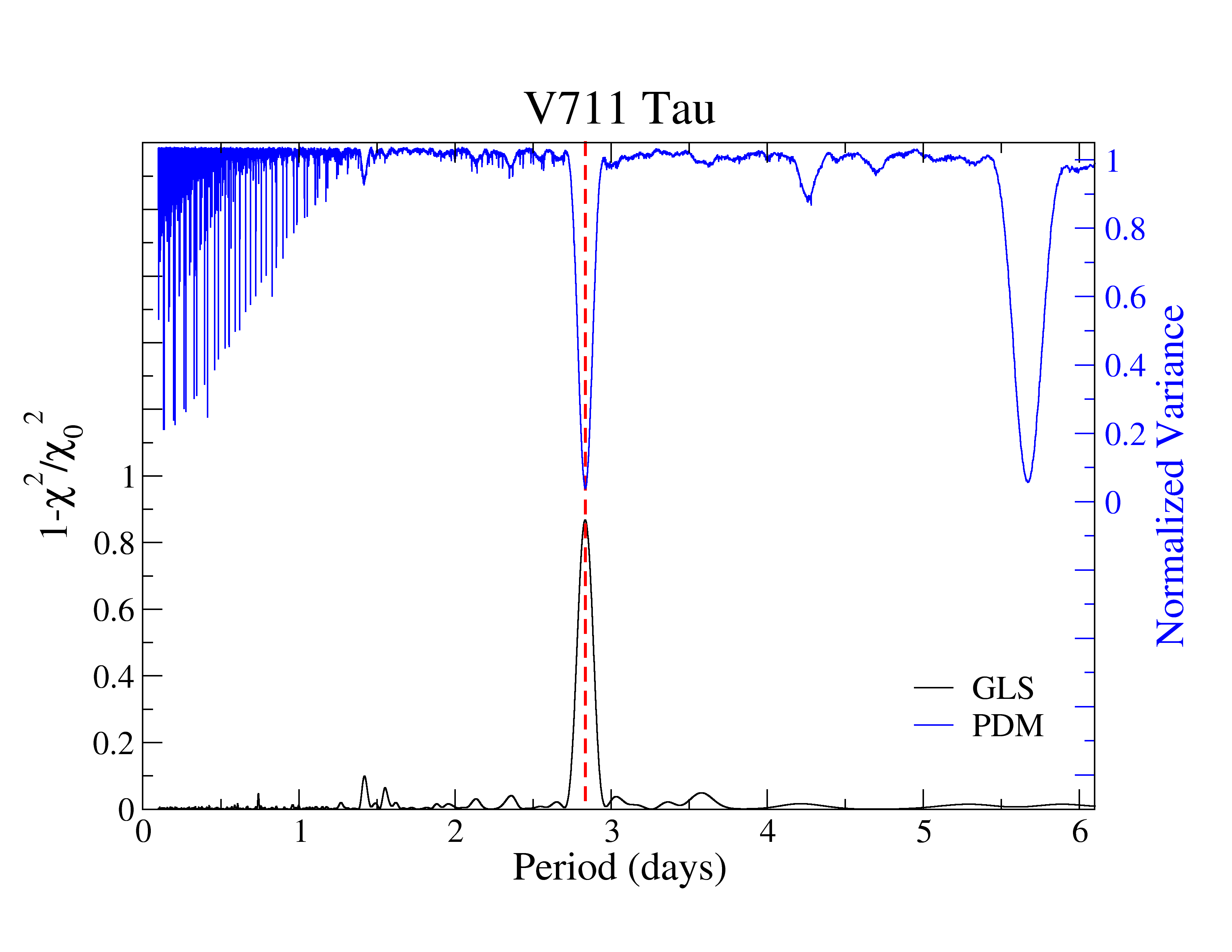}\includegraphics[clip,angle=0,width=60mm]{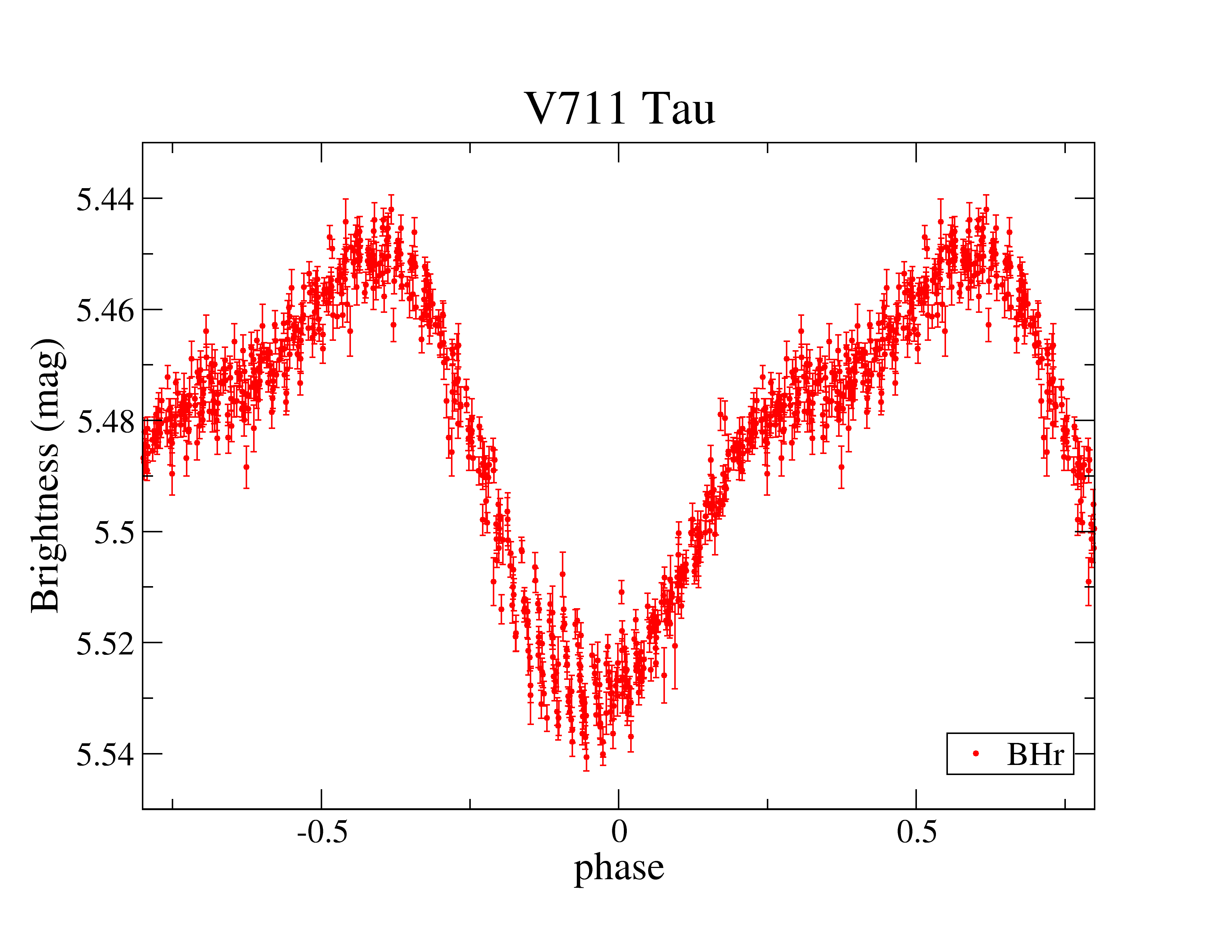}
\caption{Photometry of \HR\ from BRITE. Panel a) Full data sample from BHr. Panel b) Periodograms. Panel c) Phased light curve with the 2.836 d period. Otherwise as in Fig.~\ref{Fcap}.  }\label{Fv711}
\end{figure*}

\begin{figure*}[tbh]
{\bf a.} \hspace{60mm} {\bf b.} \hspace{60mm} {\bf c.}\\
\includegraphics[clip,angle=0,width=60mm]{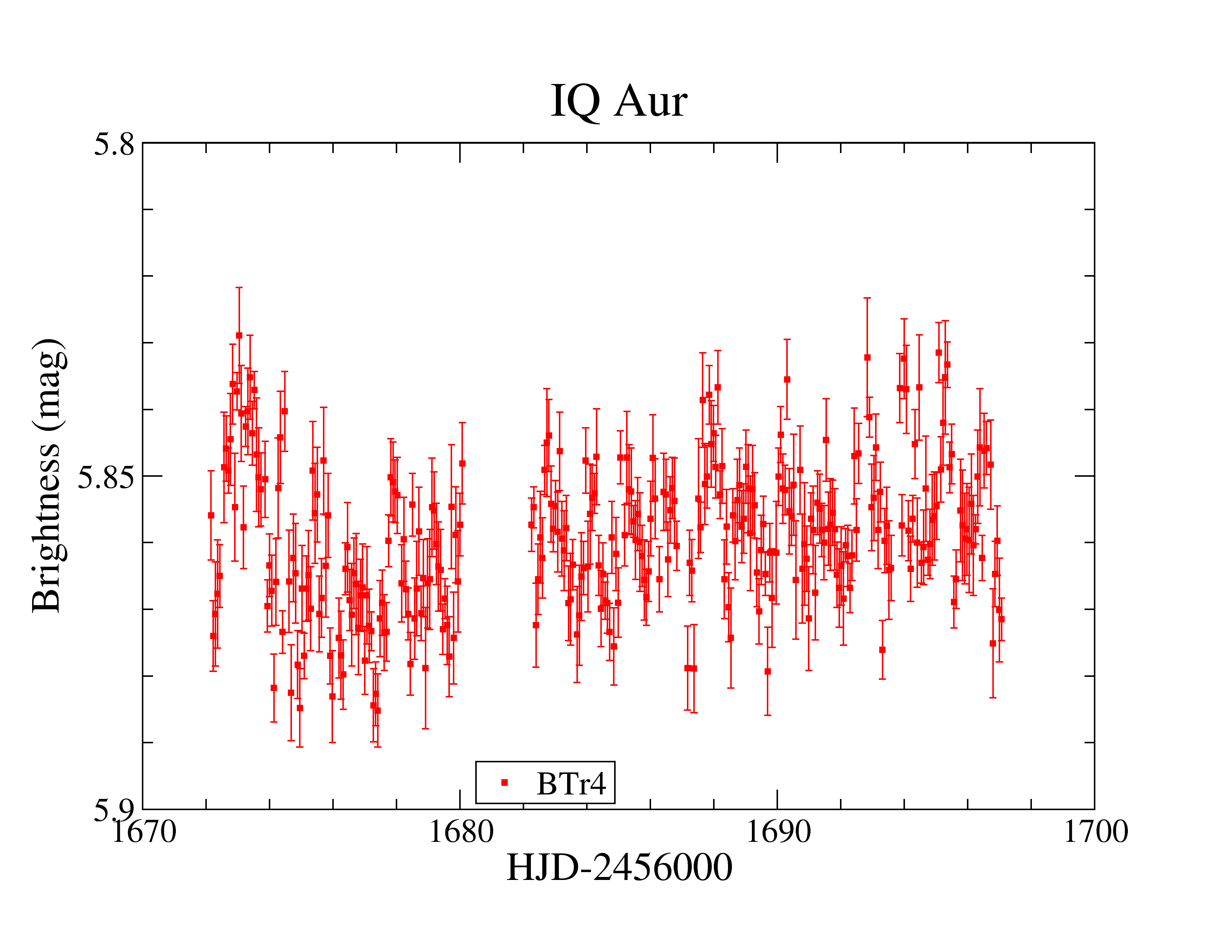}
\includegraphics[clip,angle=0,width=60mm]{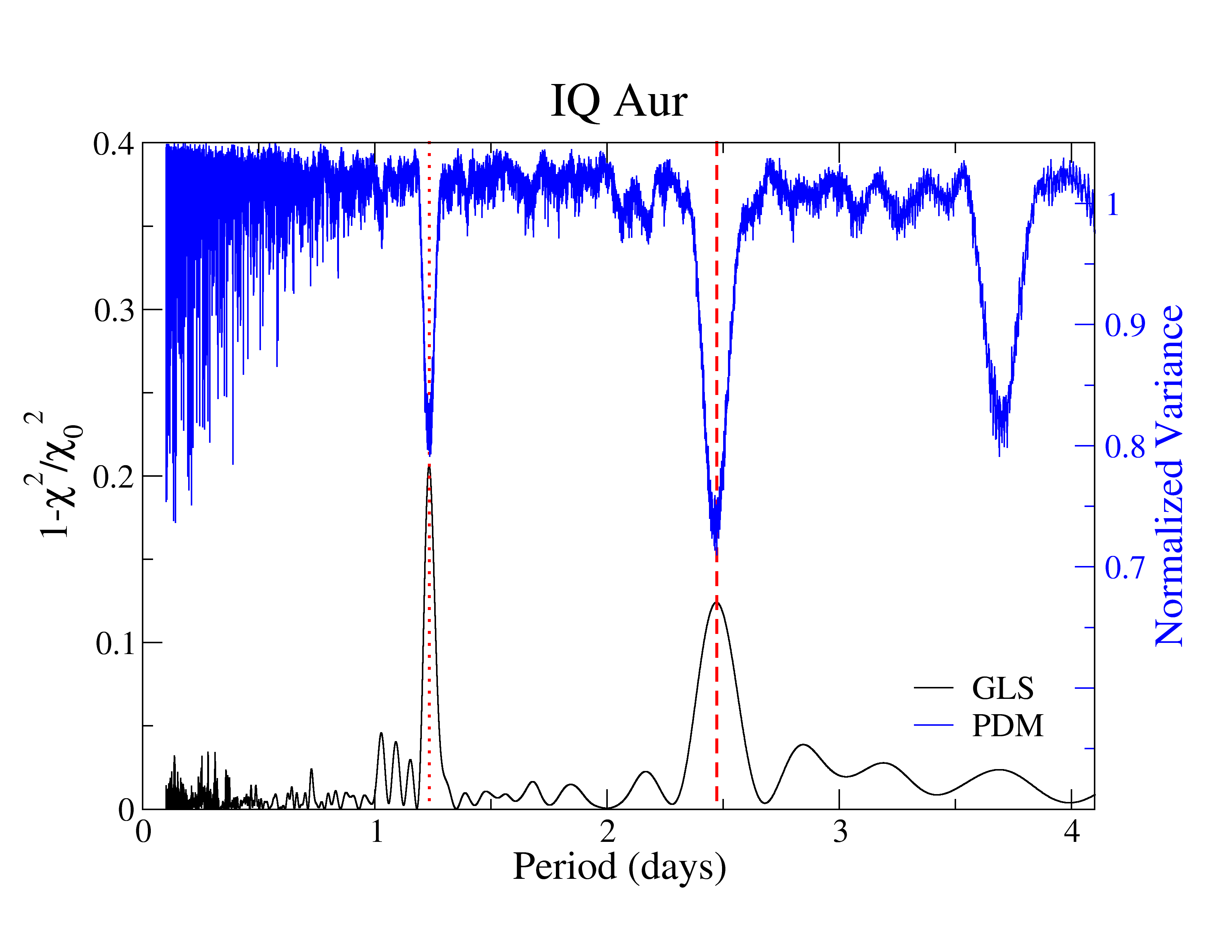}
\includegraphics[clip,angle=0,width=60mm]{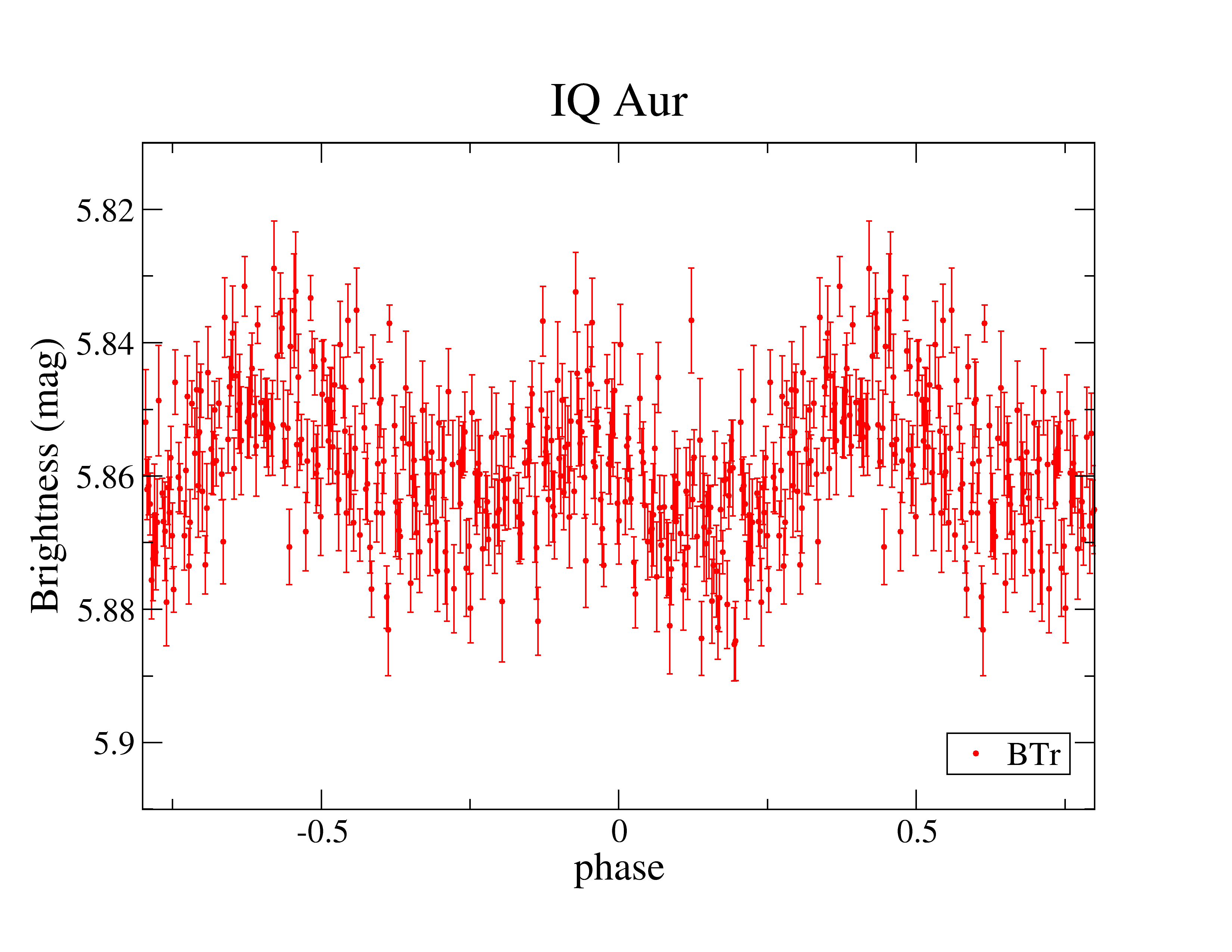}
\caption{Photometry of \iqAur\ from BRITE . Panel a) Full data sample from BTr. Panel b) Periodograms. Panel c) Phased light curve with the 2.46 d period. Otherwise as in Fig.~\ref{Fcap}.  }\label{FIQAur}
\end{figure*}

{$\zeta$ Aur}. As for $\beta$ Aur, this target was observed for 176 days by BTr with only a single 10 d gap around TJD\,1770 (Fig.~\ref{FzAur}a). In addition, BLb monitored the target for $\approx$30 days at the beginning and $\approx$15 days close to the end of the BTr coverage. The earlier of these two BLb data sets had rather large scatter and was discarded. The scatter of the latter set was three times smaller, but was still approximately three times larger than the contemporaneous BTr red-filter data. The red BTr light curve exhibits semi-regular variability with up to 0.1\,mag without a clear and convincing periodicity within our time coverage. Both periodograms (Fig.~\ref{FzAur}b) converged on a best-fit period of 70\,d that stem from the two dips that were covered. This is clearly seen in the phased light curve in Fig.~\ref{FzAur}c. The 70 d period may be reminiscent of the proposed tidal-induced, nonradial pulsations of the ellipsoidal supergiant (Eaton et al. \cite{eat}). By the end of our time coverage we caught the ingress of an atmospheric eclipse of the B star (but unfortunately not the egress). The total dimming in the BRITE red bandpass was $\approx$0.4\,mag. However, our light curve consisted of two different slopes at that time; initially a shallower one related to the apparent change of the ellipsoidal shape of the K giant, and a steeper one starting at TJD\,1809, which is the actual eclipse ingress. The eclipse full depth amounts to 0.14\,mag, which compares to the 0.15\,mag of a previous eclipse covered in $V$ on the ground by Eaton et al. \cite{eat}. Precise eclipse timings of $\zeta$\,Aur are rare because of the length of the totality of around 37\,d and the long orbital period of 972\,d. Griffin \cite{griff} noted that times of mid-eclipse had slipped over the past decades with respect to the spectroscopic orbital period and suggested that accurate eclipse timings could solve the discrepancy. Griffin's spectroscopic ephemeris predict times of conjunctions at MJD $2,445,191.90 + 972.164(\pm 0.041)\times n$, suggesting a mid-eclipse time for $n=13$ at 2,457,830.032 in agreement with our observations. Although our data do not cover the full totality,  they still allow a reasonable determination of the onset of the eclipse at 2,457,810.0$\pm$0.4.

{$\eta$ Aur}. While red-filter (BTr) data are of good quality, the blue-filter (BLb) data are scarce and the first part is affected by jumps with one-day intervals that we do not understand (especially in the BLb3 setup). Therefore, blue-filter data from BLb before the gap around TJD\,1770 were discarded. Its earlier data were rather noisy and were not used in the analysis either. The rms of the entire BTr light curve is comparably large and indicates unresolved low-amplitude variability (Fig.~\ref{FetaAur}a). Both periodograms in Fig.~\ref{FetaAur}b recover a dominating period of 1.28\,d from red and blue data independently. A least-squares fit based on this period alone suggests amplitudes of around 10\,mmag, where the blue light curve exhibits a marginally larger  amplitude than the red curve and is also indicative of containing more than one period. This is also seen in both periodograms with additional well-defined periods of 1.23\,d and 6.6\,d.

We therefore performed a more detailed period search with the software Period04 (Lenz \& Breger \cite{period}), involving multifrequency fitting and pre-whitening techniques. This search was carried out on the BTr data only, as the useful BLb data have a time base that is too short to resolve the individual signals. We detected a total of six periods at a significant level (amplitude S/N$>$4 in the periodogram; Breger et al. \cite{breger}), and conjecture that the longer 6.6 d period corresponds to a combination frequency difference. The results of the analysis are listed in Table~\ref{Teta} and clearly indicate that \etaAur\ is a newly discovered SPB star, and hence one of the three brightest representatives of its kind (the others are $\kappa$\,Cen and $\kappa$\,Vel; Daszynska-Daszkiewicz et al. \cite{das2}).

\begin{table}[tb]
\caption{Signals detected in the BTr photometry of \etaAur .}\label{Teta}\begin{tabular}{lllll}
\hline \hline \noalign{\smallskip}
ID & Frequency  & Period & Amplitude & S/N \\
   & (d$^{-1}$) & (d)    & (mmag)    & \\
\noalign{\smallskip} \hline \noalign{\smallskip}
f1   &0.7753(1) &1.2900(1) &3.6(1) &15.8 \\
f2   &0.8034(2) &1.2447(3) &1.7(1) &7.8 \\
f3   &0.6842(3) &1.4615(5) &1.3(1) &5.6 \\
f4   &0.5932(2) &1.6858(7) &1.4(1) &5.9 \\
f5   &0.4403(2) &2.2711(12)&1.5(1) &6.0 \\
f6   &0.8105(2) &1.2338(4) &1.4(1) &6.4 \\
f5--f4&0.1529    &6.543     &2.3(1) &9.0 \\
\noalign{\smallskip} \hline
\end{tabular}
\tablefoot{Values in parentheses are the errors expressed in last digit numbers. The last line represents frequency f5 minus f4.}
\end{table}

{$\theta$ Aur}. Figure~\ref{FtAur}a shows the full light curve with its impressive variability. The BTr data are good, BLb data are also good but scarce and of lower quality than the BTr data. The data set BLb1 was rejected because of its very short duration. The BTr observed for 176 days starting TJD\,1645 until 1823 with only a single 10 d gap around TJD\,1770. The BLb pointed to this field for two short epochs at TJD\,1655--1690 and 1792--1820. Both red-filter (BTr) and blue-filter (BLb) data show continuous modulation with a peak-to-peak amplitude of 34\,mmag and a clear periodicity of 3.6189$\pm$0.0001\,d (Fig.~\ref{FtAur}b). The phased light curves in Fig.~\ref{FtAur}c reveal significant deviations from a pure sinusoidal shape, but are indicative of the (known) asymmetric distribution of its surface chemical elements.

{$\nu$ Aur}. There are four setups of BTr data and no BLb data. Setup BTr7 was split into two subsets because of a 3\,mmag jump in the middle of the run at TJD\,1785, which could not be de-correlated. The data are otherwise of reasonable quality. The BTr observed for the full $\approx$176 d time span, starting from TJD\,1645 until 1823 with only the single 10 d gap around TJD\,1770. These data are shown in Fig.~\ref{FnuAur}a. The long-term trend with an amplitude range of $\approx$10\,mmag is obvious but the 176-d observing window is still too small to cover a full cycle. We fitted a quadratic polynomial and removed the trend. The residuals were then subjected to the same periodogram analysis as for the other stars and are shown in Fig.~\ref{FnuAur}b. Both algorithms converged on a 19.16$\pm$0.08\,d period with a full amplitude from a least-squares fit of 1.8\,mmag (Fig.~\ref{FnuAur}c). However, this is not significant given the overall rms of 2.7\,mmag, and is not seen in the STELLA RV data either.

{$\iota$ Aur}. The BLb had the same instrumental problems that were experienced for other stars. The BLb1 setup was too short and was rejected. Only parts of the BLb3 and BLb4 data were used because of poor quality (large scatter). The first setup on BTr1 was also too short and was rejected, only setups BTr2, 4, 5, and 7 were used. The BTr3 data were of lower quality at the beginning because the exposure time was much shorter for this setup than for the other setups (0.15\,s in comparison to 3\,s). Exposure times were then re-optimized for the brightest stars in the field; this is visible in the second half of the light curve of \iAur\ (Fig.~\ref{FiAur}a). Data from both satellite show variability on several timescales. A long-term trend is obvious in the BTr data but, in contrast to \nAur , shows a minimum and a maximum in the light curve. However, the 176 d observing window is still too short  to call the trend periodic, although it can be fitted with a period of 432\,d plus its first harmonic. Because this period  is too speculative, we decided to just fit a quadratic polynomial and then remove the trend for the period analysis. The periodograms on the residuals are shown in Fig.~\ref{FiAur}b. Both algorithms detect a clear 9.071$\pm$0.002\,d period from the red BTr data but not from the blue BLb data. Its length is far too short for a rotation period for an inactive hybrid K3 bright giant. The full amplitude is 4\,mmag and is likely attributable to complex semi-regular variability.

{$\beta$~Tau = $\gamma$~Aur}. This star was not part of the original Auriga field as shown in Fig.~1, but was observed later in field 31-Tau-I in 2017 and in field 48-OriTau-II in 2019. Its data come mostly from BLb with setups 1--4 (Fig.~\ref{FbTau}a). The BTr observations in these fields were too short and/or had  an integration time that was too short and were not useable. Photometric variability, if present at all, is not seen easily, at least not in the  light curves. Nevertheless, the GLS periodogram showed a significant period at 2.737\,d while the PDM did not (Fig.~\ref{FbTau}b). We note that the two BLb data sets from 2017 and 2019 were treated separately but gave identical periods within their errors (c/o Table~\ref{T3}). The phased light curve in Fig.~\ref{FbTau}c from a least-squares fit gave a full amplitude of only 0.54\,mmag. While very uncertain, we still take it as evidence for the rotation period of this chemically peculiar B7 giant.

{$\kappa^1$\,Cet}. The BHr coverage was 90 days with two gaps of one week around TJD\,1680 and TJD\,1750 (Fig.~\ref{FkCet}a). A maximum peak-to-peak amplitude of 25\,mmag was seen around TJD\,1730 but otherwise the light curve remained double-peaked with a time-averaged amplitude of around 10\,mmag. Both periodograms (Fig.~\ref{FkCet}b) show a dominant period at 9.065\,d (its harmonic at 4.5\,d is the second strongest period). If phased with this 9 d period (Fig.~\ref{FkCet}c) the intrinsic time-dependent light-curve variations due to spot changes or redistribution mimic increased scatter.

{V711\,Tau}. As for $\kappa^1$\,Cet, only data from BHr are available. Its coverage is also $\approx$90 days with two gaps of one week around TJD\,1680 and 1750 (Fig.~\ref{Fv711}a). The full amplitude decreased from $\approx$100\,mmag at the beginning to $\approx$70\,mmag at the end of the time series. The rotation period of \HR\ has been well known for many decades and is synchronized to the orbital motion of this close binary. Our periodograms in Fig.~\ref{Fv711}b confirm this period and specify it to 2.8359$\pm$0.0003\,d. The phase curve in Fig.~\ref{Fv711}c is just for orientation because the spot-induced changes in the light curve are a function of time and alter the amplitude as well as the shape of the light curve.

{IQ\,Aur}. The BRITE coverage was only with BTr within a single setup and thus comparably short, just 25\,d in total from TJD\,1672 to 1697 (Fig.~\ref{FIQAur}a). Nevertheless, it covered ten cycles of the previously determined period. Our two periodograms in Fig.~\ref{FIQAur}b confirm this period and specify it to 2.463$\pm$0.009\,d. The phase curve with this period is shown in Fig.~\ref{FIQAur}c with an average amplitude in the (red-sensitive) BTr data of 20\,mmag. The light curve has two maxima and two minima just like the magnetic-field curve from Zeeman-sensitive spectral lines. This had been explained (Bohlender et al. \cite{bohl}) as the two extremes of the magnetic field density and the two central-meridian passages of the magnetic equator because of the tilt between the rotational and magnetic axes.

\subsection{Radial velocities and orbits from STELLA data}

Radial velocities from the STELLA spectra were determined from an order-by-order cross correlation with a synthetic template spectrum (models from ATLAS-9; Kurucz~\cite{kur}). Sixty out of the 82 echelle orders were used for the cross correlation and gave one relative velocity per echelle order. These velocities were then weighted according to the spectral region and averaged. Its rms defines our internal error ($\approx$6~\ms ). The external rms values are naturally significantly larger and strongly depend on spectral type, line broadening, and blending. All RVs in this paper are barycentric. Table~\ref{T2} lists relevant measurement results: the time range of the STELLA coverage, the number of spectra obtained ($N_{\rm spec}$), their lowest and highest signal-to-noise ratio (S/N), and the average RV and its standard deviation. The known SBs are indicated with either SB1 when single-lined or SB2 when double-lined. If the standard deviation was greater than three times the expected minimum external rms of 30\,\ms , then the velocity is marked variable ({\sl var}). We solved for the usual elements of a double-lined spectroscopic binary using the general least-squares fitting algorithm MPFIT (Markwardt \cite{mpfit}). For solutions with non-zero eccentricity, we used the prescription from Danby \& Burkardt (\cite{danby:burkardt}) for the calculation of the eccentric anomaly.

\begin{figure*}[tbh]
\includegraphics[angle=0,width=\textwidth,clip]{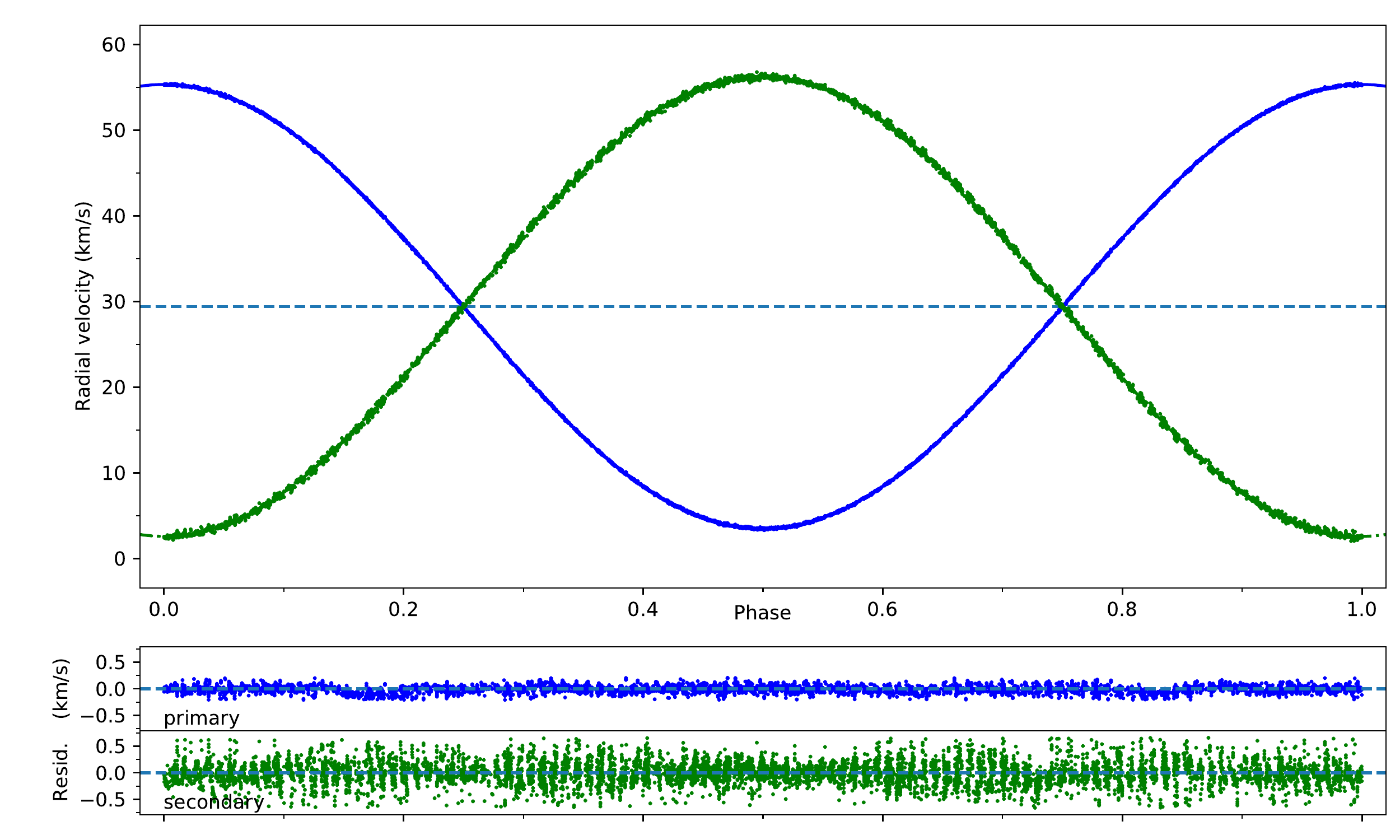}
\caption{Radial velocities from STELLA and orbit for \aAur\ = Capella. Blue dots denote the primary (G8\,III), green  dots the secondary (G0\,III). The bottom panels show the residuals from the final orbital solution with the systematics removed. Radial velocities covered the time range 2007 to 2020. }\label{Forbcap}
\end{figure*}

\subsection{Stellar parameters}

We selected five spectral orders from the STELLA/SES spectra covering the range 549--623\,nm and used these spectral orders to determine the stellar effective temperature, gravity, metallicity, and microturbulence. We applied the program ParSES (PARameters from SES), which is based on the synthetic spectrum fitting procedure laid out by Allende-Prieto et al. (\cite{all}). Model atmospheres and synthetic spectra were taken from the ATLAS-9 CD (Kurucz \cite{kur}). Synthetic spectra are pre-tabulated for large parameter ranges for a wavelength range of 380-920\,nm. All atmospheric calculations were done with a microturbulence of 2~\kms . This grid was then used to compare with the five selected echelle orders of each target spectrum. Table~\ref{T2} includes the results. We note that our limited model grid excludes the applicability of ParSES to stars with effective temperatures higher than 8000\,K.

Precise astrophysical parameters of giant stars are uncertain because there are just a few such stars that can be studied in spectroscopic binary systems and even fewer are also eclipsing binaries. Roche lobe overflow and mass exchange usually set a limit to the stellar radius in a close binary. Non-interacting binaries with giant components therefore mostly have very long orbital periods and are difficult to observe within the lifetime of an astronomer. There are, however, a few systems that have sufficiently short orbital periods (for a summary see, e.g., Andersen \cite{and}). Capella is such a system and consists of an active G0\,III and a G8\,III component in a relatively short 104 day orbit.

\subsection{Detailed STELLA results}

All but one star in our STELLA sample showed variable RVs. Only \kCet\ exhibits a constant velocity with a rms close to the expected external RV rms of $\approx$30\,\ms\ (Strassmeier et al.~\cite{orbits}). Four of the RV-variable stars were already known spectroscopic binaries (\aAur , \bAur , \epsAur , and \HR ), three of which are SB2. Our data for these stars  add to the pool of long-term velocities, in particular for \epsAur\ with its 27 yr orbital period.

{\aAur\ = Capella}. For our prime target, we used all our new and previously published STELLA RVs for a re-computation of its orbital elements, in particular for the difficult secondary star (dubbed the B component)  with its large rotational broadening. The final orbital solution is presented in Table~\ref{T4} and shown in Fig.~\ref{Forbcap} versus phase, and in the Appendix in Fig.~\ref{FAorbcap} versus HJD. This solution shows excellent agreement with the updated combined orbit by Torres et al. (\cite{torres}), which included an earlier subset of our STELLA data. For our present solution, a total of $\approx$9600 RVs per binary component over a time span of 12.9 years are available compared to 430 used by Weber \& Strassmeier (\cite{capella}). This long time coverage allows us to refine the orbital period just based on this homogeneous and high-precision data set alone. Our revised orbital period of 104.021572$\pm$0.000056\,d is 25\,s longer than that derived by Torres et al. (\cite{torres}) from data spread over a hundred years including the original Weber \& Strassmeier (\cite{capella}) STELLA data. The new period is well within the Torres et al. (\cite{torres}) 3$\sigma$ limit, but more precise by almost a factor of three.

Deriving an orbit from such a huge data set requires knowing and mitigating the systematic errors. For extracting the STELLA RVs, we applied the same two-dimensional cross-correlation technique as in our earlier paper (Weber \& Strassmeier \cite{capella}). Contrary to our initial orbital solution in 2011, the difference between the systemic velocities of the two components is now just 3\,\ms\ on average (derived from two SB1 solutions of the A and B components separately) and literally zero at the quadrature phases, as it should be. This strengthens our belief that the present solution is vastly superior than the solution presented earlier. The difference that we found in the 2011 solution was 59\,\ms\ and we had to simply correct for it without knowing the cause. The bottom panel in Fig.~\ref{Forbcap} shows the remaining residuals for both components, 63\,\ms\ for the primary and 198\,\ms\ for the secondary, albeit deviations are hardly visible anymore to the naked eye. For comparison,  the plot of RV versus HJD in Fig.~\ref{FAorbcap} shows the full residuals including the systematics.

\begin{figure*}[tbh]
{\bf a. \ \bAur} \hspace{80mm} {\bf b. \ \epsAur}\\
\includegraphics[clip,angle=0,width=87mm]{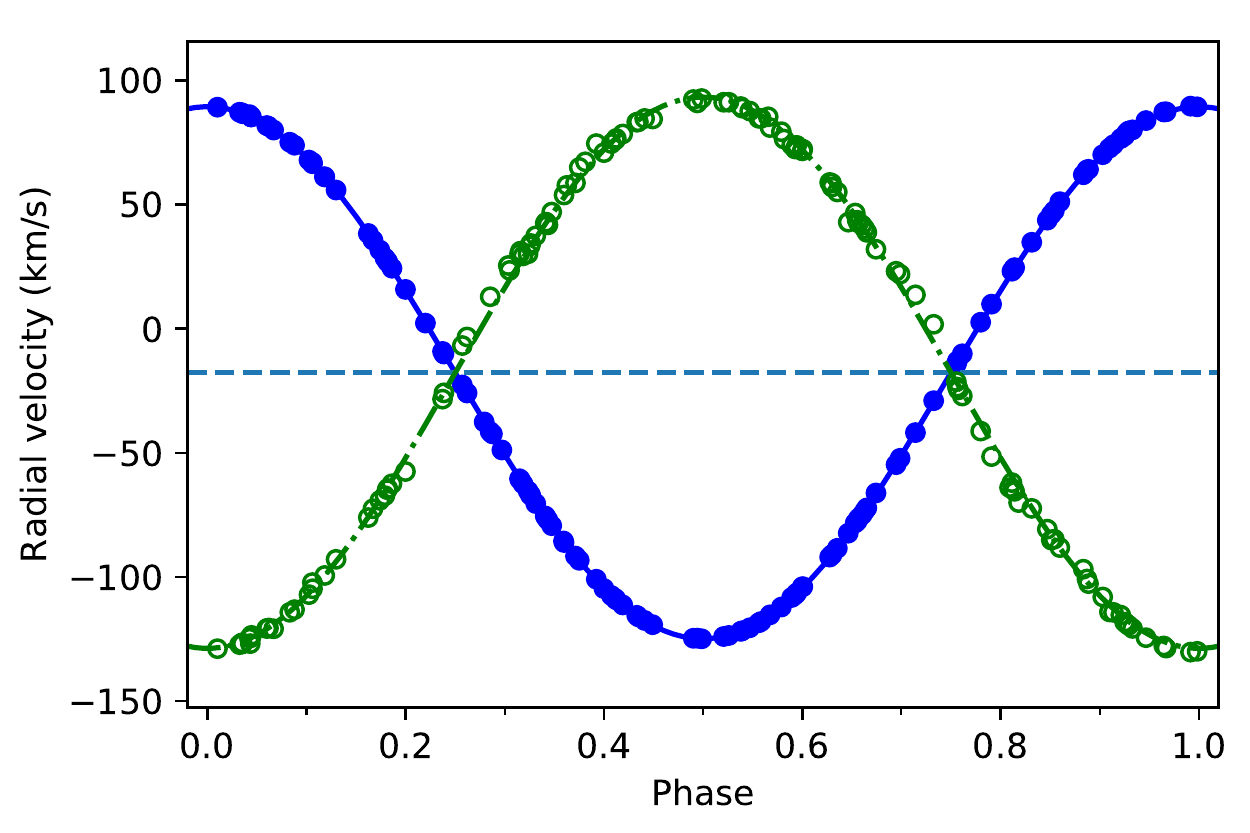}
\includegraphics[clip,angle=0,width=87mm]{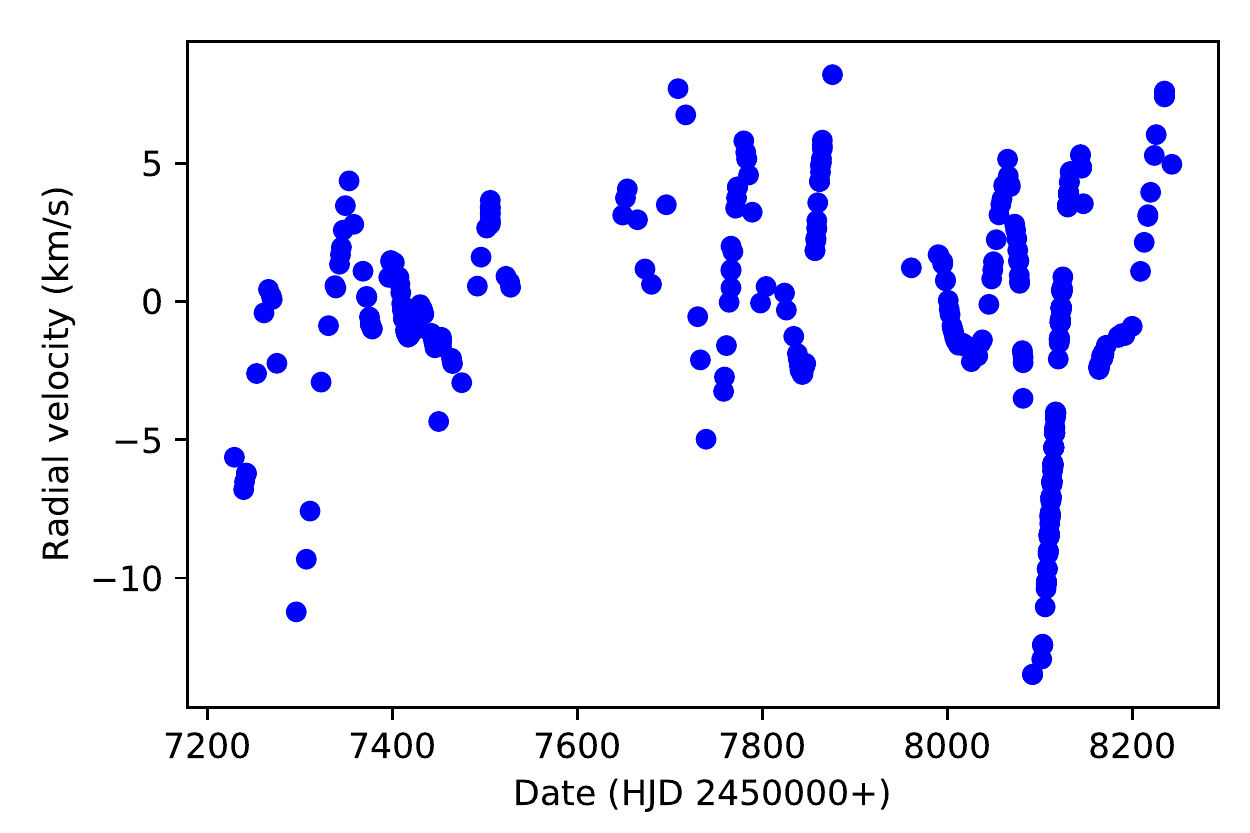}

{\bf c. \ \etaAur} \hspace{80mm} {\bf d. \ \tAur}\\
\includegraphics[clip,angle=0,width=87mm]{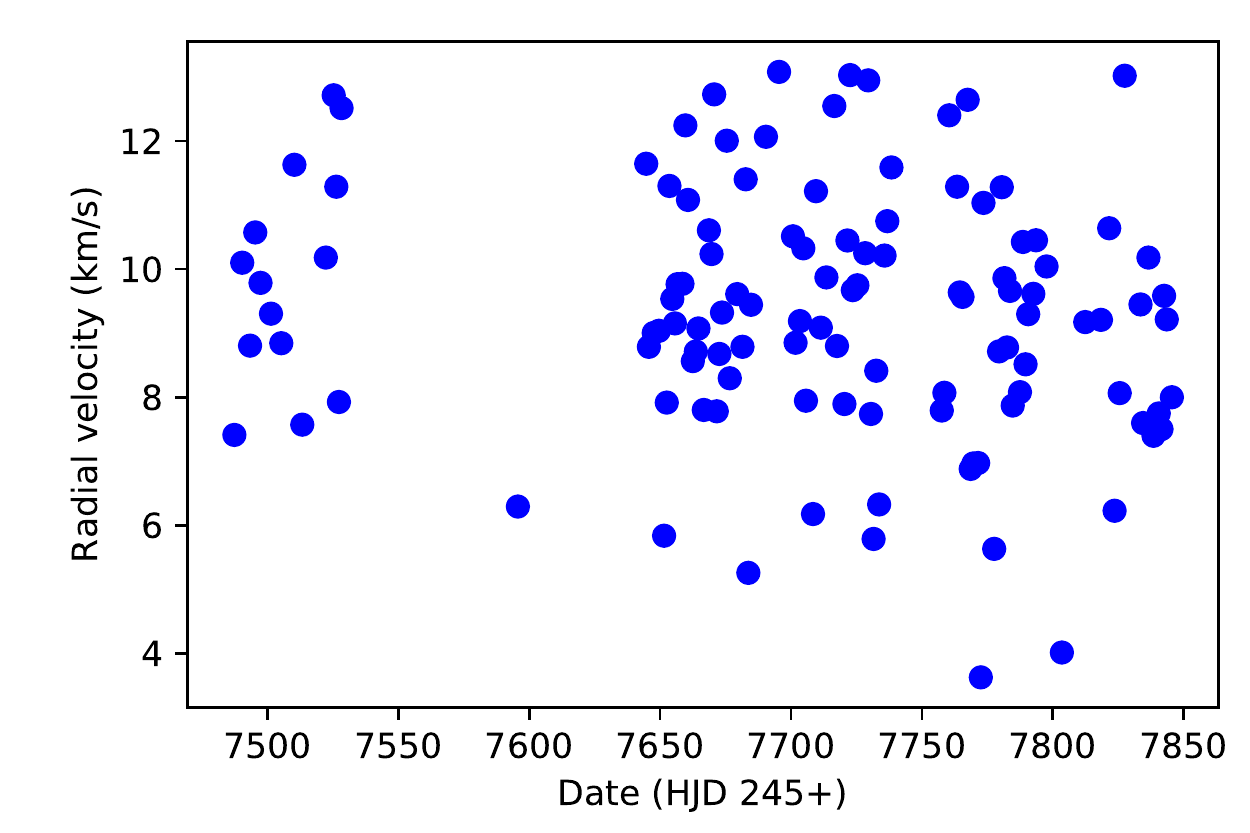}
\includegraphics[clip,angle=0,width=87mm]{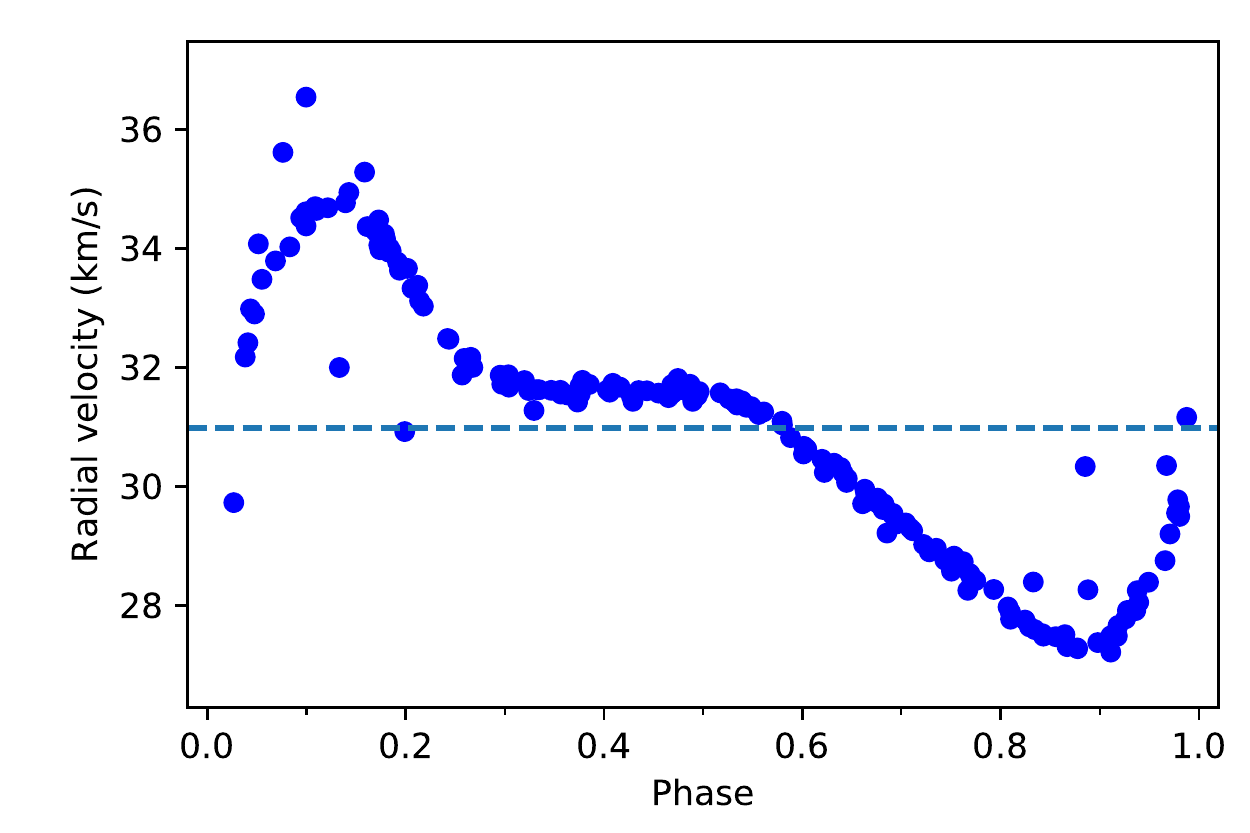}

{\bf e. \ \nAur} \hspace{80mm} {\bf f. \ \iAur}\\
\includegraphics[clip,angle=0,width=87mm]{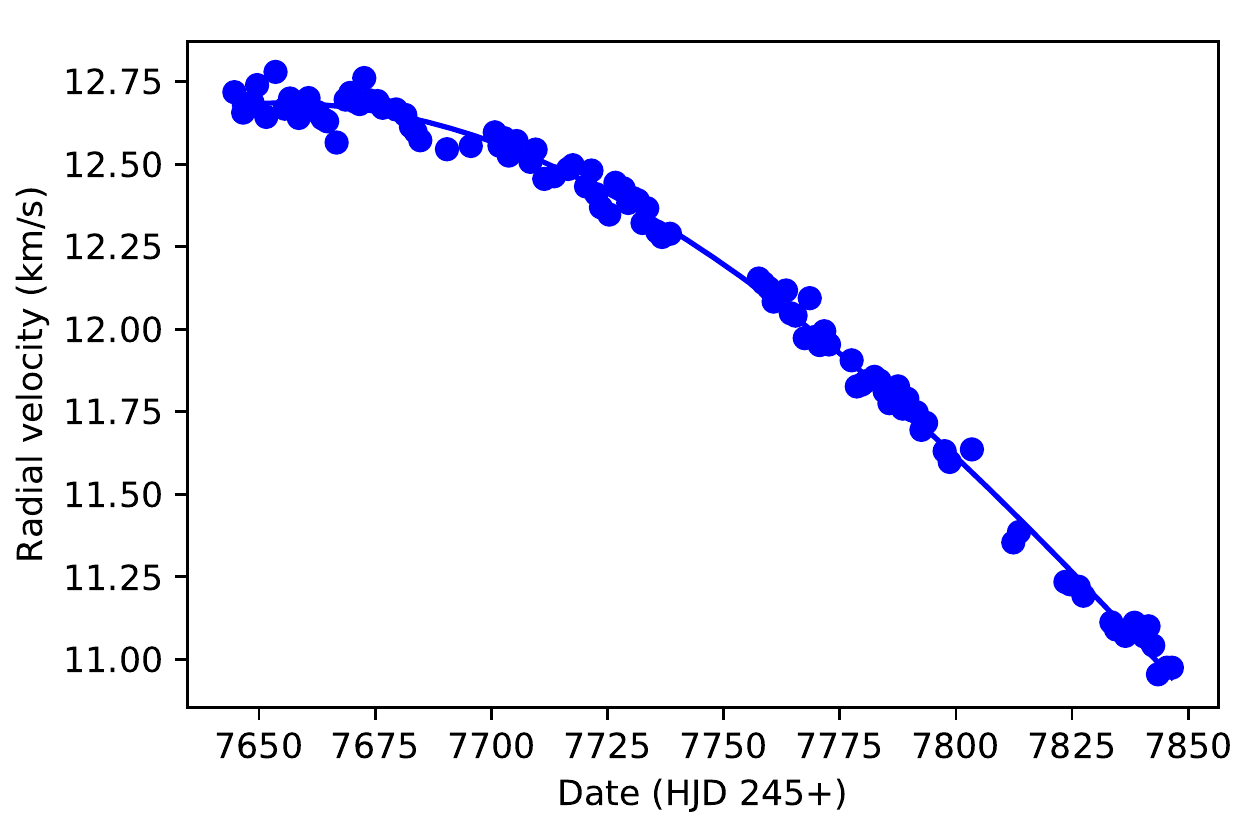}
\includegraphics[clip,angle=0,width=87mm]{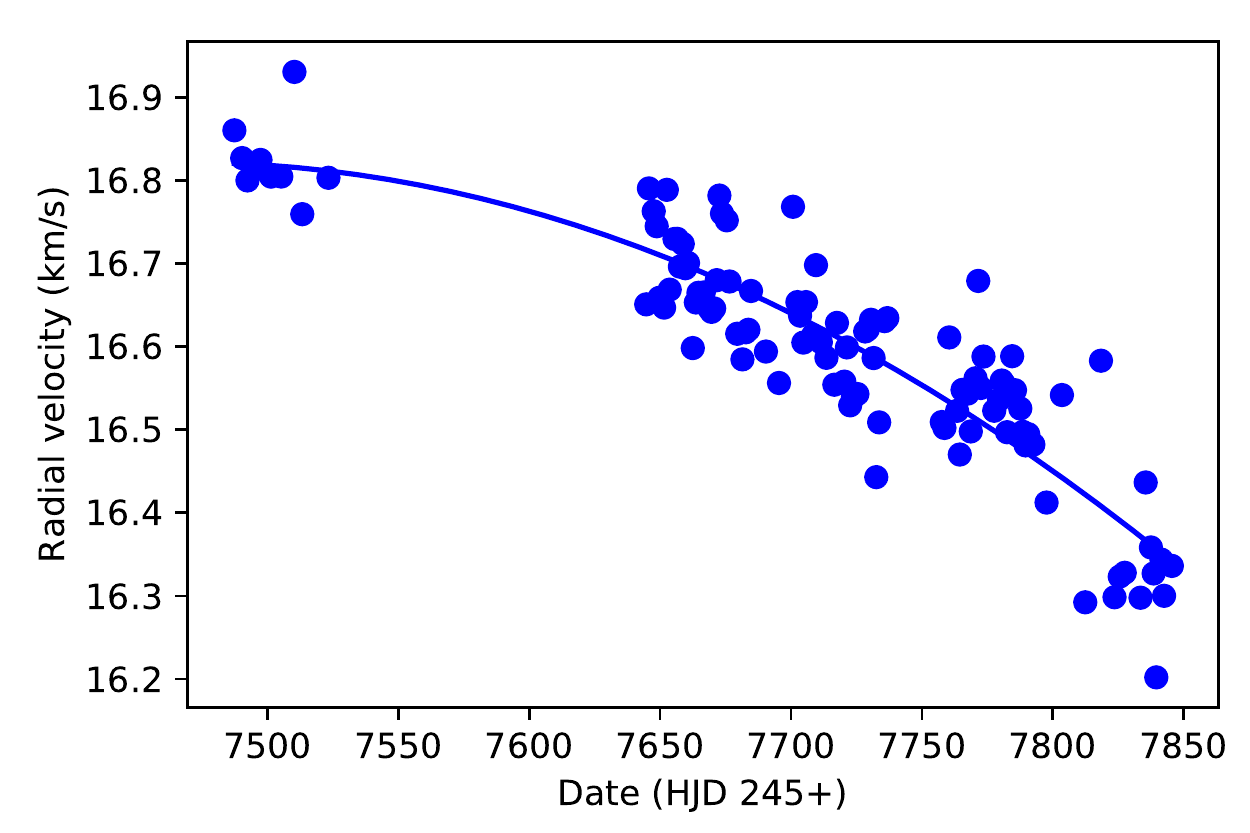}
\caption{Radial velocities of the program stars by STELLA. The data for stars with periodic variations are plotted vs. the phase. The others are plotted vs. the heliocentric Julian date. If a binary and an orbital solution were obtained then the lines represent the orbital elements. If an SB2 solution was obtained, the closed and open symbols refer to the primary and secondary component, respectively. The horizontal dashed line is the systemic velocity. }\label{Fx}\end{figure*}
\addtocounter{figure}{-1}
\begin{figure*}[tbh]
{\bf g. \ \kCet} \hspace{80mm} {\bf h. \ \HR}\\
\includegraphics[clip,angle=0,width=87mm]{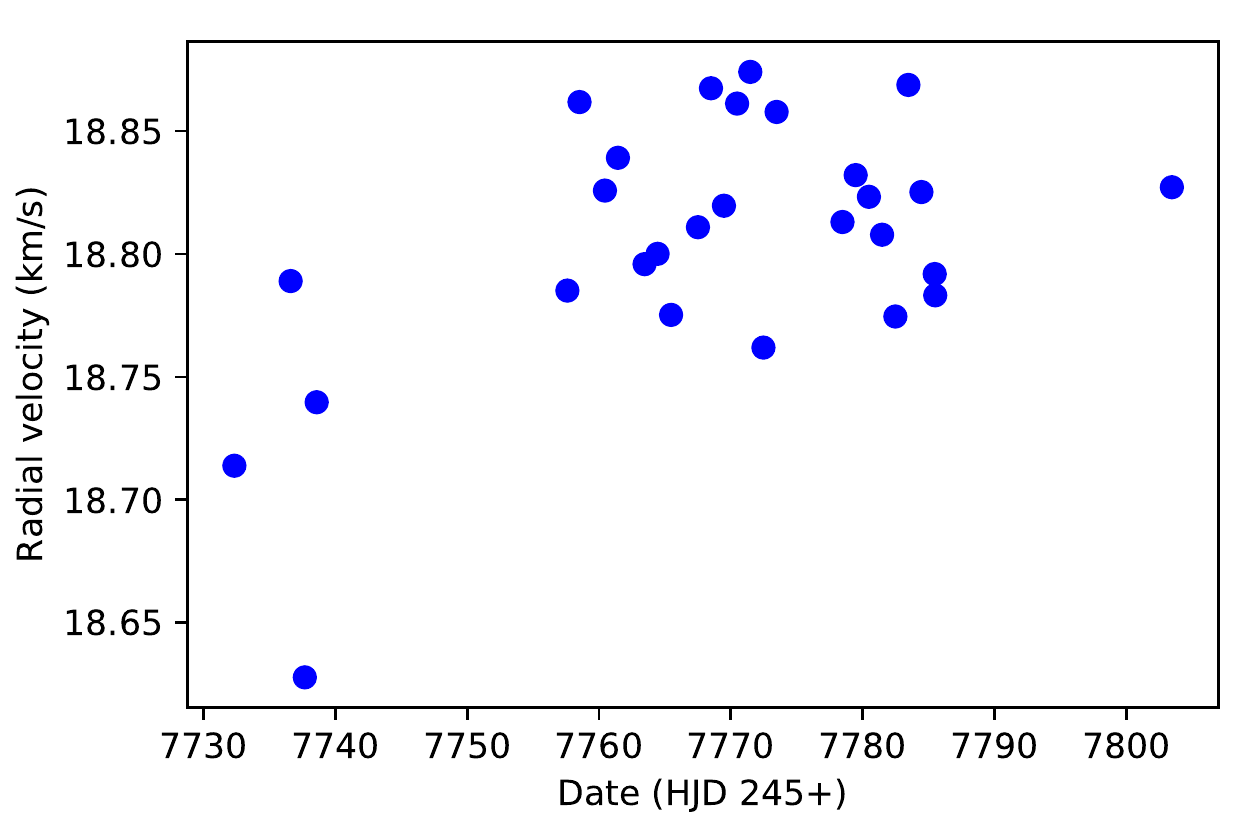}
\includegraphics[clip,angle=0,width=87mm]{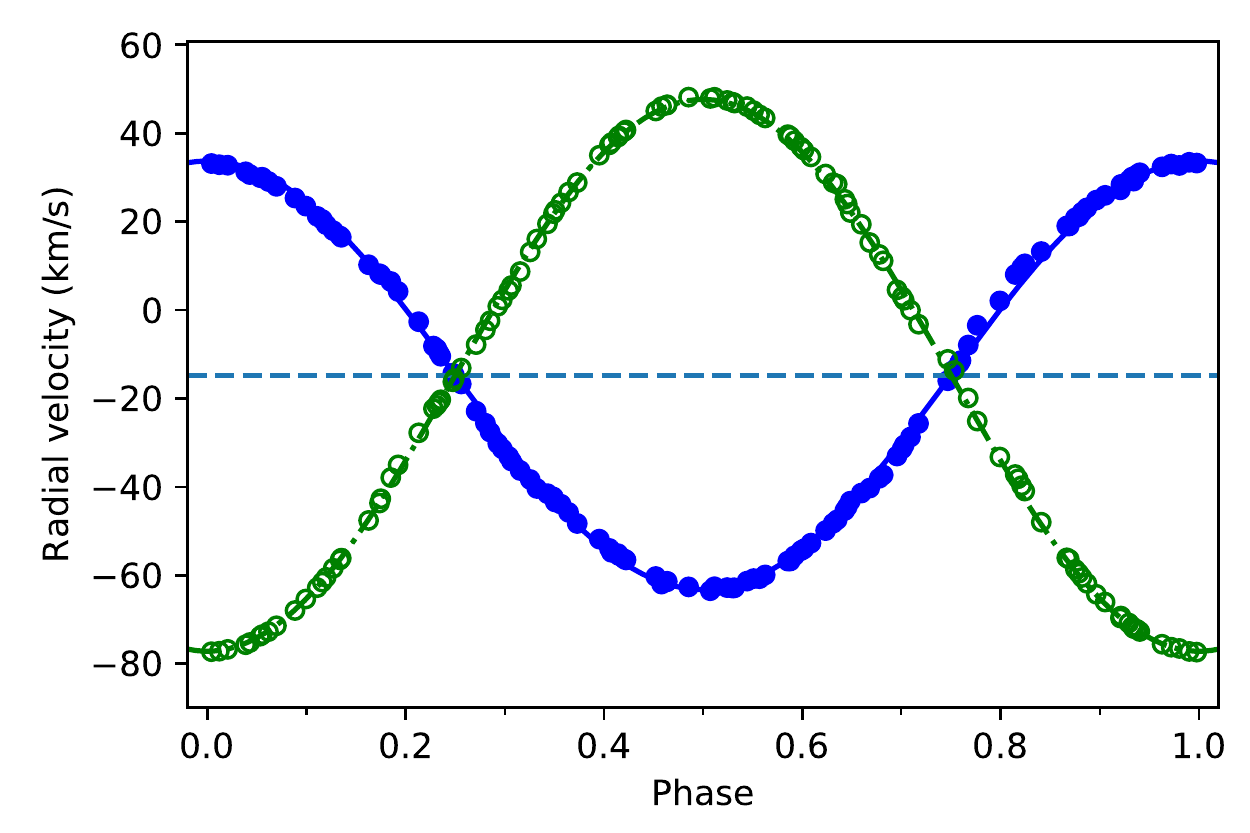}\\
\caption{(continued)}
\end{figure*}

Our new mass ratio $M_B/M_A$ is 0.96669$\pm$0.00072. With an inclination of the orbit of 137.156$\pm$0.046\,deg (Torres et al. \cite{torres}) the component masses for A and B are 2.5579$\pm$0.0013\,M$_\odot$ and 2.4727$\pm$0.0007\,M$_\odot$, respectively, not taking into account the error of the inclination. If the inclination error is added the mass errors increase to $\pm$0.0079 for A and $\pm$0.0071 for B. These masses are precise (and hopefully comparably accurate) to $\approx$0.3\%. Nevertheless, our masses are smaller by 1\%\ than the equally precise masses obtained by Torres et al. (\cite{torres}) including the astrometric data. While tiny, this value is formally more than 3$\sigma$ and thus significant. We note that above masses practically do not change if we adopt a circular  solution. The formal eccentricity is 0.000397$\pm$0.000024 with $\omega$=317\degr , and thus only half our previous value and that given by Torres et al. (\cite{torres}). The fit quality in terms of deviation per RV data point is only better by 0.5\,\ms\ for the primary, but worse by 0.3\,\ms\ for the secondary with respect to the circular solution. Table~\ref{T4} gives the eccentric solution.

\begin{table}[tbh]
\caption{Spectroscopic orbital elements for \aAur .}\label{T4}
\begin{tabular}{ll}
\hline \hline \noalign{\smallskip}
Parameter                    &  Value \\
\noalign{\smallskip} \hline \noalign{\smallskip}
$P$ (days)                   & 104.021572$\pm$0.000056 \\
$\gamma$ (\kms )             & 29.4232$\pm$0.0004 \\
$K_{A}$ (\kms )              & 25.9283$\pm$0.0019 \\
$K_{B}$ (\kms )              & 26.8219$\pm$0.0062 \\
$e$                          & 0.000397$\pm$0.000024 \\
$\omega$ ($\degr$)           & 317.0$\pm$3.5 \\
$T_0$ (HJD)                  & 2,454,381.4$\pm$1.0 \\
$a_A \sin i$ (10$^6$ km)     & 37.0878$\pm$0.0027 \\
$a_B \sin i$ (10$^6$ km)     & 38.3660$\pm$0.0089 \\
$M_A \sin^3 i$ (M$_{\sun})$  & 0.80430$\pm$0.00038 \\
$M_B \sin^3 i$ (M$_{\sun})$  & 0.77751$\pm$0.00021 \\
$N_A$, $N_B$                 & 9629, 9561 \\
rms$_A$ (\kms )              & 0.062 \\
rms$_B$ (\kms )              & 0.198 \\
\noalign{\smallskip} \hline
\end{tabular}
\tablefoot{Suffix A for primary (G8\,III), B for secondary (G0\,III). The parameter $N$ is the number of measurements used in the orbit computation; $T_0$ is the time of periastron, or ascending node for circular orbits.}
\end{table}

{\bAur }. This source was monitored for close to one year and thus allowed a precise redetermination of its orbital elements owing to its short orbital period of close to four days. Our circular SB2 fit achieved a rms per  individual RV data point of 240\,\ms\ for the primary and 2.5\,\kms\ for the secondary. The secondary star shows a significant Rossiter-McLaughlin effect contributing to its comparably large rms. Rotational broadening for both stars is comparable though; $v\sin i_1$=29.3$\pm$1.3\,\kms\ and $v\sin i_2$=29$\pm$2\,\kms . The data and the fits are shown in Fig.~\ref{Fx}a and the orbit is listed in Table~\ref{T5}. These results supersede the previous orbit from Smith (\cite{smith}) from 70 years ago.

\begin{table}[tbh]
\caption{Spectroscopic orbital elements for \bAur .}\label{T5}
\begin{tabular}{ll}
\hline \hline \noalign{\smallskip}
Parameter                    &  Value \\
\noalign{\smallskip} \hline \noalign{\smallskip}
$P$ (days)                   & $ 3.9598077 \pm 0.0000066 $ \\
$\gamma$ (\kms )             & $ -17.693 \pm 0.016 $ \\
$K_1$ (\kms )                & $ 107.202 \pm 0.020 $ \\
$K_2$ (\kms )                & $ 111.07 \pm 0.14 $ \\
$e$                          & $ 0.0 $ \\
$\omega$                     & $ \dots $ \\
$T_0$ (HJD)                  & $ 2,457,485.3084 \pm 0.0004 $ \\
$a_1 \sin i$ (10$^6$ km)     & $ 5.8373 \pm 0.0011 $ \\
$a_2 \sin i$ (10$^6$ km)     & $ 6.0478 \pm 0.0076 $ \\
$M_1 \sin^3i$ (M$_{\odot}$)  & $ 2.1708 \pm 0.0055 $ \\
$M_2 \sin^3i$ (M$_{\odot}$)  & $ 2.0952 \pm 0.0028 $ \\
$N_1$, $N_2$                 & 123, 126 \\
rms$_1$ (\kms )              & 0.24 \\
rms$_2$ (\kms )              & 2.50 \\
\noalign{\smallskip} \hline
\end{tabular}
\end{table}

{\epsAur }. We have been monitoring this F0 supergiant with STELLA since 2006, and these observations are still ongoing. Earlier data for the years 2006-2013 were presented and analyzed by Strassmeier et al. (\cite{epsaur}). The RVs in Fig.~\ref{Fx}b center on the BRITE observing window (TJD\,1640--1820; =2016/17) but also show the velocities of the adjacent seasons for comparison. At the time of the BRITE light curve \epsAur\ showed two pronounced light minima at TJD\,1680 and 1760. While the STELLA coverage did miss out on the first minimum, the second was covered and also traced a minimum in RV. A third minimum is seen in RVs with approximately the same time separation as for the BRITE light curve (74\,d), but this was not covered by BRITE. The RV amplitude changes with time and was $\approx$10\,\kms\ during the current observing period. The one pronounced brightness maximum at TJD\,1715, seen by BRITE between the two light minima, also coincides with the peak RV from STELLA. Thus, the disk-integrated radial pulsations of \epsAur\ seem to reverse when maximum or minimum light is reached, that is the star is most contracted when brightest and most expanded when faintest. It thus confirms the expectations for such pulsations (e.g., Hekker et al.~\cite{hek}). We then assembled all STELLA RVs of \epsAur\ from 12 years and searched for a period. For this, the RV contribution from the 27 yr orbit had been removed. Both the GLS and the PDM application revealed a best-fit period of 68.37938$\pm$0.00017\,d (FAP $\leq$10$^{-20}$), while the shorter RV subset shown in Fig.~\ref{Fx}b alone gave 70.68$\pm$0.07\,d and an amplitude of 18.7$\pm$0.4\,\kms .

\begin{figure}[tb]
\includegraphics[clip,angle=0,width=87mm]{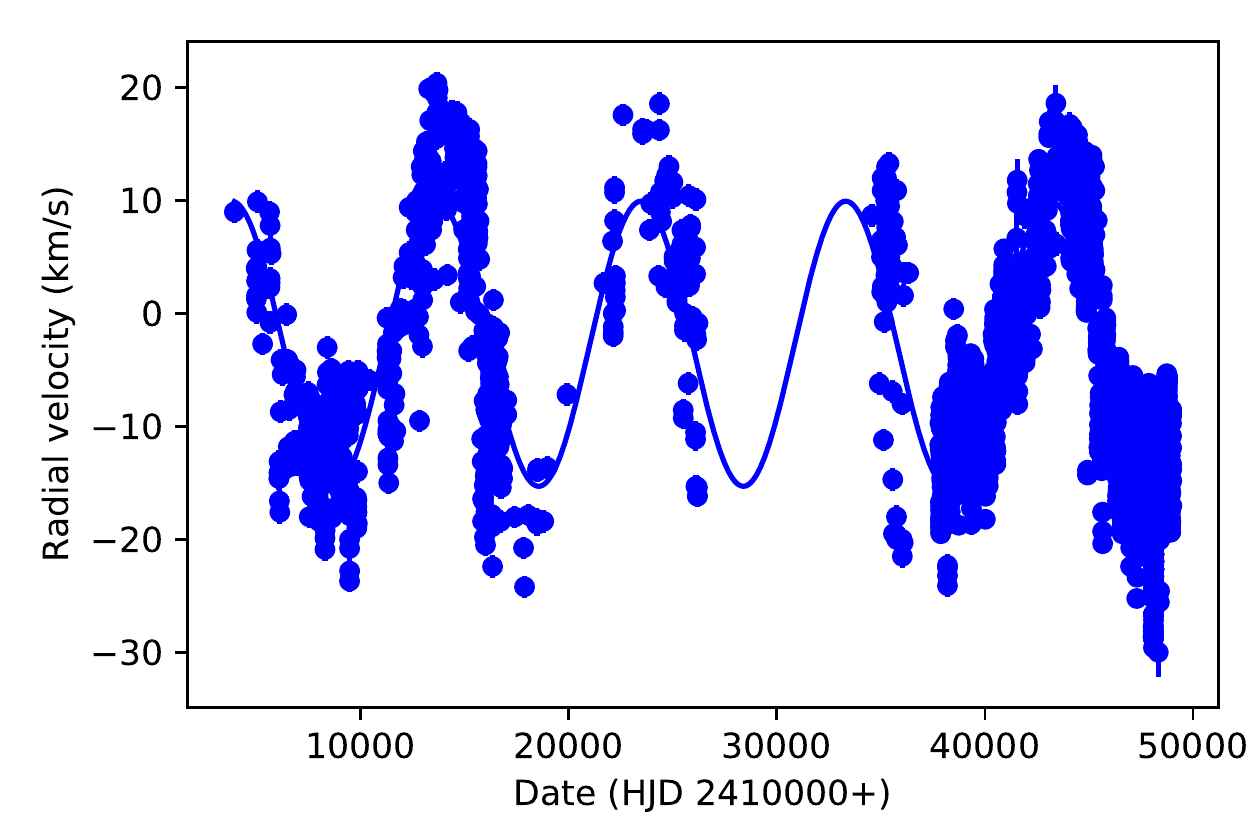}
\caption{Full set of RVs of \epsAur . Shown as dots are the data collected by Stefanik et al. (\cite{stef}) and extended by our STELLA RVs from the previous 12 years. The line is the orbital solution from Table~\ref{T6}. The time axis ranges over $\approx$130 years.}\label{F_epsorbit}
\end{figure}

Figure~\ref{F_epsorbit} shows all available RVs of \epsAur\ and presents a new orbit. The orbital solution is listed in Table~\ref{T7}. This was possible thanks to the collection of historical RVs by Stefanik et al. (\cite{stef}). We now have a total of 3034 RVs; 518 from CfA (Stefanik et al. \cite{stef}), 1714 from STELLA (see also Strassmeier et al. \cite{epsaur}), and 802 historical values dating back $\approx$130 years (individually referenced in Stefanik et al. \cite{stef}). An orbit combined with the mid times of the eclipses from photometry was given and favored by Stefanik et al. (\cite{stef}). This resulted in an orbital period longer   by 12\,d (0.12\%) compared to our pure ``Keplerian orbit'', while other parameters, such as the eccentricity and angle of the perihelion passage, agree better with the combined solution by Stefanik et al. (\cite{stef}) than with the previous pure Keplerian solutions. A trial solution of our data with the period fixed to the value from the combined solution
by  Stefanik et al. yields elements within the errors of the solution in Table~\ref{T7}. Even with the higher precision RVs from STELLA, a pure Keplerian orbit remains limited by the strong stellar RV variability due to pulsation.

\begin{table}[tb]
\caption{Spectroscopic orbital elements for \epsAur .}\label{T6}
\begin{tabular}{ll}
\hline \hline \noalign{\smallskip}
Parameter                    &  Value \\
\noalign{\smallskip} \hline \noalign{\smallskip}
$P$ (days)                  & $ 9884 \pm 10 $ \\
$\gamma$ (\kms )            & $ -2.30 \pm 0.12 $ \\
$K_{1}$ (\kms )             & $ 14.1 \pm 0.2 $ \\
$e$                         & $ 0.239 \pm 0.010 $ \\
$\omega$ ($\degr$)          & $ 37.8 \pm 2.8 $ \\
$T_0$ (HJD)                 & $ 2,434,692 \pm 70 $ \\
$a_1 \sin i$ (10$^6$ km)    & $ 1856 \pm 27 $ \\
$f(M)$ (M$_{\sun}$)         & $ 2.62 \pm 0.12 $ \\
$N$                         & 2926 \\
rms (\kms )                 & 4.7 \\
\noalign{\smallskip} \hline
\end{tabular}
\end{table}

{\etaAur }. The RVs of this source show apparently irregular values with scatter significantly larger than the expected RV uncertainties (Fig.~\ref{Fx}c). The spectral lines of the star are broadened by $v\sin i\approx 70$\,\kms\ and thus suggest fairly rapid rotation. Because \etaAur\ is a chemically peculiar star of spectral classification B3\,V (with a nominal radius of $\approx$5\,R$_\odot$), we would expect a rotation period of several days. A periodogram of the RV data shows several peaks of similar amplitude and with S/N on the verge of a detection. Intriguingly, the three strongest of those peaks occur, within the errors, at frequencies already identified with stellar pulsations from the BRITE photometry and can therefore be believed as real. In Table~\ref{TetaRV}, we list those signals with their identifications (IDs) adopted from Table~\ref{Teta}, and we also provide the phase shift $\Delta\phi$ between the RV and the photometric phase of these signals. These phase lags are similar, and therefore are also the ratios of the RV to photometric amplitude (between 0.77 and 0.93\,\kms\ per mmag). Furthermore, both of these observables are in the range of theoretically predicted amplitude ratios and phase lags of low-degree gravity modes in SPB stars  (Daszynska-Daszkiewicz \cite{das}), leaving little doubt that \etaAur\ is a new member of this class. The absence of the strongest photometric signal in the RV data may be a consequence of the type of oscillationmode seen; some may have amplitude ratios that are too small for a detection in our data. We note that the longest
photometric period that is also absent in the RV data cannot correspond to the stellar rotation period; with $R\approx 5$\,R$_\odot$ and $P = 6.543$\,d, the projected rotational velocity would be $\approx$39\,\kms , which is about half the measured $v\sin i$. The absence of any changes aside from the weak pulsation signals implies that it is likely a single star.

\begin{table}[tb]
\caption{Signals detected in the STELLA RVs of \etaAur .}\label{TetaRV}
\begin{tabular}{lllll}
\hline \hline \noalign{\smallskip}
ID & Frequency  & Period & Amplitude & $\Delta\phi$ \\
   & (d$^{-1}$) & (d)    & (\kms )    & (rad) \\
\noalign{\smallskip} \hline \noalign{\smallskip}
f2 &0.8034(3) &1.2447(5) &1.3(3) &4.2(2) \\
f3 &0.6842(4) &1.4615(8) &1.1(3) &4.9(2) \\
f4 &0.5932(3) &1.6858(9) &1.3(3) &4.6(2) \\
\noalign{\smallskip} \hline
\end{tabular}
\tablefoot{IDs are from the photometry signals in Table~\ref{Teta}. The values in parentheses are the errors expressed in last digit numbers; $\Delta\phi$ is the shift between photometric and RV phases. }
\end{table}

{\tAur }. This target is a well-known rapidly rotating CP variable. We obtain a 3.6177$\pm$0.0006 day period from the STELLA RV variations, which agrees well with the previously determined photometric rotation period of 3.6188\,d (Adelman \cite{adel}, Krticka et al. \cite{krt}). It also matches our BRITE period of 3.6189$\pm$0.0001\,d. The velocity curve in Fig.~\ref{Fx}d however appears strongly non-sinusoidal, much more than the light curve, which is atypical for rotational modulation and resembles more a typical high-eccentricity binary. The peak-to-peak RV amplitude of almost 7~\kms\ is also atypically large for rotational modulation. The shape and amplitude remained constant for the time of our observations of $\approx$200\,days. Nonetheless, the cause of the RV variations are the very asymmetric line profiles due to chemically enhanced and/or depleted star spots (see, e.g., Kochukhov et al. \cite{koch}).

{\nAur }. The K0.5\,III giant \nAur\ shows a nonlinear trend of its RVs, suggesting that it could be a long-period eccentric SB1 binary with an undetected low-mass companion. The STELLA RVs in 2017 decline over a range of 1.8\,\kms\ in $\approx$200\,d, but were 4\,\kms\ lower and constant over the same time range in 2019 when we re-observed the target. Our data from 2017 are shown in Fig.~\ref{Fx}e. Adding the eight velocities from Massarotti et al. (\cite{mas:lam}) from between 2001--2006 and three velocities from the ELODIE (Moultaka et al. \cite{elodie}) and CORAVEL (Famay et al. \cite{coravel}) archives allows an eccentric orbit prediction with the following provisional elements for guidance: $P$=7370$\pm$310\,d, $T_0$=2,457,802$\pm$10, $K_1$=2.14$\pm$0.03~\kms , $\gamma$=+10.51$\pm$0.08~\kms , $e$=0.71$\pm$0.01, $\omega$=62$\pm$4\degr, $a_1\sin i$=152$\pm$7\,10$^6$ km, and $f(m)$ = 0.0026$\pm$0.0002 with a rms per RV data point of 133\,\ms\ from $N$=253 data points. We do not give a plot of the orbit because it is too preliminary. We note that the BRITE light curve in Fig.~\ref{FnuAur}a shows a contemporaneous weak brightening until TJD\,1770 and a fading thereafter. The period of 19.16\,d from the residual BRITE data after removing the trend is not seen in the STELLA RVs. It would be too short for the rotation period of this otherwise (magnetically) inactive K0.5 giant. Clearly, not only is a longer base line required for determination of a real orbit, but a more frequent sampling as well to fully exclude nonradial pulsations as the cause of the variability.

{\iAur }. This star is a hybrid giant of classification K3\,II-III and shows RVs with a similar nonlinear trend as \nAur\ but even lower in amplitude. Our RV data cover two observing seasons 2016/17 and 2019; the 2016/17 season is shown in Fig.~\ref{Fx}f. Its RVs decreased by 0.6\,\kms\ within the observed $\approx$200\,d in 2016/17, but increased by 0.1\,\kms\ within $\approx$110\,d in 2019, which suggests a period of $\approx$4~yrs (1450\,d). This is comparable with the period of 1586\,d listed by Hekker et al. (\cite{hek}) from their long-term RVs at Lick Observatory. On the short cadence, we note that the seasonal STELLA RV residuals are $\approx$75\,\ms\ and thus too high for a star with $v\sin i$ of 7.5\,\kms\ (we expect half smaller rms). A periodogram of the residual RVs after removing the trend shows several peaks; the strongest is also the shortest with $P=0.90$\,d and a least-squares amplitude of 64\,\ms\ with a rms of approximately the same amount. This could be the signature of nonradial pulsations and makes the long-period, low-mass companion scenario adopted by Hekker et al. (\cite{hek}) questionable. At this point we were provided with new RV data from the Lick Survey from 2001--2012 by Sabine Reffert from the Landessternwarte Heidelberg (a coauthor of the Hekker et al. paper), which clearly showed the semi-periodic character of its RVs. We thus confirm the suggestion of Reffert (2020, priv. comm.) that the star is not the host of a low-mass stellar companion but an oscillating star. Like for \nAur , the BRITE period of 9.07\,d is not seen in the STELLA RVs (nor the Lick RVs) and would be too short for the rotation period of a large giant.

{\kCet }. No periodicity could be found in the RV data shown in Fig.~\ref{Fx}g. We note that the one RV data point at HJD\,2,457,737 that is lower by about 0.1\,\kms\ than the rest of the data is the one spectrum with the lowest S/N of all. STELLA likely caught a cloud passage during its integration. If removed and the other points given equal weight, the rms scatter is 35\,\ms , which is what we expect from STELLA. \kCet\ is thus a single star and its photometric period indicates its rotation.

\begin{table}[tbh]
\caption{Spectroscopic orbital elements for \HR .}\label{T7}
\begin{tabular}{ll}
\hline \hline \noalign{\smallskip}
Parameter                    &  Value \\
\noalign{\smallskip} \hline \noalign{\smallskip}
$P$ (days)                   & $ 2.837711 \pm 0.000066 $ \\
$\gamma$ (\kms )             & $ -14.76 \pm 0.07 $ \\
$K_1$ (\kms )                & $ 48.47 \pm 0.13 $ \\
$K_2$ (\kms )                & $ 62.41 \pm 0.20 $ \\
$e$                          & $ 0.0 $ \\
$\omega$                     & $ \dots $ \\
$T_0$ (HJD)                  & $ 2,457,729.7084 \pm 0.0017 $ \\
$a_1 \sin i$ (10$^6$ km)     & $ 1.8915 \pm 0.0050 $ \\
$a_2 \sin i$ (10$^6$ km)     & $ 2.4355 \pm 0.0078 $ \\
$M_1 \sin^3i$ (M$_{\odot}$)  & $ 0.2256 \pm 0.0016 $ \\
$M_2 \sin^3i$ (M$_{\odot}$)  & $ 0.1752 \pm 0.0011 $ \\
$N_1$, $N_2$                 & 119, 118 \\
rms$_1$ (\kms )              & 1.22 \\
rms$_2$ (\kms )              & 0.84 \\
\noalign{\smallskip} \hline
\end{tabular}
\end{table}

{\HR\ = HR\,1099}. The RVs for this SB2 RS\,CVn binary are rotationally modulated owing to large starspots on the K-subgiant primary (see, e.g., Lanza et al. \cite{lan} and references therein). The rotation period of this source is synchronized to the orbital period within the typical measurement errors. Previous RVs revealed a long-term sinusoidal modulation of the orbital period with 36.3$\pm$1.9\,yr (Muneer et al. \cite{mun}) and it is thus interesting to repeat orbital solutions from time to time. A new SB2 solution is presented in Table~\ref{T7} from the present STELLA data. The solution is also shown in graphical form in Fig.~\ref{Fx}h. Our orbit takes into account the RV modulations due to spot jitter. The primary star induces systematic RV variations of up to 2.0$\pm$0.2\,\kms\ with a main period of 0.945\,d. A periodogram for the secondary-star RV residuals also identified a peak at a period of below one day (0.567\,d), but with an even smaller amplitude of 1.3$\pm$0.2\,\kms . Both of these periods are just apparent periods due to respective rotational modulation from multiple spots crossing the disk and emerge and fade away at the same time. We removed this RV jitter by simply convolving the residuals with a smoothing function (Blackman function) and then subtracting this function from the RV data before solving for the orbital elements. The data in Fig.~\ref{Fx}h are the spot-uncorrected RVs.

\section{Summary and conclusions}\label{S5}

Time series photometry with the BRITE satellites and contemporaneous high-resolution STELLA spectroscopy of 12 bright stars was presented and analyzed as part of a search for rotation periods in evolved stars. We had selected a field in the constellation Auriga (termed 20-AurPer-I-2016 in BRITE jargon) with a total of 9 bright stars. Three more targets from other BRITE fields were later added to the sample. These targets exhibit a variety of phenomena: rotational modulation, radial and nonradial pulsations, and eclipses. Our analysis focused on finding the periodicity from the light curves together with RVs from the ground-based spectra. We present the first space-based photometry of famous Capella. It was found to be strikingly constant in the red bandpass with an rms over 176\,d of just 1\,mmag, but showed a 10.1$\pm$0.6\,d periodicity in the blue-bandpass light curve, which we interpret to be the rotation period of its active and hotter secondary star. From ground-based \Halpha\ photometry we had found earlier that the G8\,III primary rotates synchronously, while the rapidly rotating G0\,III secondary showed evidence for a $\approx$8.6\,d period (Strassmeier et al.~\cite{cap1}). Because the difference of the two secondary-star periods, ground-based versus space-based, is significant, and because the BRITE period is of very low amplitude of just 0.9\,mmag, we consider the 10.1 d period still preliminary.

We also present a new orbital solution for the two components of Capella from 12.9 yr of high-precision STELLA RV data. This orbit is based on an outstanding collection of 9600 echelle spectra from the STELLA robot and, together with the inclination of the orbital plane from Torres et al. (\cite{torres}), resulted in giant-star masses with a precision of $\approx$0.3\%. The individual masses are 2.5579$\pm$0.0079\,M$_\odot$ and 2.4727$\pm$0.0071\,M$_\odot$ for the primary and secondary, respectively, with the errors dominated by the error of the inclination of the orbital plane. These masses are 1\% smaller than those derived by Torres et al. (\cite{torres}) from a combined spectroscopic and astrometric solution. Together with the Torres et al. (\cite{torres}) analysis, the new solution for Capella is the most precise orbit for any giant star in the sky. The present position of the G0 component in the Hertzsprung gap suggests ongoing changes in its internal structure. We expect that this has a profound impact on the visible surface and can explain its fast rotation. Precise rotation periods of stars in this evolutionary stage may thus help to further understand the angular-momentum loss in late-type stars.

Results for the other targets included the detection of the main pulsation period of the F0 supergiant \epsAur\ from a multi-harmonic fit of the light curve of 152\,d. This is noteworthy because the STELLA RVs revealed a clear 68 d period from 12 years of high-precision data. Although the light curve showed two minima separated by 74\,d, a single period of that duration would not fit the data adequately. The contemporaneous RVs indicated that the (stellar) disk-integrated pulsations seem to revert when maximum or minimum light is reached, that is, the star is apparently most contracted when brightest and most expanded when faintest. We also combined our STELLA RVs with all previously published RVs and presented a new orbital solution with a period of 9884$\pm$10\,d. This now includes RV data over an epoch of $\approx$130 years. The main limitation for the orbit is the pulsation of the F0 supergiant.

An ingress of an eclipse of the \zAur\ binary system was covered with BRITE and a precise timing for its eclipse onset at HJD 2,457,810.0$\pm$0.4 derived. A possible 70 d period from the outside-eclipse light fits the proposed tidal-induced, nonradial pulsations of this ellipsoidal K4 supergiant. \etaAur\ is identified as a SPB star with a main period of 1.289$\pm$0.001\,d. Five more periods are seen in the photometry and three of these are also seen in the RV data. The amplitude ratios as well as the phase lags between photometric and RV periods are in agreement with expected low-degree gravity modes of SPB stars. Therefore, we interpret these to be due to nonradial pulsations. \etaAur\ appears to be among the brightest SPB stars discovered so far. The rotation period of the famous magnetic Ap star \tAur\ is easily seen from the photometry, but also from the spectroscopy with a period of 3.6189$\pm$0.0001\,d and 3.6177$\pm$0.0006\,d, respectively. The RVs of this star show a striking non-sinusoidal shape with a large amplitude of 7\,\kms , which is likely due to the line-profile deformations from the inhomogeneous surface distribution of its chemical elements. Such a non-sinusoidal shape may explain the small period difference and suggest that the two periods are in agreement within its errors. Photometric rotation periods are also confirmed for the magnetic Ap star \iqAur\ of 2.463\,d and for the solar-type star \kCet\ of 9.065\,d, and also for the B7 HgMn giant \bTau\ of 2.74\,d. The latter period remains uncertain because it was reconstructed only from the GLS periodogram and with the very small amplitude of 0.54\,mmag. Revised orbital solutions are derived for the eclipsing SB2 binary \bAur , which replaces the initial orbit from Smith (\cite{smith}), and for the RS~CVn binary \HR\ for which a spot-corrected orbital solution was achieved. The two K giants \nAur\ and \iAur\ are found with long-term trends in both the light curve and the RVs. \nAur\ could be a long-period eccentric SB1 system with a low-mass companion for which we predict a provisional orbital solution with a period of 20\,yr and an eccentricity of 0.7. The RV variations of the hybrid giant \iAur\ are of even lower amplitude (0.7\,\kms) but shorter period ($\approx$4\,yrs) and are more likely due to surface oscillations. Long-term brightness trends were seen for both stars and appear related with the RVs.

\begin{acknowledgements}
STELLA was made possible by funding through the State of Brandenburg (MWFK) and the German Federal Ministry of Education and Research (BMBF). The facility is a collaboration of the AIP in Brandenburg with the IAC in Tenerife. This work has made use of the VALD database, operated at Uppsala University, the Institute of Astronomy RAS in Moscow, and the University of Vienna as well as of NASA's Astrophysics Data System and of CDS's Simbad database which we both gracefully acknowledge. APi acknowledges support from the National Science Centre (NCN) grant 2016/21/B/ST9/01126. APo was responsible for image processing and automation of photometric routines of BRITE nanosatellite data and was supported by Silesian University of Technology: Higher Education Funding for Statutory Activities and Rector Grant 02/140/RGJ20/0001. AFJM thanks NSERC (Canada) for financial aid. GAW acknowledges support from the Natural Sciences and Engineering Research Council (NSERC) of Canada in the form of a Discovery Grant. The Canadian BRITE team acknowledges mission support from the Director General (Space) of the Canadian Department of National Defence. GH acknowledges financial support by the Polish NCN grant 2015/18/A/ST9/00578.
\end{acknowledgements}

\appendix

\section{Orbit of Capella}

\begin{figure*}[htb]
\center
\includegraphics[angle=0,width=165mm,clip]{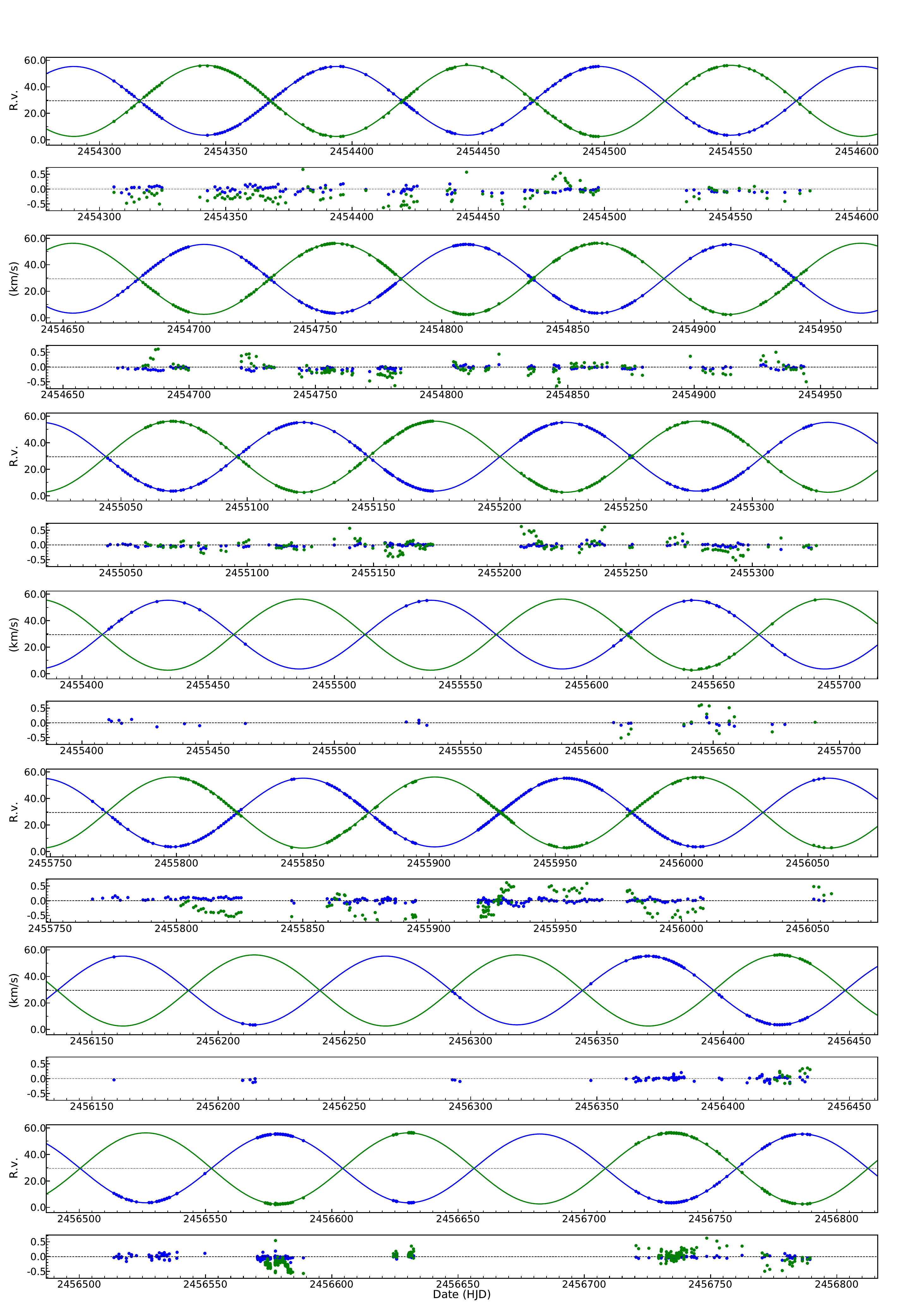}
\caption{Capella RVs from STELLA compared with our newly computed orbit as a function of time. Points of different color denote the cool and hot component, respectively (blue = primary, green = secondary). The lower of the respective subpanels show the residuals from the orbital fit, same symbols. The residuals in this plot include the systematic errors that were corrected for in the final solution. All panels show RV vs. HJD. }
 \label{FAorbcap}
\end{figure*}

\setcounter{figure}{0}
\begin{figure*}[htb]
\center
\includegraphics[angle=0,width=165mm,clip]{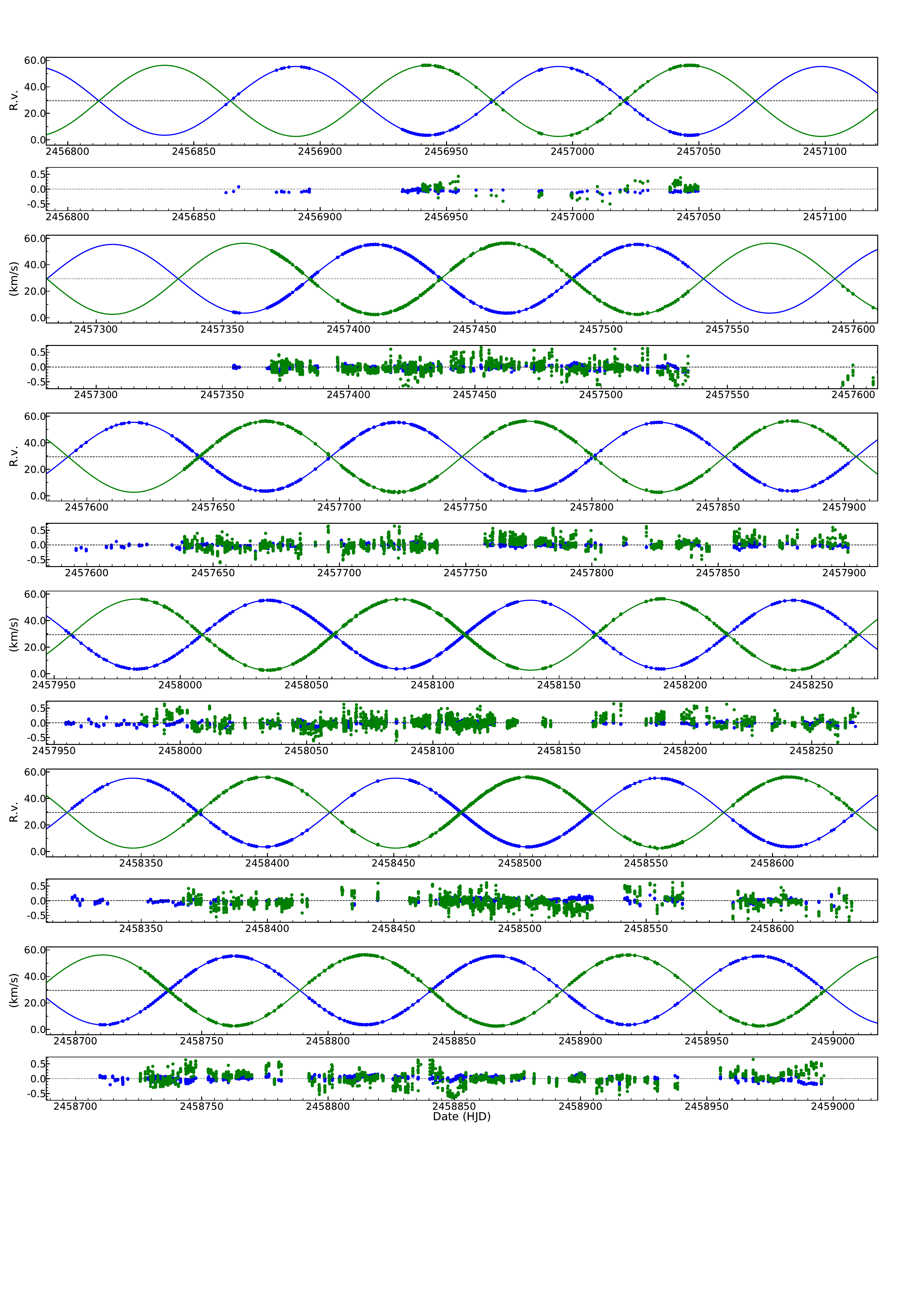}
\caption{(continued).}
\end{figure*}

\end{document}